\documentclass[12pt,twoside,a4paper,normalheadings,headsepline]{scrbook}

\usepackage{axocolor}
\usepackage{graphicx}
\usepackage{amsmath,amssymb}
\usepackage{pennames}
\usepackage{bibspacing}
\usepackage{multicol}
\usepackage{tdr_master}
\usepackage{tdrdefinitions}

\newcommand{\href}[2]{\hspace*{-3mm}{#2}}

\begin{document}

\setcounter{part}{6}
\setcounter{chapter}{0}
\setcounter{secnumdepth}{3}
\setcounter{tocdepth}{2}

\thispagestyle{empty}
\pagestyle{empty}
\begin{picture}(150,180)\unitlength 1mm

{\sfb
\put(0,40){\Huge TESLA}
\put(0,15){\begin{minipage}{14cm}
       \LARGE\bf
        \begin{flushleft}{\sfb\LARGE
        The Superconducting
        Electron Positron \\Linear Collider
        with an Integrated\\
        X-Ray Laser Laboratory\\
        }\end{flushleft}\end{minipage}}
\put(0,-10){\Huge Technical Design Report}
\put(5,-25){\LARGE Part VI: Appendices}
\put(5,-37){\LARGE Chapter 1: Photon Collider at TESLA }
\put(15,-130){\includegraphics[height=8cm]{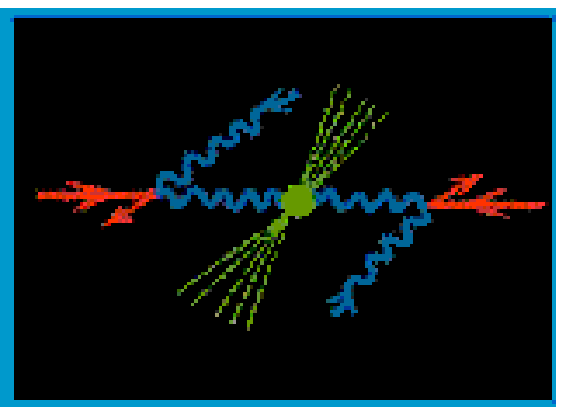}}
\put(-5,-150){{\sfb\bf\large DESY-2001-011, ECFA-2001-209}}
\put(-5,-155){{\sfb\bf\large TESLA-2001-23, TESLA-FEL-2001-05}}

\put(131,-150){\sfb \large March}
\put(135,-155){\sfb \large 2001}}
\end{picture}

\newpage

\thispagestyle{empty}

\begin{picture}(150,180)\unitlength 1mm
  \put(-10,-65){\includegraphics[height=2.5cm]{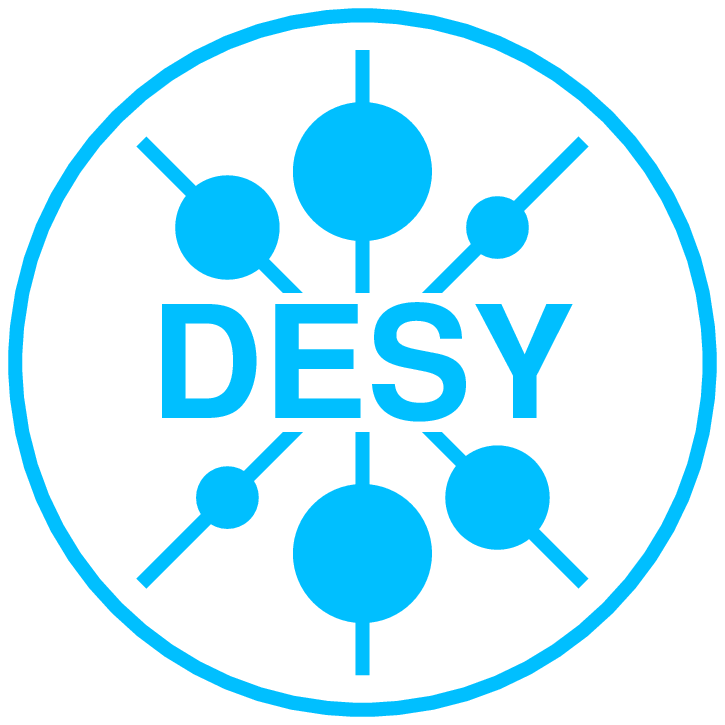}}
  \put(-10,-120){\begin{minipage}{10cm}{
   {\bf\noindent Publisher:} \\
   \noindent DESY\\
   \noindent Deutsches Elektronen-Synchroton\\
   \noindent Notkestra{\ss}e 85, D-22607 Hamburg\\
   \noindent Germany\\
   \noindent \url{http://www.desy.de}\\
   \noindent E-mail: desyinfo@desy.de\\

   \noindent Member of the Hermann von Helmholtz Association\\
   \noindent of National Research Centers (HGF)\\

   \noindent Reproduction including extracts is permitted \\
   \noindent subject to crediting the source.\\


   {\bf \noindent Copy deadline:} March 2001 \\
   ISBN 3-935702-00-0\\
   ISSN 0418-9833
}
\end{minipage}}

\end{picture}



\cleardoublepage

\pagestyle{empty}
\begin{picture}(150,180)\unitlength 1mm

{\sfb
\put(0,55){\Huge TESLA}

\put(0,40){\Huge Technical Design Report}
}

{\large
\put(-5,15)   {{\bf\Large PART I: Executive Summary }}
      \put(25,7){Editors: \begin{minipage}[t]{10cm}{F.Richard, J.R.Schneider, D.Trines, A.Wagner}\end{minipage}}
\put(-5,-5) {{\bf \Large PART II: The Accelerator}}
      \put(25,-13){Editors: \begin{minipage}[t]{9cm}{R.Brinkmann,
      K.Fl\"ottmann, J.Rossbach,\hfil P.Schm\"user, N.Walker, H.Weise}\end{minipage}}
\put(-5,-25) {{\bf \Large PART III: Physics at an e$^+$e$^-$ Linear Collider}}
      \put(25,-33){Editors: \begin{minipage}[t]{10cm}{R.D.Heuer,
      D.Miller, F.Richard, P.Zerwas}\end{minipage} }
\put(-5,-45) {{\bf \Large PART IV: A Detector for TESLA}}
      \put(25,-53){Editors: \begin{minipage}[t]{10cm}{T.Behnke, S.Bertolucci,
      R.D.Heuer, R.Settles }\end{minipage}}
\put(-5,-65) {{\bf \Large PART V: The X-Ray Free Electron Laser Laboratory}}
      \put(25,-73){Editors:
      \begin{minipage}[t]{10cm}{G.Materlik, T.Tschentscher}\end{minipage}}
\put(-5,-85) {{\bf \Large PART VI: Appendices}}
      \put(25,-93){Editors: \begin{minipage}[t]{10cm}{
        R.Klanner\\
        Chapter 1: V.Telnov\\
        Chapter 2: U.Katz, M.Klein, A.Levy\\
        Chapter 3: R.Kaiser, W.D.Nowak\\
        Chapter 4: E.DeSanctis, J.-M.Laget, K.Rith
}\end{minipage}}
}
\end{picture}

\clearpage

\begin{picture}(150,180)\unitlength 1mm

  \put(0,30){\sfb \LARGE Part VI}
\put(0,15) {\sfb \LARGE Chapter 1: Photon Collider at TESLA }

  \put(0,0){\sfb \Large Editor: V. Telnov}

\end{picture}

\clearpage

\pagestyle{headings}
\pagenumbering{arabic}
\setcounter{page}{1}
\setcounter{tocdepth}{2}

\chapter{The Photon Collider at TESLA}
\pagestyle{headings}
\vspace*{-5mm}
%
%
\vspace*{-2mm}
\begin{flushleft}
\noindent 
B.~Badelek$^{43}$$^{ }$,
C.~Bl\"{o}chinger$^{44}$$^{ }$,
J.~Bl\"{u}mlein$^{12}$$^{ }$,
E.~Boos$^{28}$$^{ }$,
R.~Brinkmann$^{12}$$^{ }$,
H.~Burkhardt$^{11}$$^{ }$,
P.~Bussey$^{17}$$^{ }$,
C.~Carimalo$^{33}$$^{ }$,
J.~Chyla$^{34}$$^{ }$,
A.K.~\c{C}ift\c{c}i$^{4}$$^{ }$,
W.~Decking$^{12}$$^{ }$,
A.~De~Roeck$^{11}$$^{ }$,
V.~Fadin$^{10}$$^{ }$,
M.~Ferrario$^{15}$$^{ }$,
A.~Finch$^{24}$$^{ }$,
H.~Fraas$^{44}$$^{ }$,
F.~Franke$^{44}$$^{ }$,
M.~Galynskii$^{27}$$^{ }$,
A.~Gamp$^{12}$$^{ }$,
I.~Ginzburg$^{31}$$^{ }$,
R.~Godbole$^{6}$$^{ }$,
D.S.~Gorbunov$^{28}$$^{ }$,
G.~Gounaris$^{39}$$^{ }$,
K.~Hagiwara$^{22}$$^{ }$,
L.~Han$^{19}$$^{ }$,
R.-D.~Heuer$^{18}$$^{ }$,
C.~Heusch$^{36}$$^{ }$,
J.~Illana$^{12}$$^{ }$,
V.~Ilyin$^{28}$$^{ }$,
P.~Jankowski$^{43}$$^{ }$,
Yi~Jiang$^{19}$$^{ }$,
G.~Jikia$^{16}$$^{ }$,
L.~J\"{o}nsson$^{26}$$^{ }$,
M.~Kalachnikow$^{8}$$^{ }$,
F.~Kapusta$^{33}$$^{ }$,
R.~Klanner$^{12,18}$$^{ }$,
M.~Klasen$^{12}$$^{ }$,
K.~Kobayashi$^{41}$$^{ }$,
T.~Kon$^{40}$$^{ }$,
G.~Kotkin$^{30}$$^{ }$,
M.~Kr\"{a}mer$^{14}$$^{ }$,
M.~Krawczyk$^{43}$$^{ }$,
Y.P.~Kuang$^{7}$$^{ }$,
E.~Kuraev$^{13}$$^{ }$,
J.~Kwiecinski$^{23}$$^{ }$,
M.~Leenen$^{12}$$^{ }$,
M.~Levchuk$^{27}$$^{ }$,
W.F.~Ma$^{19}$$^{ }$,
H.~Martyn$^{1}$$^{ }$,
T.~Mayer$^{44}$$^{ }$,
M.~Melles$^{35}$$^{ }$,
D.J~Miller$^{25}$$^{ }$,
S.~Mtingwa$^{29}$$^{ }$,
M.~M\"{u}hlleitner$^{12}$$^{ }$,
B.~Muryn$^{23}$$^{ }$,
P.V.~Nickles$^{8}$$^{ }$,
R.~Orava$^{20}$$^{ }$,
G.~Pancheri$^{15}$$^{ }$,
A.~Penin$^{12}$$^{ }$,
A.~Potylitsyn$^{42}$$^{ }$,
P.~Poulose$^{6}$$^{ }$,
T.~Quast$^{8}$$^{ }$,
P.~Raimondi$^{37}$$^{ }$,
H.~Redlin$^{8}$$^{ }$,
F.~Richard$^{32}$$^{ }$,
S.D.~Rindani$^{2}$$^{ }$,
T.~Rizzo$^{37}$$^{ }$,
E.~Saldin$^{12}$$^{ }$,
W.~Sandner$^{8}$$^{ }$,
H.~Sch\"{o}nnagel$^{8}$$^{ }$,
E.~Schneidmiller$^{12}$$^{ }$,
H.J.~Schreiber$^{12}$$^{ }$,
S.~Schreiber$^{12}$$^{ }$,
K.P.~Sch\"{u}ler$^{12}$$^{ }$,
V.~Serbo$^{30}$$^{ }$,
A.~Seryi$^{37}$$^{ }$,
R.~Shanidze$^{38}$$^{ }$,
W.~Da~Silva$^{33}$$^{ }$,
S.~S\"{o}ldner-Rembold$^{11}$$^{ }$,
M.~Spira$^{35}$$^{ }$,
A.M.~Stasto$^{23}$$^{ }$,
S.~Sultansoy$^{5}$$^{ }$,
T.~Takahashi$^{21}$$^{ }$,
V.~Telnov$^{10,12}$$^{ }$,
A.~Tkabladze$^{12}$$^{ }$,
D.~Trines$^{12}$$^{ }$,
A.~Undrus$^{9}$$^{ }$,
A.~Wagner$^{12}$$^{ }$,
N.~Walker$^{12}$$^{ }$,
I.~Watanabe$^{3}$$^{ }$,
T.~Wengler$^{11}$$^{ }$,
I.~Will$^{8,12}$$^{ }$,
S.~Wipf$^{12}$$^{ }$,
\"{O}.~Yava\c{s}$^{4}$$^{ }$,
K.~Yokoya$^{22}$$^{ }$,
M.~Yurkov$^{12}$$^{ }$,
A.F.~Zarnecki$^{43}$$^{ }$,
P.~Zerwas$^{12}$$^{ }$,
F.~Zomer$^{32}$$^{ }$.
\end{flushleft}
%
%
\vspace*{-5mm}
\bigskip 
{\footnotesize {\begin{multicols}{2} 
\noindent 
 $ \mbox{\,} $ $ ^{1}$ \begin{minipage}[t]{7cm} RWTH Aachen, Germany \end{minipage} \\ 
 $ \mbox{\,} $ $ ^{2}$ \begin{minipage}[t]{7cm} PRL, Ahmedabad, India \end{minipage} \\ 
 $ \mbox{\,} $ $ ^{3}$ \begin{minipage}[t]{7cm} Akita KeizaiHoka University, Japan \end{minipage} \\ 
 $ \mbox{\,} $ $ ^{4}$ \begin{minipage}[t]{7cm} Ankara University, Turkey \end{minipage} \\ 
 $ \mbox{\,} $ $ ^{5}$ \begin{minipage}[t]{7cm} Gazi  University, Ankara, Turkey \end{minipage} \\ 
 $ \mbox{\,} $ $ ^{6}$ \begin{minipage}[t]{7cm} Indian Institute of Science, Bangalore, India \end{minipage} \\ 
 $ \mbox{\,} $ $ ^{7}$ \begin{minipage}[t]{7cm} Tsinghua University, Beijing, P.R.~China \end{minipage} \\ 
 $ \mbox{\,} $ $ ^{8}$ \begin{minipage}[t]{7cm} Max--Born--Institute,  Berlin, Germany \end{minipage} \\ 
 $ \mbox{\,} $ $ ^{9}$ \begin{minipage}[t]{7cm} BNL, Upton, USA \end{minipage} \\ 
$ ^{10}$ \begin{minipage}[t]{7cm} Budker INP, Novosibirsk, Russia \end{minipage} \\ 
$ ^{11}$ \begin{minipage}[t]{7cm} CERN, Geneva, Switzerland \end{minipage} \\ 
$ ^{12}$ \begin{minipage}[t]{7cm} DESY, Hamburg and Zeuthen, Germany \end{minipage} \\ 
$ ^{13}$ \begin{minipage}[t]{7cm} JINR, Dubna, Russia \end{minipage} \\ 
$ ^{14}$ \begin{minipage}[t]{7cm} University of Edinburgh, UK \end{minipage} \\ 
$ ^{15}$ \begin{minipage}[t]{7cm} INFN-LNF, Frascati,  Italy \end{minipage} \\ 
$ ^{16}$ \begin{minipage}[t]{7cm} Universit{\"{a}}t Freiburg, Germany \end{minipage} \\ 
$ ^{17}$ \begin{minipage}[t]{7cm} University of Glasgow, UK \end{minipage} \\ 
$ ^{18}$ \begin{minipage}[t]{7cm} Universit{\"{a}}t Hamburg, Germany \end{minipage} \\ 
$ ^{19}$ \begin{minipage}[t]{7cm} CUST, Hefei, P.R.China  \end{minipage} \\ 
$ ^{20}$ \begin{minipage}[t]{7cm} University of Helsinki, Finland \end{minipage} \\ 
$ ^{21}$ \begin{minipage}[t]{7cm} Hiroshima University, Japan \end{minipage} \\ 
$ ^{22}$ \begin{minipage}[t]{7cm} KEK, Tsukuba, Japan \end{minipage} \\ 
$ ^{23}$ \begin{minipage}[t]{7cm} INP, Krakow, Poland \end{minipage} \\ 
$ ^{24}$ \begin{minipage}[t]{7cm} University of Lancaster, UK \end{minipage} \\ 
$ ^{25}$ \begin{minipage}[t]{7cm} UCL, London, UK \end{minipage} \\ 
$ ^{26}$ \begin{minipage}[t]{7cm} University of Lund, Sweden \end{minipage} \\ 
$ ^{27}$ \begin{minipage}[t]{7cm} Inst. of Physics, Minsk, Belarus \end{minipage} \\ 
$ ^{28}$ \begin{minipage}[t]{7cm} Moscow State University, Russia \end{minipage} \\ 
$ ^{29}$ \begin{minipage}[t]{7cm} N.Carolina Univ., USA \end{minipage} \\ 
$ ^{30}$ \begin{minipage}[t]{7cm} Novosibirsk State University, Russia \end{minipage} \\ 
$ ^{31}$ \begin{minipage}[t]{7cm} Inst. of Math., Novosibirsk, Russia \end{minipage} \\ 
$ ^{32}$ \begin{minipage}[t]{7cm} LAL, Orsay, France \end{minipage} \\ 
$ ^{33}$ \begin{minipage}[t]{7cm} Universit\'e de Paris VI--VII, France \end{minipage} \\ 
$ ^{34}$ \begin{minipage}[t]{7cm} IP, Prague, Czech Republic \end{minipage} \\ 
$ ^{35}$ \begin{minipage}[t]{7cm} PSI, Villingen, Switzerland \end{minipage} \\ 
$ ^{36}$ \begin{minipage}[t]{7cm} UCSC, Santa Cruz,  USA \end{minipage} \\ 
$ ^{37}$ \begin{minipage}[t]{7cm} SLAC, Stanford, USA \end{minipage} \\ 
$ ^{38}$ \begin{minipage}[t]{7cm} Tbilisi State University, Georgia \end{minipage} \\ 
$ ^{39}$ \begin{minipage}[t]{7cm} University of Thessaloniki, Greece \end{minipage} \\ 
$ ^{40}$ \begin{minipage}[t]{7cm} Seikei University, Tokyo, Japan \end{minipage} \\ 
$ ^{41}$ \begin{minipage}[t]{7cm} Sumimoto Heavy Industries, Tokyo, Japan \end{minipage} \\ 
$ ^{42}$ \begin{minipage}[t]{7cm} Polytechnic Institute, Tomsk, Russia \end{minipage} \\ 
$ ^{43}$ \begin{minipage}[t]{7cm} Warsaw  University,  Poland \end{minipage} \\ 
$ ^{44}$ \begin{minipage}[t]{7cm} Universit{\"{a}}t W{\"{u}}rzburg, Germany \end{minipage} \\ 

%
%

\end{multicols}  }  } 

\cleardoublepage
\tableofcontents
\markboth{The Photon Collider at TESLA}{Introduction}
\cleardoublepage
\label{appendix_gammagamma}
\renewcommand{\lsim}{\raisebox{-0.07cm}{$\,\stackrel{<}{{\scriptstyle\sim}}\,$}} 
\renewcommand{\gsim}{\raisebox{-0.07cm}{$\,\stackrel{>}{{\scriptstyle\sim}}\,$}} 

\providecommand{\WP}{\ensuremath{W^+}}
\providecommand{\WM}{\ensuremath{W^-}}
\providecommand{\EL}{\ensuremath{e}}
\providecommand{\G}{\ensuremath{\gamma}}
\renewcommand{\Z}{\ensuremath{Z}}
\renewcommand{\W}{\ensuremath{W}}
\renewcommand{\h}{\ensuremath{h}}
\renewcommand{\H}{\ensuremath{H}}
\renewcommand{\A}{\ensuremath{A}}
\providecommand{\HPM}{\ensuremath{H^\pm}}
\newcommand{\hZ}{\ensuremath{h^0}} 
\providecommand{\HZ}{\ensuremath{H^0}}
\providecommand{\AZ}{\ensuremath{A^0}} 
\providecommand{\HP}{\ensuremath{H^+}}
\providecommand{\HM}{\ensuremath{H^-}} 
\providecommand{\EE}{\ensuremath{\EL\EL}}
\providecommand{\GG}{\ensuremath{\G\G}}
\providecommand{\GE}{\ensuremath{\G\EL}}
\providecommand{\ZZ}{\ensuremath{\Z\Z}}
\providecommand{\WW}{\ensuremath{\W\W}}
\providecommand{\ccbar}{\ensuremath{c\bar c}}
\providecommand{\bbbar}{\ensuremath{b\bar b}}
\providecommand{\qqbar}{\ensuremath{q\bar q}} 
\providecommand{\ttbar}{\ensuremath{t\bar t}}
\providecommand{\ffbar}{\ensuremath{f\bar f}} 
\providecommand{\ccbarg}{\ensuremath{c\bar cg}}
\providecommand{\bbbarg}{\ensuremath{b\bar bg}} 
\providecommand{\pair}[1]{\ensuremath{#1 \bar{#1}}}
\providecommand{\mass}[1]{\ensuremath{M_{#1}}}
\providecommand{\CP}{\mbox{${\cal CP}$}}
\providecommand{\C}{\mbox{${\cal C}$}}
\providecommand{\NC}{\mbox{${\cal NC}$}}
\providecommand{\CC}{\mbox{${\cal CC}$}}
\providecommand{\EP}{\ensuremath{\EL^+}}
\providecommand{\EM}{\ensuremath{\EL^-}}
\providecommand{\EPEM}{\ensuremath{\EP\EM}}
\providecommand{\EMEM}{\ensuremath{\EM\EM}}
\renewcommand{\fb}{\ensuremath{\,\mathrm{fb}}}
\renewcommand{\pb}{\ensuremath{\,\mathrm{pb}}}
\newcommand{\nb}{\ensuremath{\,\mathrm{nb}}}
\providecommand{\TEV}{\ensuremath{\,\mathrm{TeV}}}
\providecommand{\TEVI}{\ensuremath{\,{TeV}}}
\providecommand{\GEV}{\ensuremath{\,\mathrm{GeV}}}
\providecommand{\GEVI}{\ensuremath{\,{GeV}}}
\providecommand{\MEV}{\ensuremath{\,\mathrm{MeV}}}
\providecommand{\KEV}{\ensuremath{\,\mathrm{keV}}}
\providecommand{\eV}{\ensuremath{\,\mathrm{eV}}}
\providecommand{\LGG}{\ensuremath{L_{\GG}}}
\providecommand{\LGE}{\ensuremath{L_{\GE}}}
\providecommand{\LEE}{\ensuremath{L_{\EMEM}}}
\providecommand{\WGG}{\ensuremath{W_{\GG}}}
\providecommand{\EV}{\ensuremath{\,\mathrm{eV}}}
\providecommand{\M}{\ensuremath{\,\mathrm{m}}}
\providecommand{\CM}{\ensuremath{\,\mathrm{cm}}}
\providecommand{\MM}{\ensuremath{\,\mathrm{mm}}}
\providecommand{\NM}{\ensuremath{\,\mathrm{nm}}}
\providecommand{\MKM}{\ensuremath{\,\mu\mathrm{m}}}
\providecommand{\SEC}{\ensuremath{\,\mathrm{s}}}
\providecommand{\msec}{\ensuremath{\,\mathrm{ms}}}
\providecommand{\MKS}{\ensuremath{\,\mu\mathrm{s}}}
\providecommand{\ns}{\ensuremath{\,\mathrm{ns}}}
\providecommand{\ps}{\ensuremath{\,\mathrm{ps}}}
\providecommand{\CMS}{\ensuremath{\,\mathrm{cm^{-2}s^{-1}}}}
\providecommand{\MRAD}{\ensuremath{\,\mathrm{mrad}}}
\providecommand{\MRADI}{\ensuremath{\,{mrad}}}
\providecommand{\RAD}{\ensuremath{\,\mathrm{rad}}}
\providecommand{\TW}{\ensuremath{\,\mathrm{TW}}}
\providecommand{\GW}{\ensuremath{\,\mathrm{GW}}}
\providecommand{\MW}{\ensuremath{\,\mathrm{MW}}}
\providecommand{\kW}{\ensuremath{\,\mathrm{kW}}}
\providecommand{\Watt}{\ensuremath{\,\mathrm{W}}}
\providecommand{\kHz}{\ensuremath{\,\mathrm{kHz}}}
\providecommand{\Hz}{\ensuremath{\,\mathrm{Hz}}}
\providecommand{\kJ}{\ensuremath{\,\mathrm{kJ}}}
\providecommand{\J}{\ensuremath{\,\mathrm{J}}}
\providecommand{\T}{\ensuremath{\,\mathrm{T}}}
\providecommand{\G}{\ensuremath{\,\mathrm{G}}}
\providecommand{\kG}{\ensuremath{\,\mathrm{kG}}}
\providecommand{\nC}{\ensuremath{\,\mathrm{nC}}}
\providecommand{\kA}{\ensuremath{\,\mathrm{kA}}}
\providecommand{\E}{\mbox{$\epsilon$}}
\providecommand{\EN}{\mbox{$\epsilon_n$}}
\providecommand{\EI}{\mbox{$\epsilon_i$}}
\providecommand{\ENI}{\mbox{$\epsilon_{ni}$}}
\providecommand{\ENX}{\mbox{$\epsilon_{nx}$}}
\providecommand{\ENY}{\mbox{$\epsilon_{ny}$}}
\providecommand{\EX}{\mbox{$\epsilon_x$}}
\providecommand{\EY}{\mbox{$\epsilon_y$}}
\providecommand{\BI}{\mbox{$\beta_i$}}
\providecommand{\BX}{\mbox{$\beta_x$}}
\providecommand{\BY}{\mbox{$\beta_y$}}
\providecommand{\SX}{\mbox{$\sigma_x$}}
\providecommand{\SY}{\mbox{$\sigma_y$}}
\providecommand{\SZ}{\mbox{$\sigma_z$}}
\providecommand{\SI}{\mbox{$\sigma_i$}}
\providecommand{\SIP}{\mbox{$\sigma_i^{\prime}$}}
\def\b{\beta}
\def\g{\gamma}
\def\SM{SM}
\def\MSSM{MSSM}
\def\2HDM{2HDM}

\def\BR{\mbox{BR}}
\def\be\begin{equation}
\def\ee\end{equation}
\providecommand{\bea}{\begin{eqnarray}}
\providecommand{\eea}{\end{eqnarray}}
\providecommand{\bear}{\begin{equation}\begin{array}}
\providecommand{\eear[1]}{\end{array}{#1}\end{equation}}
\providecommand{\dst}{\displaystyle}
\providecommand{\fordef}{\stackrel{def}{=}}
\providecommand{\bm}{\boldmath}
\providecommand{\fr}[2]{\frac{{\displaystyle #1}}{{\displaystyle #2}}}
\providecommand{\nn}{\nonumber}
\providecommand{\pa}{\partial}
\providecommand{\la}{\langle}
\providecommand{\ra}{\rangle}
\providecommand{\fn}[1]{\footnote{{\normalsize #1}}}
\def\emline#1#2#3#4#5#6{%
       \put(#1,#2){\special{em:moveto}}%
       \put(#4,#5){\special{em:lineto}}}

\newenvironment{Itemize}{\begin{list}{$\bullet$}%
{\setlength{\topsep}{0.2mm}\setlength{\partopsep}{0.2mm}%
\setlength{\itemsep}{0.2mm}\setlength{\parsep}{0.2mm}}}%
{\end{list}}
\newcounter{enumct}
\newenvironment{Enumerate}{\begin{list}{\arabic{enumct}.}%
{\usecounter{enumct}\setlength{\topsep}{0.2mm}%
\setlength{\partopsep}{0.2mm}\setlength{\itemsep}{0.2mm}%
\setlength{\parsep}{0.2mm}}}{\end{list}}
\providecommand{\sw}{\mbox{$\sin\Theta_W\,$}}
\providecommand{\cw}{\mbox{$\cos\Theta_W\,$}}
\providecommand{\epe}{\mbox{$e^+e^-\,$}}
\providecommand{\ggam}{\mbox{$\gamma\gamma\,$}}
\providecommand{\egam}{\mbox{$e\gamma\,$}}
\providecommand{\gewnu}{\mbox{$e\gamma\to W\nu\,$}}
\providecommand{\eeww}{\mbox{$e^+e^-\to W^+W^-\,$}}
\providecommand{\ggww}{\mbox{$\gamma\gamma\to W^+W^-\,$}}
\providecommand{\ggzz}{\mbox{$\gamma\gamma\to ZZ\,$}}
\providecommand{\egeh}{\mbox{$e\gamma\to eH\,$}}
\providecommand{\geeww}{\mbox{$e\gamma\to e W^+W^-\,$}}
\providecommand{\beq}{\begin{equation}}
\providecommand{\eeq}{\end{equation}}
\providecommand{\beqn}{\begin{eqnarray}}
\providecommand{\eeqn}{\end{eqnarray}}
\providecommand{\lum}[1]{{\rm luminosity} $ #1 $ cm$^{-2}$ s$^{-1}\,$}
\providecommand{\intlum}[1]{{\rm annual luminosity} $ #1$  fb$^{-1}\,$}
\providecommand{\mw}{\mbox{$M_W\,$}}
\providecommand{\mww}{\mbox{$M_W^2\,$}}
\providecommand{\mh}{\mbox{$M_H\,$}}
\providecommand{\mhh}{\mbox{$M_H^2\,$}}
\providecommand{\mz}{\mbox{$M_Z\,$}}
\providecommand{\mzz}{\mbox{$M_Z^2\,$}}
\providecommand{\sigmaw}{\mbox{$\sigma_W\,$}}
\providecommand{\sww}{\mbox{$\sin^2\Theta_W\,$}}
\providecommand{\cww}{\mbox{$\cos^2\Theta_W\,$}}
\providecommand{\ptr}{\mbox{$p_{\bot}\,$}}
\providecommand{\ptrs}{\mbox{$p_{\bot}^2\,$}}
\providecommand{\lgam}{\mbox{$\lambda_{\gamma}$}}
\providecommand{\lga}[1]{\mbox{$\lambda_{#1}$}}
\providecommand{\lggam}[2]{\mbox{$\lambda_{#1}\lambda_{#2}$}}
\providecommand{\lgg}{\lambda_1\lambda_2}
\providecommand{\lel}{\mbox{$\lambda_e$}}
\providecommand{\ggh}{\mbox{$\gamma\gamma\to hadrons$}}
%

\section{Introduction} \label{s1}

In addition to the \EPEM\ physics program, the TESLA linear collider will
provide a
unique opportunity to study \GG\ and \GE\ interactions at energies and
luminosities comparable to those in \EPEM\
collisions~\cite{GKST81,GKST83,GKST84}. High 
energy photons for \GG, \GE\ collisions can be obtained using Compton
backscattering of laser light off the high energy electrons.  Modern
laser technology provides already the laser systems for the \GG\ and \GE\ 
collider (``Photon Collider''). 

The physics potential of the Photon Collider is very
rich and complements in an essential way the physics program of the TESLA
\EPEM\ mode. The
Photon Collider will considerably contribute to the detailed
understanding of new phenomena (Higgs boson,
supersymmetry, quantum gravity with extra dimensions  etc.).  In some scenarios the Photon Collider is
the best instrument for the 
discovery of elements of New Physics.  Although many particles can be produced
both at \EPEM\ and \GG, \GE\  collisions, the reactions are different and
will give complementary information about new physics phenomena.  A
few examples: 

\begin{itemize}
\item 
  The study of charged parity $\C=-$ resonances  in \EPEM\ collisions led to
  many fundamental 
  results. In \GG\ collisions, resonances with $\C=+$ are 
  produced directly. One of the most important examples is the Higgs
  boson of the Standard Model. The precise knowledge of its two--photon width
  is of 
  particular importance.  It is sensitive to heavy virtual charged
  particles.  Supersymmetry predicts three neutral Higgs
  bosons.  Photon colliders can produce the heavy
  Higgs bosons with masses about 1.5 times higher than  in \EPEM\ 
  collisions at the same collider and allow to measure their \GG\ widths.
  Moreover, the photon collider will allow us to study
  electroweak symmetry breaking (EWSB) in both the weak--coupling and
  the strong--coupling scenarios.  
\item 
  A \GG\ collider can produce  pairs of any charged particles
  (charged Higgs, supersymmetric particles etc.) with a cross section
  about one order of magnitude higher than those in \EPEM\ collisions.
  Moreover, the cross sections depend in a different form on various
  physical parameters.  The polarisation of the photon beams
  and the large  cross sections allow to obtain valuable
  information on these particles and their interactions.
\item 
  At a \GE\ collider charged particles can be produced with masses higher
  than in pair production of  \EPEM\ collisions (like a new $\W^{\prime}$ boson
  and a neutrino or a supersymmetric scalar electron plus a neutralino).
\item
  Photon colliders offer unique possibilities for measuring the \GG\ fusion of
  hadrons for probing the hadronic structure of the photon.
\end{itemize}
Polarised photon beams, large
cross sections and sufficiently large luminosities allow to
significantly enhance the discovery limits of many new particles in
SUSY and other models and to substantially improve the accuracy of the
precision measurements of anomalous \W\ boson and top quark couplings
thereby complementing and improving the measurements at the \EPEM
mode of the TESLA.

In order to make this new field of particle physics accessible, the
Linear Collider needs  two interaction regions (IR): one for
\EPEM\ collisions and the other one for \GG\ and \GE\ collisions.

In the following we describe the physics programme of photon colliders, the
basic principles of a photon collider and its characteristics, the
requirements for the lasers and possible laser and optical schemes, the
expected \GG\ and \GE\ luminosities, and accelerator, interaction
region, background and detector issues specific for photon colliders.

The second interaction region for \GG\ and \GE\ collisions is
considered in the TESLA design and the special accelerator
requirements are taken into account. The costs however are not included in the
Technical Design Report.

\subsection{Principle of a photon collider}\label{s1.1}

\begin{figure}[!hbt]
  \centering 
  \vspace*{0.2cm}
  \epsfig{file=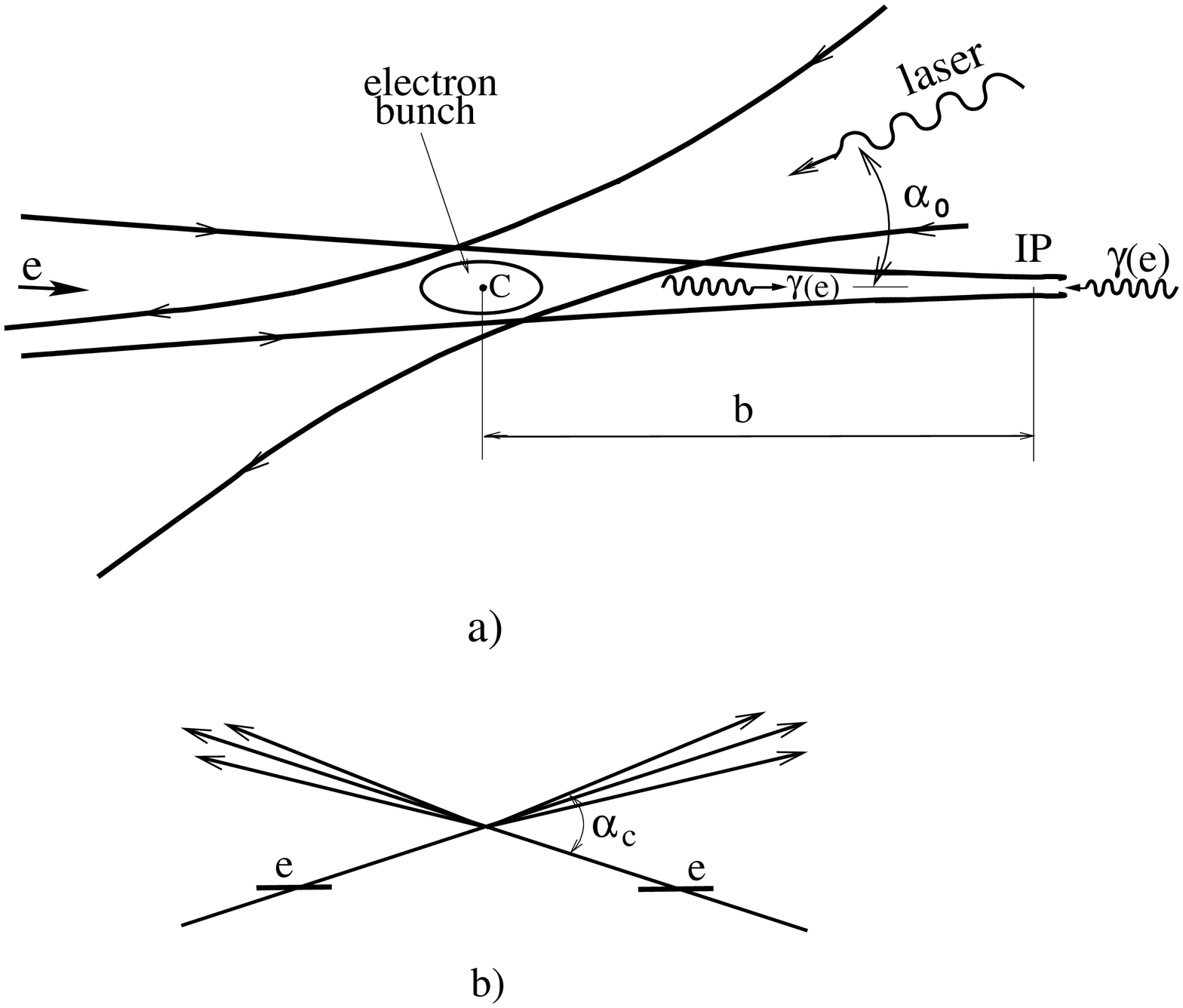,width=9.5cm,angle=0} 
\caption{Scheme of  \GG, \GE\ collider.}
\label{ggcol}
\end{figure}

The basic scheme of the  Photon Collider is shown in Fig.~\ref{ggcol}.
Two electron beams of energy $E_0$ after the final focus system travel
towards the interaction point (IP) and at a distance $b$ of about 1--5\MM\ from the IP collide with the focused laser beam.  After scattering,
the photons have an energy close to that of the initial electrons and
follow their direction to the interaction point (IP) (with  small
additional angular spread of the order of $1/\G$, where $\G =
E_0/mc^2$), where they collide with a similar opposite beam of high energy
photons or electrons.  Using a laser with a flash
energy of several Joules one can ``convert'' almost all electrons to
high energy photons. The photon spot size at the IP will be almost
equal to that of the electrons at the IP and therefore the total luminosity of \GG,
\GE\ collisions will be similar to the ``geometric''
luminosity of the basic \EMEM\ beams (positrons are not necessary for
photon colliders).  To avoid background from the disrupted beams, a
crab crossing scheme is used (Fig.~\ref{ggcol}).

The maximum energy of the scattered photons is~\cite{GKST81,GKST83}

\begin{equation}
\omega_m=\frac{x}{x+1}E_0; \;\;\;\;
x \approx \frac{4E_0\omega_0}{m^2c^4}
 \simeq 15.3\left[\frac{E_0}{\TEV}\right]\left[\frac{\omega_0}{\EV}\right]=
 19\left[\frac{E_0}{\TEV}\right]
\left[\frac{\MKM}{\lambda}\right],
\label{x}
\end{equation}
where $E_0$ is the electron beam energy and $\omega_0$ the energy of
the laser photon. For example, for $E_0 =250 \GEV$, $\omega_0
=1.17\eV$ ($\lambda=1.06\MKM$) (Nd:Glass and other powerful lasers) we
obtain $x=4.5$ and $\omega_m = 0.82\, E_0 = 205 \GEV$ (it will be
somewhat lower due to nonlinear effects in Compton scattering
(Section~\ref{s3})).

For increasing values of $x$ the high energy photon spectrum becomes
more peaked towards maximum energies.  The value $x\approx 4.8$ is a
good choice for photon colliders, because for $x > 4.8$ the
produced high energy photons create QED \EPEM\ pairs in collision with
the laser photons, and as result the \GG\ luminosity is 
reduced~\cite{GKST83,TEL90,TEL95}.  Hence, the maximum centre of mass
system (c.m.s.) energy in \GG\ collisions is about 80\%, and in \GE\ 
collisions 90\% of that in \EPEM\ collisions. If for some study lower
photon energies are needed, one can use the same laser and decrease
the electron beam energy. The same laser with $\lambda \approx 1 \MKM$ 
can be used for all TESLA energies. At $2E_0 = 800\GEV$ the parameter 
$x\approx 7$, which is larger than 4.8. But  nonlinear effects at the 
conversion region effectively increase the threshold for \EPEM\ production,
so that \EPEM\ production is significantly reduced.
 
The luminosity distribution in \GG\ collisions has a high energy
peak and a low energy part (Section~\ref{s4}).  The peak has a width at half
maximum of about 15\%. The photons in the peak can have a high degree
of circular polarisation. This peak region is the most useful for
experimentation. When comparing event rates in \GG\ and \EPEM\ 
collisions we will use the value of the \GG\ luminosity in this peak
region $z>0.8z_m$ where $z=W_{\gamma\gamma}/2E_0$ ($W_{\GG}$ is the
\GG\ invariant mass) and $z_m=\omega_m/E_0$.

The energy spectrum of high energy photons becomes most peaked if the
initial electrons are longitudinally polarised and the laser photons
are circularly polarised (Section~\ref{s3.1}). This gives almost a
factor of 3--4 increase of the luminosity in the high energy peak.
The average degree of the circular polarisation of the photons within the
high-energy peak amounts to 90--95\%.  The sign of the polarisation
can easily be changed by changing the signs of electron and laser
polarisations.

A linear polarisation $l_{\gamma}$ of the high energy photons can be
obtained by using linearly as well as circular polarised laser light~\cite{GKST84}. The
degree of the linear polarisation at maximum energy depends on $x$, it is
0.334, 0.6, 0.8 for $x=4.8,2,1$ respectively (Section~\ref{s3}). Polarisation
asymmetries are proportional to $l_{\gamma}^2$, therefore low $x$
values are preferable. The study of Higgs bosons
with linearly polarised photons constitutes a very important part of
the physics program at photon colliders.

The luminosities expected at the TESLA Photon Collider are presented
in Table~\ref{tabtel}, for comparison the \EPEM\ luminosity is also
included (a more detailed table is given is Section~\ref{s4.5.2}).

\begin{table}[!hbtp]
{\renewcommand{\arraystretch}{1.45}
\begin{center}
\begin{tabular}{l r c c} \hline
 2$E_0$, \GEV & $200$ & $500$ & $800$ \\
$ L_{geom}, 10^{34} \CMS $ & 4.8 & 12.0 &  19.1 \\ \hline
 $W_{\GG,\,max}$, \GEV & $122$ & $390$ & $670$ \\  
$ \LGG (z>0.8z_{m,\GG\ }),10^{34} \CMS $ & 0.43 &
1.1 & 1.7 \\ \hline
 $W_{\GE,\,max}$, \GEV& $156$ & $440$ & $732$ \\
$ L_{\EL\G} (z>0.8z_{m,\GE\ }),10^{34} \CMS $ &
0.36 & 0.94 & 1.3 \\ \hline \hline
$L_{\EPEM}, 10^{34} \CMS $ & 1.3 & 3.4 & 5.8 \\

\end{tabular} \end{center} }
\caption{Parameters of the Photon Collider based on TESLA. 
\GG, \GE\ luminosities are given for $z>0.8z_m$.
The laser wave length $\lambda=1.06\MKM$ and nonlinear effects in 
Compton scattering are taken into account. The 
luminosity of the basic \EPEM\ collider is given in the last line.
 \label{tabtel}}
\end{table}
One can see that {\it for the same beam  parameters and energy}~\footnote{in
\EPEM\ collisions at $2E_0 = 800\GEV$ beams are somewhat different} 

\begin{equation}
 \LGG (z>0.8z_m) \approx \frac{1}{3} \LEE.
  \label{lgge+e-}
\end{equation}
The \GG\ luminosity in the high energy luminosity peak for TESLA is
just proportional to the geometric luminosity $L_{geom}$ of the
electron beams: $\LGG(z>0.8z_m) \approx 0.09 L_{geom}$.  The latter can
be made larger for \GG\ collisions than the \EPEM\ luminosity because
beamstrahlung and beam repulsion are absent for photon beams.  It is
achieved using beams with smallest possible emittances and stronger
beam focusing in the horizontal plane (in \EPEM\ collisions beams
should be flat due to beamstrahlung).
Thus, using electron beams with  smaller emittances one can reach
higher \GG\ luminosities than \EPEM\ luminosities, which
are restricted by beam collision effects.

The laser required must be in the micrometer wave length region, with few
Joules of flash energy, about one picosecond duration and, very large,
about $100\kW$ average power. The optical scheme with multiple use of
the same laser pulse allows to reduce the necessary average laser
power at least by one order of magnitude.  Such a laser can be a solid
state laser with diode pumping, chirped pulse amplification and
elements of adaptive optics. All this technologies are already
developed for laser fusion and other projects.  It corresponds to
a large-room size laser facility. A special tunable FEL is another option
(Section~\ref{s5.2}).

\subsection[Particle production in high energy $\gamma\gamma$,
  $\gamma$e collisions]
{Particle production in high energy $\bgamma\bgamma$,
  $\bgamma$e collisions}\label{s1.2} 
 
\begin{figure}[htb]
  \begin{center}
    \epsfig{file=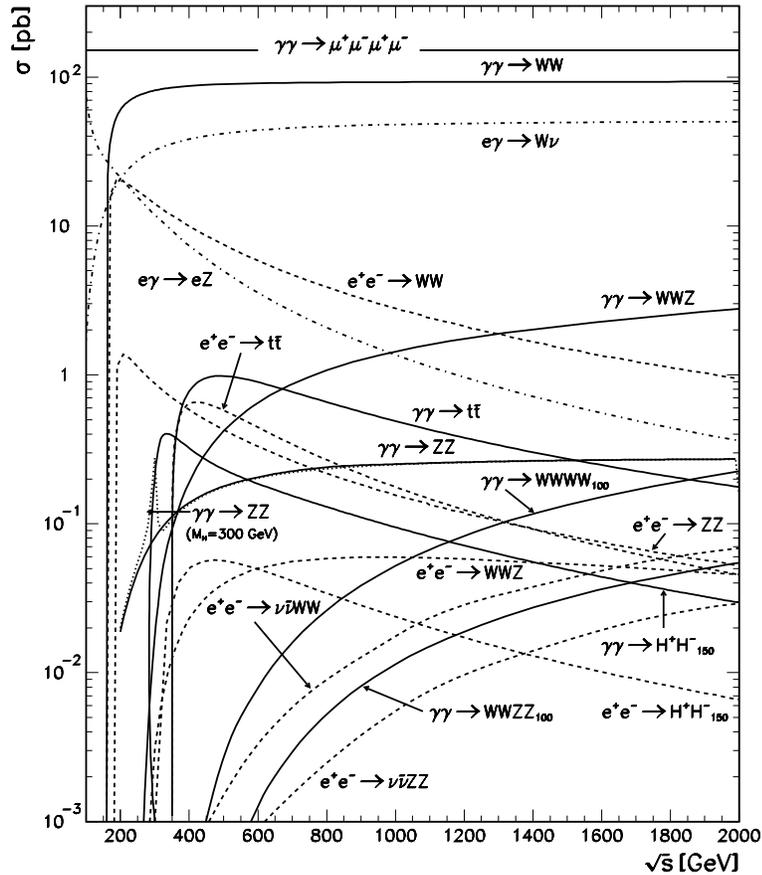,width=0.7\textwidth
}
  \end{center}
  \caption{ Typical  cross sections in \GG , \GE\ 
    and \EPEM\ collisions. The polarisation is assumed to be zero. Solid,
    dash--dotted and dashed curves correspond to
    \GG , \GE\ and \EPEM\ modes respectively. Unless
    indicated otherwise the neutral Higgs mass was taken to be $100\GEVI$. For
    charged Higgs pair production, $\mass{\HPM} = 150\GEVI$ was assumed.}
  \label{fig:cs}
\end{figure}

In the collision of photons any charged particle can be produced due
to direct coupling. Neutral particles are produced via loops built up by
charged particles ($\GG \to$ Higgs, $\GG, \ZZ$).  The
comparison of cross--sections for some processes in \EPEM\ and \GG,
\GE\ collisions is presented in Fig.~\ref{fig:cs}~\cite{CS}.

\begin{figure}[!thb]
  \centering
  \hspace*{-1.cm} \epsfig{file=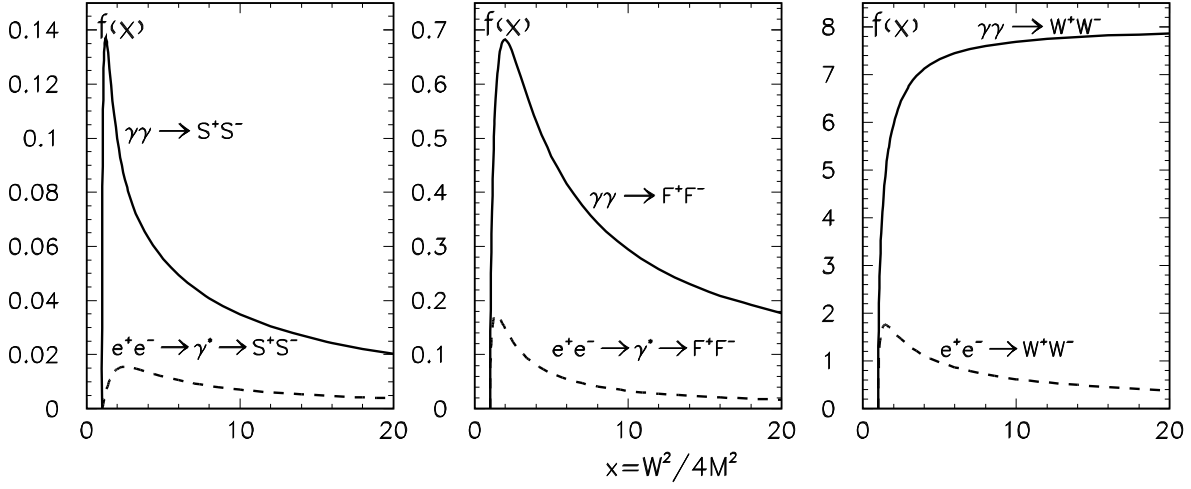,width=18.cm}
  \caption{Comparison between cross sections for charged pair production in
    unpolarised \EPEM\ and \GG\ collisions. S (scalars), F (fermions), W
    (\W\ bosons); $\sigma = (\pi\alpha^2/M^2)f(x)$, $M$ is the particle
    mass, W is the invariant mass (c.m.s. energy of colliding beams),
    $f(x)$ are shown.  Contribution of Z boson for production of S and F
    in \EPEM\ collisions was not taken into account, it is less than 10\%} 
  \label{charged}
\end{figure}

\begin{figure}[!thb]
  \centering
  \hspace*{-0.5cm} \epsfig{file=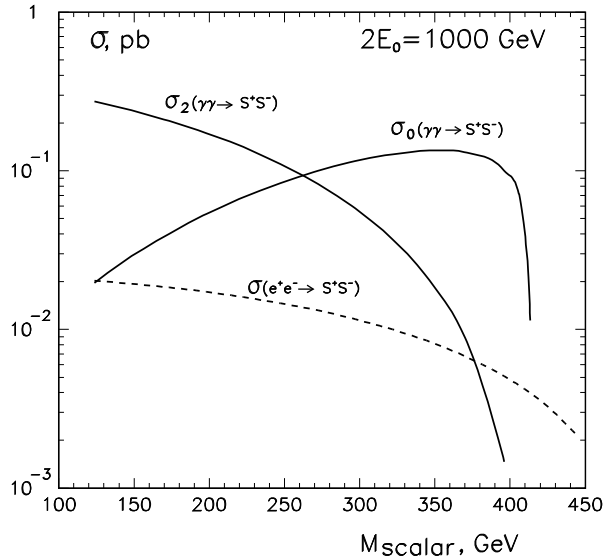,width=11.cm}
  \caption{Pair production cross sections for charged scalars 
    in \EPEM\ and \GG\ collisions at $2E_0 = 1 \TEVI$ collider (in \GG\ 
    collision $W_{max}\approx 0.82\TEVI $ ($x=4.6$)); $\sigma_0$ and
    $\sigma_2$ correspond to the total \GG\ helicity 0 and 2
    respectively.  Comparison is valid for other beam energies if
    masses are scaled proportionally.} 
  \label{crossel}
\end{figure}

The cross
sections for pairs of scalars, fermions or vector particles are all
significantly  larger (about one order of magnitude) in \GG\ 
collisions compared with \EPEM\ collisions, as shown in
Fig.~\ref{charged}~\cite{TEL90,TEL95,GSwinter,GINSD}. For example, the maximum
cross section for \HP\HM\ production with unpolarised photons is 
about 7 times higher than that in \EPEM\ collisions (see
Fig.~\ref{fig:cs}). With polarised photons and not far from
threshold it is even larger by a factor of 20,
Fig.~\ref{crossel}~\cite{Tfrei}. Using the luminosity given in the
Table~\ref{tabtel} the event rate is 8 times higher.  

\begin{figure}[!thb]
  \centering
  \hspace*{-0.5cm} \epsfig{file=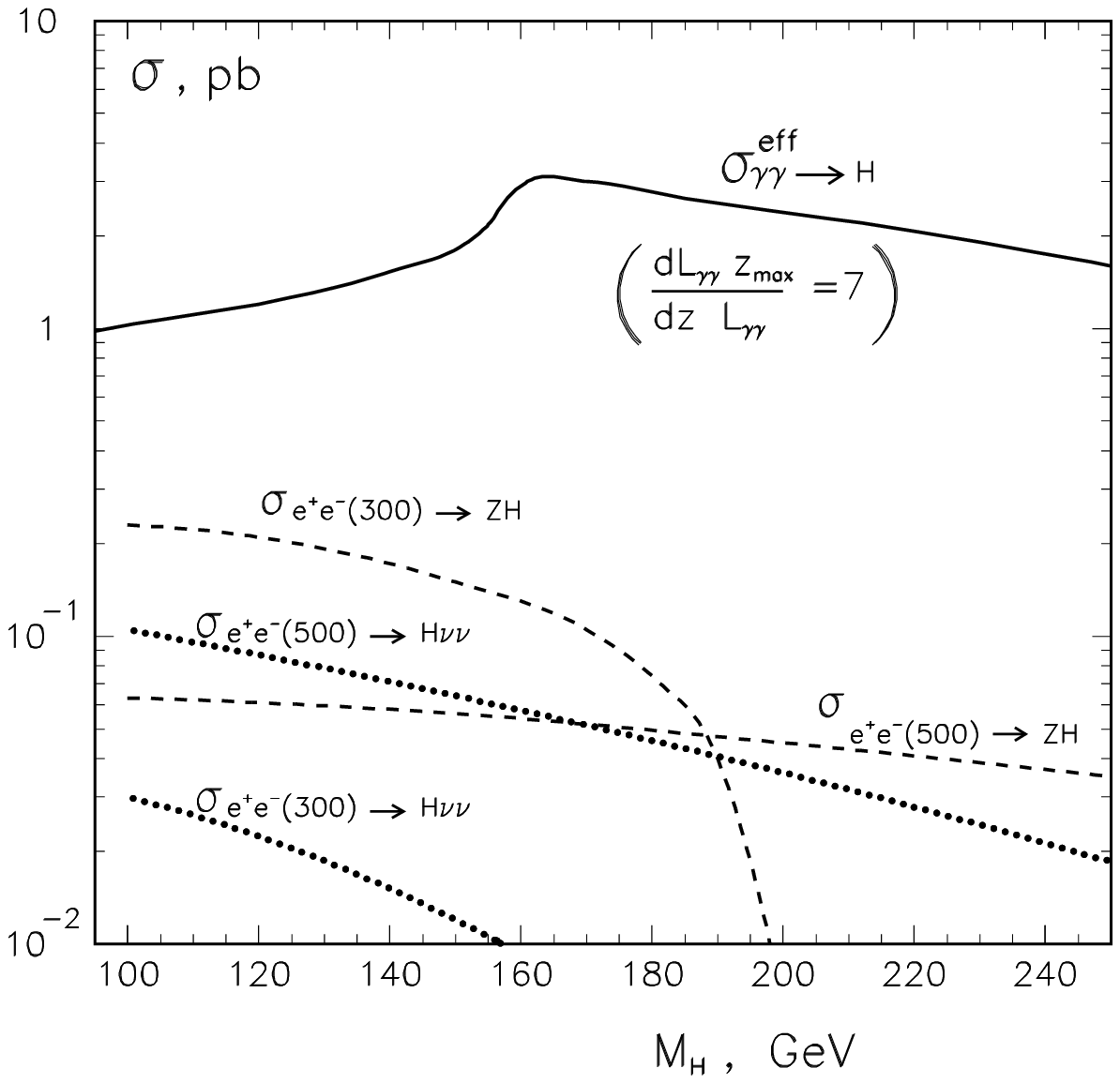,width=11.cm}
  \caption{Total cross sections of the Higgs boson production in
    \GG\ and \EPEM\ collisions. To obtain the Higgs boson
    production rate at the photon collider the cross section should be
    multiplied by the luminosity in the high energy peak
    $L_{\GG}(z>0.65)$ given in the Table~\protect\ref{tabtel}.}
  \label{hcross}
\end{figure} 
The two--photon production of pairs of charged particles is a pure QED
process, while the cross section for pair production in \EPEM\ collision is
mediated by \G\ and \Z\ exchange so that it 
depends also on the weak isospin of the produced particles. The \EPEM\ process
may also be affected by the exchange of new
particles in the t--channel.  Therefore, measurements of pair production
both in \EPEM\ and \GG\ collisions help to disentangle
different couplings of the charged particles.

Another example is the direct resonant production of the Higgs boson
in \GG\ collisions. It is evident from Fig.~\ref{hcross}~\cite{ee97}, that the
cross section at the photon collider is 
several times larger than the Higgs production cross section in
\EPEM\ collisions. Although the \GG\ luminosity is smaller
than the \EPEM\ luminosity (Table~\ref{tabtel}), the production rate
of the Standard Model (\SM) Higgs boson with mass between $130$ and $250\GEV$ in
\GG\ collisions is nevertheless 1--10 times the rate in \EPEM\ collisions 
at $2E_0 = 500\GEV$.

Photon colliders used in the \GE\ mode can produce  particles
which are kinematically not accessible at the same collider in the
\EPEM\ mode.  For example, in \GE\ collisions one can produce a
heavy charged particle in association with a light neutral one, such as
supersymmetric selectron plus neutralino, 
$\GE\to \tilde{\EL} \tilde{\chi} ^0$ or a new \W $'$ boson and neutrino, 
$\GE\to \W '\nu$. In this way the discovery limits can be extended. 

Based on these arguments alone, and without knowing {\it a priori} the
particular scenario of new physics, there is a strong complementarity
for \EPEM\ and \GG\ or \GE\ modes for new physics searches.  

The idea of \GE\ and \GG\ collisions at linear colliders via
Compton backscattering has been proposed by the Novosibirsk group~\cite{GKST81,GKST83,GKST84}. 
Reviews of further developments can be
found in~\cite{TEL90,TEL95,CS,GINSD,Tfrei,ee97,Skrinsky,TEL91,BBC,TEL92,BRODBER,BAIL94,MILLERM,JLCgg,TESLAgg,Sessler,telnov} and the conceptual(zero) design reports~\cite{NLC,TESLA,JLC} and
references therein.

A review of the physics potential and available technologies of \GG,
\GE\ colliders, can be found in the proceedings of workshops on
photon colliders held in 1995 at Berkeley~\cite{BERK} and in 2000
at DESY~\cite{GG2000}.

\section{The Physics } \label{s2}

\subsection{Possible scenarios} \label{s2.1}

The two goals of studies at the next generation of colliders are the proper
understanding of electroweak symmetry breaking, associated with the problem of
mass, and
the discovery of new physics beyond the Standard Model (\SM).
Three scenarios are possible for future experiments~\cite{Ginzfrei}:\\ 

\begin{itemize}

\item
  New particles or interactions will be directly discovered at the TEVATRON
  and LHC. A Linear Collider (LC) in the \EPEM\ and \GG\ modes will then play
  a crucial role 
  in the detailed and 
  thorough study of these new phenomena and in the reconstruction of the
  underlying fundamental theories. 
\item
  LHC and LC will
  discover and study in detail the Higgs boson but no spectacular
  signatures of new physics or new particles will be observed.  In this case
  the precision studies of the deviations of the properties of the Higgs
  boson, electroweak gauge bosons and the top quark from their Standard Model
  (\SM)  
  predictions can provide clues to the physics beyond the Standard
  Model.
\item
  Electroweak symmetry breaking (EWSB) is a dynamical phenomenon. The
  interactions of 
  \W\ bosons and $t$ quarks must then be studied at high energies to explore new
  strong interactions at the \TEV\ scale.
\end{itemize}

Electroweak symmetry breaking in the \SM\ is based on the Higgs mechanism, which introduces one elementary Higgs boson.
The model agrees with the present data, partly at the per--mille level, and
the recent global analysis of 
precision electroweak data in the framework of the \SM~\cite{osaka}
suggests that the Higgs boson is lighter than $200\GEV$. A Higgs boson in this
mass range is expected to be discovered at the TEVATRON or the LHC. However,
it will be the LC in all its modes that tests 
whether this particle is indeed the \SM\ Higgs boson or whether it is
eventually 
one of the Higgs states in extended models like the two Higgs doublets
(\2HDM) or the minimal supersymmetric generalisation of the \SM , e.g.
\MSSM . At least five Higgs bosons are predicted in supersymmetric models,
\hZ, \HZ, \AZ, \HP, \HM. Unique opportunities are offered by the Photon
Collider to search for the heavy Higgs bosons in areas of SUSY parameter space
not accessible elsewhere.

\subsection{Higgs boson physics} \label{s2.2}

The Higgs boson plays an essential role in the EWSB
mechanism and the origin of mass.  The lower bound on
$\mass{\h}$ from direct searches at LEP is presently $113.5\GEV$ at
$95\%$~confidence level (CL)~\cite{igokemenes}. A surplus of events at LEP
provides tantalising indications of a Higgs boson with 
$\mass{\h}=115^{+1.3}_{-0.7}\GEV$ (90\%~CL) at a level of
2.9$\sigma$~\cite{igokemenes,higgs1151,higgs1152}.  Recent global 
analyses of precision electroweak data~\cite{osaka} suggest that the
Higgs boson is light, yielding at 95\%~CL that 
$\mass{\h}=62^{+53}_{-30}\GEV$.  There is remarkable agreement
with the well known upper bound of $\sim 130\GEV$ for the lightest
Higgs boson mass in the minimal version of supersymmetric theories,
the \MSSM~\cite{MSSM1,MSSM2}. Such a Higgs boson should definitely be
discovered at the LHC if not already at the TEVATRON.

Once the Higgs boson is discovered, it will be crucial 
to determine the mass, the total width, spin, parity, \CP--nature and the
tree--level and one--loop induced couplings in a model independent way.
Here the \EPEM\ and \GG\ modes of the LC should play a central role.
The \GG\ collider option of a LC offers the unique possibility
to produce the Higgs boson as an s--channel
resonance~\cite{Barklow,GunionHaber1,GunionHaber2,BBC1}:
$$
\GG\to \hZ\to \bbbar ,\WW^*, \ZZ, \tau \tau, gg, \GG\ 
\ldots\,.
$$
The total width of the Higgs boson at masses below $400\GEV$ is much smaller
than the characteristic width of the \GG\ luminosity spectra (FWHM 
$\sim 10$--15\%), so that the Higgs production rate is proportional to
$d\LGG/d\WGG$: 
\begin{equation}
\dot{N}_{\GG \to \h}
=\LGG \times \frac{d\LGG \mass{\h}}{d\WGG \LGG}\frac{4\pi^2\Gamma_{\GG}
(1+\lambda_1 \lambda_2)}{\mass{\h}^3}\equiv \LGG \times \sigma^{eff}.
\end{equation}
$\Gamma_{\GG}$ is the the two--photon width of the Higgs boson and $\lambda_i$
are the photon helicities.  

The search and study of the Higgs boson can be carried out best by exploiting
the high energy peak of the \GG\ luminosity 
energy spectrum where $d\LGG/d\WGG$ has a maximum and the photons have
a high degree of circular polarisation. The effective cross section for
$(d\LGG/d\WGG) (\mass{\h}/\LGG)=7$ and 
$1+\lambda_1 \lambda_2=2$ is presented in Fig.~\ref{hcross}.  The luminosity
in the high energy 
luminosity peak ($z>0.8z_m$) was defined in Section~\ref{s1.1}.
For the luminosities given in Table~\ref{tabtel} the ratio of the Higgs
rates in \GG\ and \EPEM\ collisions is about 1 to 10 for
$\mass{\h}=100$--$250\GEV$. 

The Higgs boson at photon colliders can be detected as a peak in the
invariant mass distribution or (and) it can be searched for by 
scanning the  energy using the sharp high--energy edge of the luminosity
distribution~\cite{ee97,ohgaki}. The scanning allows also to determine 
backgrounds. A cut on the acollinearity angle between
two jets from the Higgs decay (\bbbar\ for instance) allows to select
events with a narrow (FWHM $\sim 8\%$) distribution of the invariant
mass~\cite{Tfrei,TKEK}.


The Higgs \GG\ partial width $\Gamma(\h\to\GG)$ is of special interest,
since it is generated at the one--loop level including all heavy charged
particles with masses generated by the Higgs mechanism. In this
case the heavy particles do not in general decouple. As a result the
Higgs cross section in \GG\ collisions is sensitive to 
contributions of new particles with masses beyond the energy
covered directly by accelerators. Combined
measurements of $\Gamma(\h\to\GG)$ and the branching ratio $\BR(\h\to\GG)$ at the
\EPEM\ and \GG\ LC provide a model--independent measurement of the
total Higgs width~\cite{snow96}.

The required accuracy of the $\Gamma(\h\to\GG)$ measurements 
in the SUSY sector can be inferred from the results of the studies of the
coupling of the 
lightest SUSY Higgs boson to two photons in the decoupling regime~\cite{decoupling1,decoupling2}. It was shown that in the decoupling limit, where all
other Higgs bosons and the supersymmetric particles are very heavy,
chargino and top squark loops can 
generate a sizable difference between the standard and the SUSY
two--photon Higgs couplings. Typical deviations are at the few percent
level. Top squarks heavier than $250\GEV$ can induce deviations 
larger than $\sim 10\%$ if their couplings to the Higgs boson are large.

The ability to control the polarisations of the back--scattered photons
provides a powerful tool for exploring the \CP\  properties of any
single neutral Higgs boson that can be produced with reasonable rate
at the Photon Collider~\cite{GF92,GK94,KKSZ94}. The \CP--even
Higgs bosons $\hZ$, $\HZ$ couple to the combination
$\vec{\varepsilon_1}\cdot \vec{\varepsilon_2}$,
while the \CP--odd Higgs boson $\AZ$ couples to
$[\vec{\varepsilon_1}\times \vec{\varepsilon_2}]\cdot\vec{k_\gamma}$,
where the $\vec{\varepsilon_i}$ are the photon polarisation vectors.  The
\CP--even Higgs bosons couple to linearly polarised photons with a maximal
strength for parallel polarisation vectors, the \CP--odd Higgs boson for
perpendicular polarisation vectors:
\begin{equation}
\sigma \propto 1 \pm l_{\G 1}l_{\G 2}\cos{2\phi},
\label{clin}
\end{equation}
The degrees of linear polarisation are denoted by $l_{\G i}$ and
$\phi$ is the angle between $ \vec{l_{\G 1}}$ and $ \vec{l_{\G 2}}$; the $\pm$
signs correspond to $\CP =\pm 1$ scalar 
particles.

\begin{figure}[htb]
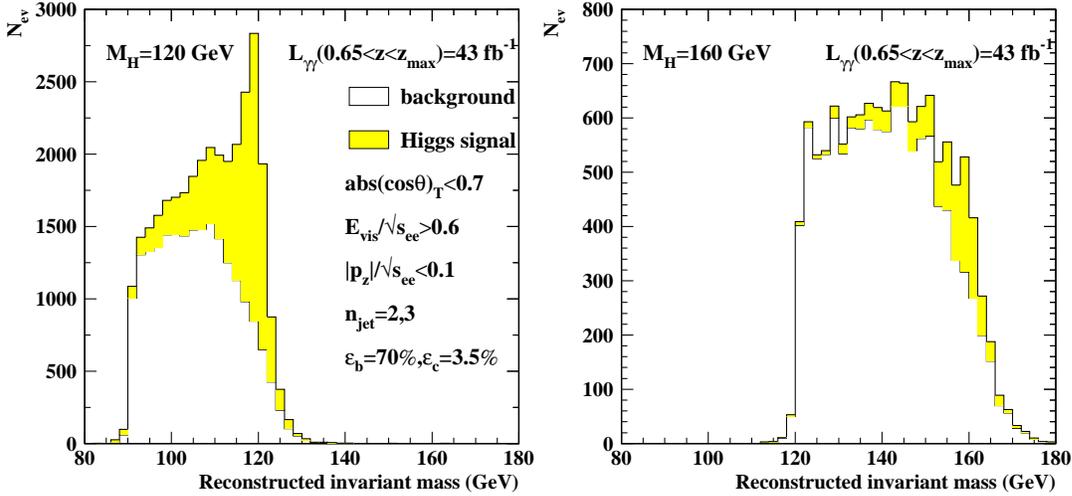

\epsfig{file=fig5_120.epsi,width=7cm} \epsfig{file=fig5_160.epsi,width=7cm} 
\caption{Mass distributions for the Higgs signal and heavy quark
background for a) $\mass{\h}=120\GEVI$ and b) $160\GEVI$. The Compton 
parameter $x = 4.8$ was assumed. The text in the figure shows 
cuts on the jets parameters~\protect\cite{Soldner1,Soldner2}.}
\label{fig:higgs}
\end{figure}

\subsubsection{Light \SM\ and \MSSM\ Higgs boson}

A light Higgs boson \h\ with mass below the \WW\ threshold can be detected
in the \bbbar\ decay mode. Simulations of this process have been
performed in~\cite{JLCgg,Soldner1,Soldner2,BBC1,Ohgaki1997,Ohgaki1998,Melles1,Melles2}.
 The main background to the \h\ boson production is the
continuum production of \bbbar\ and \ccbar\ pairs. A high degree of circular
polarisation of the photon beams is crucial in this case, since for equal
photon helicities $(\pm\pm)$, which produce the
spin--zero resonant states, the $\GG\to \qqbar$ QED Born
cross section is suppressed by a factor
$\mass{q}^2/\W_{\GG}^2$~\cite{Barklow,Ispirian1,Ispirian2}.

A Monte Carlo simulation of $\GG\to \h\to \bbbar$ for $\mass{\h} = 120$
and $160\GEV$ has been performed for an integrated luminosity in the high
energy peak of $\LGG(0.8z_m<z<z_m)=43\fb^{-1}$ in~\cite{Soldner1,Soldner2,JikiaSoldner}. Real and virtual gluon corrections for
the Higgs signal and the
backgrounds~\cite{Melles1,Melles2,JikiaSoldner,JikiaTkabladze1,JikiaTkabladze2,Khhiggs1,Khhiggs2,melles-dl1,melles-dl2,melles-dl3} have
been taken into account.

The results for the invariant mass distributions for the
combined $\bbbar (\G)$ and $\ccbar (\G)$ backgrounds, after cuts, and for the
Higgs 
signal are shown in Fig.~\ref{fig:higgs}~\cite{Soldner1,Soldner2}.  Due to the
large charm production cross--section in \GG\ collisions,
excellent $b$ tagging is required~\cite{Soldner1,Soldner2,Melles1,Melles2}. A $b$ tagging
efficiency of $70\%$ for \bbbar\ events and residual efficiency
of $3.5\%$ for \ccbar\ events were used in these studies.
A relative statistical error of  
\begin{equation}
\frac{\Delta[\Gamma(\h\to \GG)\BR(\h\to
\bbbar)]}{[\Gamma(\h\to \GG)\BR(\h\to\bbbar)]}\approx 2\%
\label{ggamh}
\end{equation}
can be achieved in the Higgs mass range between $120$ and $140\GEV$~\cite{Soldner1,Soldner2}.

It has been shown that the $\h\to \bbbar$ branching ratio can be measured at
the LC in \EPEM\ (and \GG) collisions with an accuracy of
$1\%$~\cite{battaglia}, the partial two--photon Higgs width can then be
calculated using the relation 
$$
\Gamma(\h\to \GG)=\frac{[\Gamma(\h\to \GG) 
  \BR(\h\to\bbbar)]}{[\BR(\h\to\bbbar)]} 
$$ 
with almost the same accuracy
as in eq.~(\ref{ggamh}).  Such a high precision for the
$\Gamma(\h\to \GG)$ width can only be achieved at the \GG\ mode
of the LC. On this basis it should be possible to
discriminate between the \SM\ Higgs particle and the lightest scalar
Higgs boson of the \MSSM\  or the \2HDM~\cite{decoupling1,decoupling2}, and contributions
of new heavy particles should become apparent.

The \SM\ Higgs boson with mass $135 < \mass{\H} < 190\GEV$ is expected to decay
predominantly into $\WW^*$ or $\WW$ pairs ($\W^*$ is a virtual $\W$
boson). This decay mode should permit the detection of the Higgs boson
signal below and slightly above the threshold of $\WW$ pair production~\cite{WW*1,WW*2,WW*31,WW*32}.  In order to determine the two--photon Higgs
width in this case one can use the same relation as above after replacing the
b quark by the real/virtual \W\ boson.

The branching ratio $\BR(\WW^*)$ is obtained from Higgs--strahlung.  It
was shown~\cite{WW*31,WW*32} that for $\mass{\h} = 160\GEV$ the product
$\Gamma(\h\to \GG) \BR (\h\to \WW^*)$ can be measured at the Photon Collider
with the 
statistical accuracy better than 2\% at the integrated \GG\ luminosity of
$40\fb^{-1}$ in the high energy peak.  The accuracy of 
$\Gamma(\h\to\GG)$ will be determined by the accuracy of the 
$\BR (\h\to\WW^*)$ measurement  
in \EPEM\ collisions which is expected to be about 2\%.

Above the \ZZ\ threshold the most promising channel to detect the
Higgs signal is the reaction $\GG\to \ZZ$~\cite{ZZ1,ZZ2,ZZ3,ZZ4}.  In order to
suppress the significant background from the tree level $\WP\WM$ pair
production, leptonic ($l^+l^-\ l^+l^-$, $\BR =1\%$) or
semileptonic ($l^+l^-\ q\bar q$, $\BR=14\%$) decay modes of the $\ZZ$
pairs must be selected. Although in the \SM\ there is only a one--loop induced
continuum 
production of $\ZZ$ pairs, it represents a large irreducible background
for the Higgs signal well above the $\WW$ threshold~\cite{ZZ1,ZZ2,ZZ3,ZZ4}. Due to
this background the intermediate mass Higgs boson signal can be
observed at the $\GG$ collider in the $\ZZ$ mode if the Higgs
mass lies below $350$--$400\GEV$.

Hence, the two--photon \SM\ Higgs width can be measured at the photon
collider, either in $\bbbar$, $\WW^*$ or $\ZZ$ decay modes, up to the
Higgs mass of $350$--$400\GEV$.  Other decay modes, like $\h\to \tau\tau,
\GG$, may also be exploited at photon colliders, but no studies have
been done so far.

Assuming that in addition to the
measurement of the $\h\to \bbbar$ branching ratio also 
the $\h\to\GG$ branching ratio can be measured (with an accuracy of $10$--$15\%$)
at TESLA~\cite{brhgg1,brhgg2}, the total width of the Higgs boson can be determined in
a model--independent way to an accuracy as dominated by the error on
$\BR(\h\to\GG)$
$$ 
\Gamma_{\h}=\frac{[\Gamma(\h\to \GG) 
  \BR (\h\to\bbbar)]}{[\BR(\h\to \GG)] [\BR(\h\to\bbbar)]}. 
$$ 
The measurement of this branching ratio at the Photon Collider (normalised to
$\BR(h\to\bbbar)$ from the \EPEM\ mode) will improve the 
accuracy of the total Higgs width.

\subsubsection{Heavy \MSSM\ and \2HDM\ Higgs bosons}

The minimal supersymmetric extension of the Standard Model 
contains two charged (\HPM) Higgs bosons and three neutral Higgs
bosons: the light \CP--even Higgs particle (\h), and heavy \CP--even
(\H) and the \CP--odd (\A) Higgs states.  If we assume a large value
of the \A\ mass, the properties of the light \CP--even Higgs boson \h\ are
similar to those of the 
light \SM\ Higgs boson, and can be detected in the
$\bbbar$ decay mode, just as the \SM\ Higgs. Its mass is
bound to $\mass{\h}\lsim 130\GEV$.  However, the masses of the heavy Higgs
bosons 
\H, \A, \HPM are expected to be of the order of the electroweak scale
up to about $1\TEV$.  The heavy Higgs bosons are nearly degenerate. The
$\WW$ and $\ZZ$ decay modes are suppressed for the heavy $\H$ case, and
these decays are forbidden for the $\A$ boson.
Instead of the $\WW$, $\ZZ$ decay modes, the $\ttbar$ decay channel
may be useful if the Higgs boson masses are heavier than $\mass{t}$, and
if $\tan\beta \ll 10$ ($\tan\beta$ is the Goldstone mixing--parameter of
\MSSM).  An 
important property of the SUSY couplings is the enhancement of the
bottom Yukawa couplings with increasing $\tan\beta$.  For moderate and
large values of $\tan\beta$, the decay mode to \bbbar~\cite{Muhlleitner1,Muhlleitner2} (and to $\tau^+\tau^-$ in some cases) is
substantial.

Extensive studies have demonstrated that, while the light Higgs boson
$\h$ of \MSSM\ can be found at the LHC, the heavy bosons $\H$ and $\A$ may
escape discovery for intermediate values of $\tan \beta$~\cite{ATLAS1,ATLAS2}.  At an
\EPEM\ LC the heavy \MSSM\ Higgs bosons can only be found in associated
production $\EPEM \to \H\A$~\cite{3b1,3b2,3b3}, with $\H$ and $\A$ having very similar
masses. In the first phase of the LC  with a total \EPEM
energy of $500\GEV$ the heavy Higgs bosons can thus be discovered for 
masses up to about $250\GEV$. The mass reach can be extended by a factor of 1.6
in the \GG\ mode of TESLA, in which the Higgs bosons $\H$, $\A$ can
be singly produced.

\begin{figure}[hbt]
\begin{center}
\epsfig{file=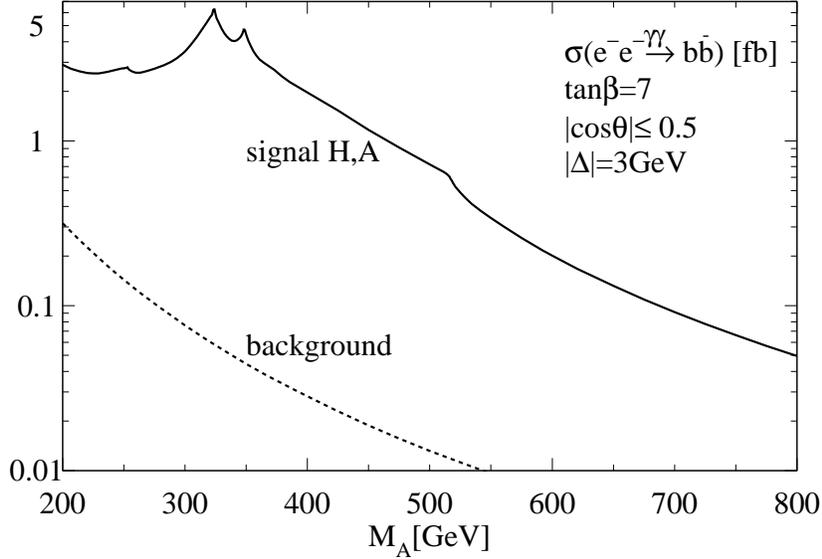,width=0.7\textwidth}
\caption{ Cross section for resonant heavy Higgs  \H, \A\ boson
  production as a function of the pseudoscalar Higgs mass $\mass{\A}$ with
  decay into \bbbar\ pairs, and the corresponding background cross
  section. The maximum of the photon luminosity in the $J_z=0$
  configuration has been tuned to coincide with $\mass{\A}$. The cross
  sections are defined in \bbbar\ mass bins of $\mass{\A}\pm 3\GEVI$ around
  the $\A$ resonance. An angular cut on the bottom production angle
  $\theta$ has been imposed: $|\cos\theta| <0.5$. The \MSSM\ 
  parameters have been chosen as $\tan\beta=7, \mass{2}=-\mu=200\GEVI$. See
  also comments in the text.}
\label{fig:bot}
\end{center}
\end{figure}

The results for the cross section of the $\H$, $\A$ signal in the \bbbar\
decay mode and the corresponding background for the value of 
$\tan\beta=7$ are shown in Fig.~\ref{fig:bot} as a function of the
pseudoscalar mass $\mass{\A}$~\cite{Muhlleitner1,Muhlleitner2}. From the figure one can
see that the background is strongly suppressed with respect to the
signal. The significance of the heavy Higgs boson signals is sufficient for
a discovery of the Higgs particles with masses up to about $70$--$80\%$ of
the LC c.m.s. energy. For $2E_0 = 500\GEV$ the \H, \A\ bosons with
masses up to about $0.8\times 2E_0 \approx 400\GEV$ can be discovered in
the \bbbar\ channel at the Photon Collider. For a LC with 
$2E_0=800\GEV$ the range can be extended to about $660\GEV$~\cite{Muhlleitner2,muehldiss}. Also the one--loop induced two--photon width of the \H, 
\A\ Higgs 
states will be measured. For heavier Higgs masses the signal becomes
too small to be detected. Note that the cross section given in
Fig.~\ref{fig:bot} takes into account the $\EL\to\G$ conversion $k^2
L_{geom} \sim 0.4L_{geom}$ ($k$ being the $\EL\to\G$ conversion
coefficient) which results in a luminosity of
$4.8\times 10^{34}\CMS \sim 1.5 L_{\EPEM}$ for $2E_0 = 500\GEV$ and which
grows proportional to the energy.

The separation of the almost degenerate $\H$ and $\A$ states may be achieved 
using the linear polarisation of the colliding photons (see
eq.~\ref{clin}). The $\H$ and $\A$ states can be produced from collisions
of parallel and  perpendicularly polarised incoming photons,
respectively~\cite{GF92,GK94,KKSZ94,Yang50,CK95,Illana}. The
possible \CP--violating mixing of  $\H$ and $\A$ can be
distinguished from the overlap of these resonances by analysing the
polarisation asymmetry in the two--photon production~\cite{GIv}.

\begin{figure*}[htb]
\begin{center}
\epsfig{file=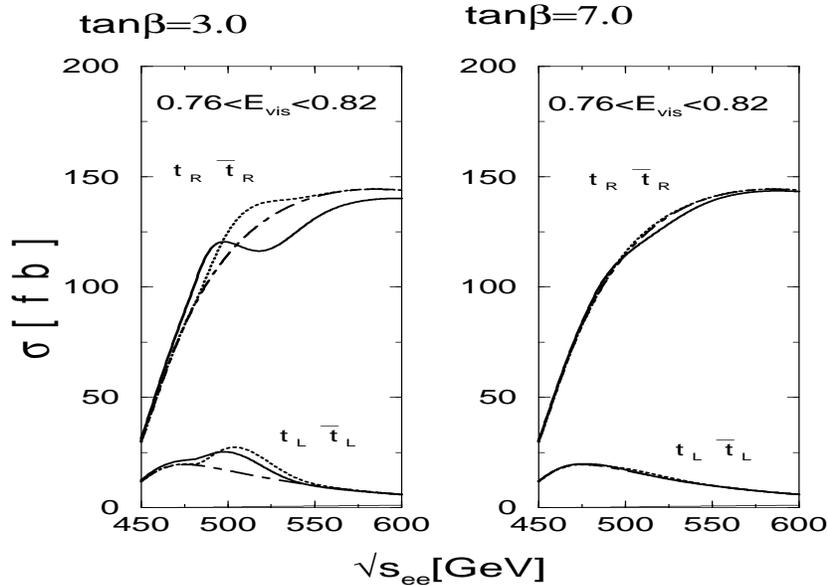,width=0.7\textwidth,height=0.5\textwidth}
\caption{The effective top pair cross sections $\GG\to\ttbar$ 
  convoluted with the $\EL\to\G$ conversion efficiency within the visible
  energy range as indicated.  The bold--solid curves correspond to the
  correct cross sections, the dotted curves are the ones neglecting
  the interference, and the dot--dashed are the continuum cross
  sections, respectively.  The upper curves are for $t_R
  \overline{t}_R$, and the lower ones for $t_L \overline{t}_L$.  The
  sum of the cross sections for $t_R \overline{t}_L$ and $t_L
  \overline{t}_R$, are also plotted as  thin--continuous line
  very near to the bottom horizontal axis.  The left figure is for
  $\tan\beta= 3$, and the right for $\tan\beta = 7$~\protect\cite{AKSW001,AKSW002}.  }
\label{F:AKSW2}
\end{center} \end{figure*}

The interference between $\H$ and $\A$ states can be also studied in the
reaction $\GG\to\ttbar$ with circularly polarised photon beams
by measuring the top quark helicity~\cite{AKSW001,AKSW002}. The corresponding
cross sections are shown in Fig.~\ref{F:AKSW2}. The effect of the
interference is clearly visible for the value of $\tan\beta=3$.
The $RR$ cross section is bigger than the $LL$ cross section ($R$($L$) is
right(left) 
helicity) due to the continuum. Large interference effects are visible
in both modes.  Without the measurement of the top quark polarisation
there still remains  a strong  interference effect between the
continuum and the Higgs amplitudes, which can be measured.

For energies corresponding to the maximum cross sections
(not far from the threshold) with proper polarisation the pair
production rate of charged Higgs $\GG\to \HP\HM$ at the TESLA Photon
Collider will be almost an order of magnitude larger than at the
\EPEM\ LC due to the much larger cross section.

Scenarios, in which all new
particles are very heavy, may be realised not only in the 
\MSSM\, but also in other extended models of the Higgs sector, for
example in models with just two Higgs doublets. In this case the two--photon
Higgs boson width, for $\h$ or $\H$, will differ from the \SM\ 
value even if all direct couplings to the gauge bosons \W/\Z\ and the fermions
are equal to the corresponding couplings in the \SM, driven by the
contributions 
of extra heavy charged particles.  In the \2HDM\ these particles are
the charged Higgs bosons.  Different realizations of the
  \2HDM\ have been discussed in~\cite{GKO1,GKO2}. Assuming that the
partial widths of the observed Higgs boson to quarks, $\Z$ or $\W$
bosons are close to their \SM\ values, three sets of possible values
of the couplings to $\GG$ can be obtained.
Fig.~\ref{fig:gaga-a,b} shows deviations of the two--photon Higgs width
from the \SM\ value for these three variants.  The shaded regions are
derived from the anticipated $1\sigma$ experimental bounds around the
\SM\ values for the Higgs couplings to fermions and gauge bosons.
Comparing the numbers in these figures with the achievable accuracy of the
two--photon Higgs width at a photon collider~(\ref{ggamh}) 
the difference between \SM\ and \2HDM\ should definitely be observable~\cite{GKO1,GKO2}. 

\begin{figure}[htb]
\hspace{1.5cm} \epsfig{file=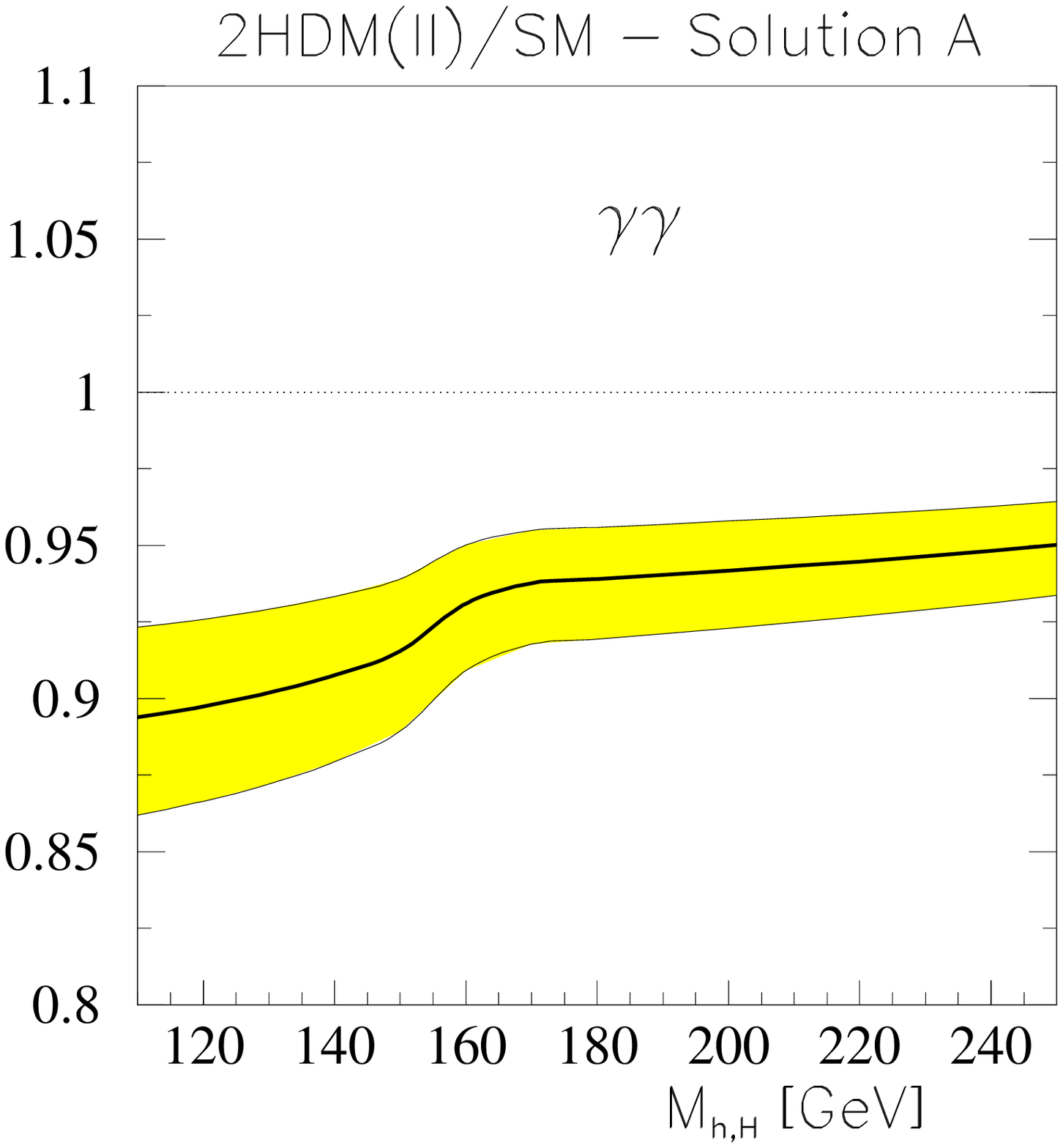,width=6.5cm}
\epsfig{file=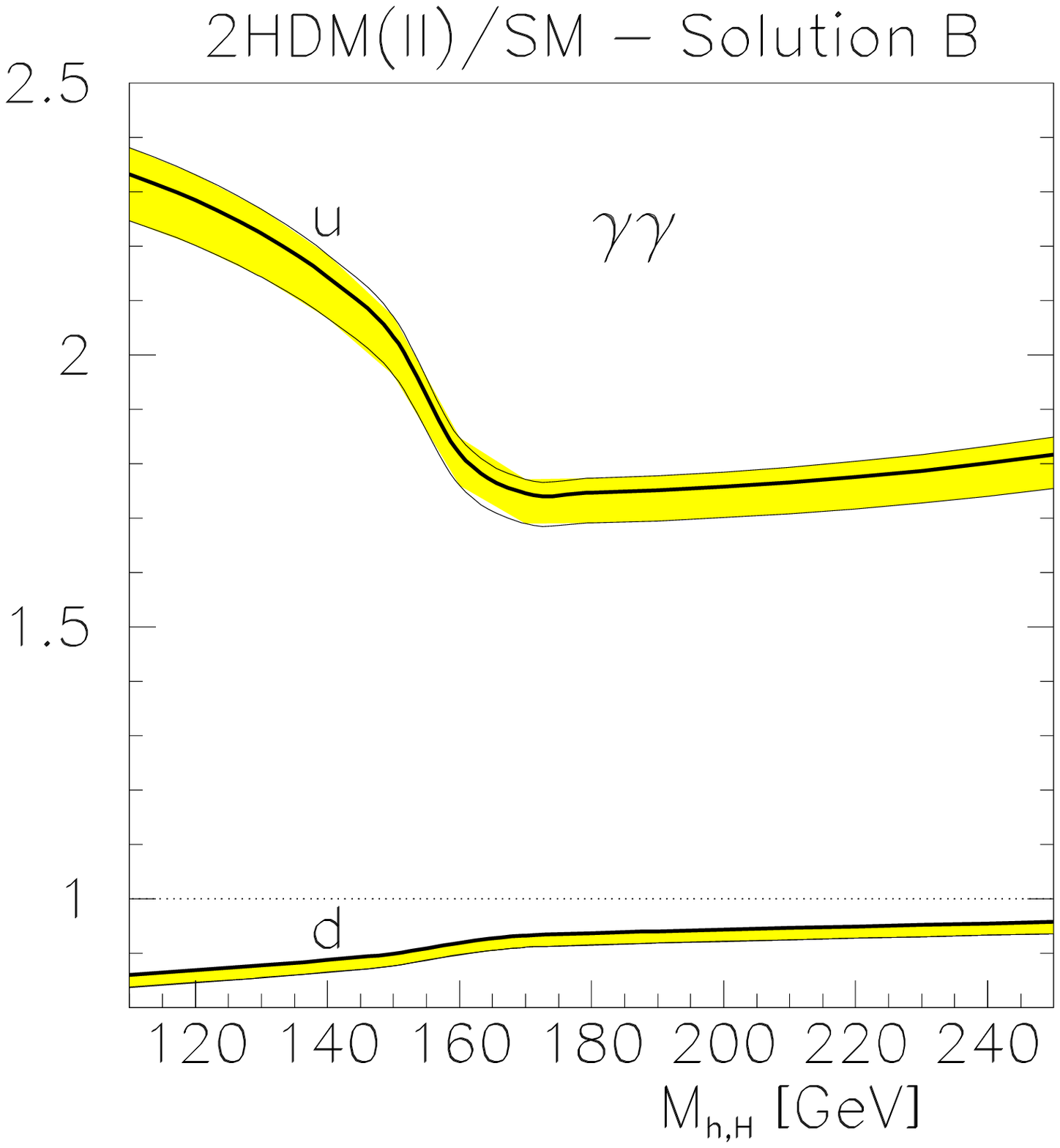,width=6.5cm}
\caption{The ratio of the two--photon Higgs width in the \2HDM\ to its
\SM\ value, for two different solutions~\protect\cite{GKO1,GKO2}.}
\label{fig:gaga-a,b}
\end{figure}

The \CP\ parity of the neutral Higgs boson can
be measured using linearly polarised photons.  Moreover, if the Higgs
boson is a mixture of \CP--even and \CP--odd states, for instance in a
general 
\2HDM\ with a \CP--violating neutral sector, the 
interference of these two terms gives rise to a \CP--violating
asymmetry~\cite{GF92,GK94,KKSZ94,GIv,ACHL00}.  Two
\CP--violating ratios could be observed to linear order in the
\CP--violating couplings:

$$
{\mathcal A_1}=\frac{|{\mathcal M}_{++}|^2-|{\mathcal M}_{--}|^2} {|{\mathcal
M}_{++}|^2+|{\mathcal M}_{--}|^2}, \quad {\mathcal A}_2=\frac{2\Im({\mathcal
M}_{--}^*{\mathcal M}_{++})} {|{\mathcal M}_{++}|^2+|{\mathcal M}_{--}|^2}.
$$
%
%
In terms of experimental values the first asymmetry can be found from
$$
T_{-} = \frac{N_{++}-N_{--}}{N_{++}+N_{--}}=
\frac{\langle\xi_2\rangle+\langle\tilde\xi_2\rangle}
{1+\langle\xi_2\tilde\xi_2\rangle}{\mathcal A}_1,
$$
where $N_{\pm\pm}$ correspond to the event rates for positive
(negative) initial photon helicities and $\xi_i$, $\tilde\xi_i$ are the 
Stokes polarisation parameters. The measurement
of the asymmetry is achieved by simultaneously flipping the helicities
of the laser beams used for production of polarised electrons and
$\G \to \EL$ conversion.  The asymmetry to be measured with linearly
polarised photons is given by
\begin{equation}
T_\psi = \frac{N(\phi=\frac{\pi}{4})-N(\phi=-\frac{\pi}{4})}
{N(\phi=\frac{\pi}{4})+N(\phi=-\frac{\pi}{4})} = 
\frac{\langle\xi_3\tilde\xi_1\rangle+\langle\xi_1\tilde\xi_3\rangle}
{1+\langle\xi_2\tilde\xi_2\rangle}{\mathcal A}_2,
\label{poldep}
\end{equation}
where $\phi$ is the angle between the linear polarisation vectors of
the photons. The asymmetries are typically larger than
10\% and they are observable for a large range of the \2HDM\ parameter
space if \CP\ violation is present in the Higgs potential.

Hence, high degrees of both circular and linear
polarisations for the high energy photon beams provide additional analysing
power for the detailed study of the Higgs sector at the $\GG$ collider.

\subsection{Supersymmetry}\label{s2.3}

In $\GG$ collisions, any kind of  charged particle can be
produced in pairs, provided the mass is below the kinematical bound.
Potential SUSY targets for a photon collider are the charged sfermions~\cite{JLCgg,Klasen}, the charginos~\cite{JLCgg,Mayer}
and the charged Higgs bosons.

For the \GG\ luminosity given in the Table~\ref{tabtel}, the production
rates for these particles will be larger than that in \EPEM\ 
collisions and detailed studies of the charged supersymmetric
particles should be possible.  In addition, the
cross sections in \GG\ collisions are given just by QED to leading
order, while in \EPEM\ collisions also $\Z$ boson and (sometimes)
t--channel exchanges contribute.  So, studying these 
processes in both channels provides complementary information about the
interactions of the charged supersymmetric particles.

The \GE\ collider could be the ideal machine for the discovery of
scalar electrons ($\tilde{\EL}$) and neutrinos ($\tilde{\nu}$) in the reactions
$\GE \to \tilde{\EL}^- \tilde{\chi}_1^0 ,\, \tilde{\W}\tilde{\nu}$~\cite{JLCgg,Cuyp1,Cuyp2,Cuyp3,GK921,GK922}.  Selectrons and neutralinos may be discovered in
\GE\ collisions up to the kinematical limit of
\begin{equation}
\mass{{\tilde{\EL}^-}} < 0.9\times 2E_0 - \mass{{\tilde{\chi}_i^0}} , 
\label{eq:sebound}
\end{equation}
where $2E_0$ is the energy of the original \EPEM\ collider.
This bound is larger than the bound obtained from $\tilde{\EL}^+ \tilde{\EL}^-$
pair production in the \EPEM\ mode, if 
$\mass{{\tilde{\chi}_i^0}} < 0.4 \times 2E_0$.

\begin{figure}[h]
\centering
\includegraphics[width=15.5cm, angle=0, bb= 78 578 521 743,clip] {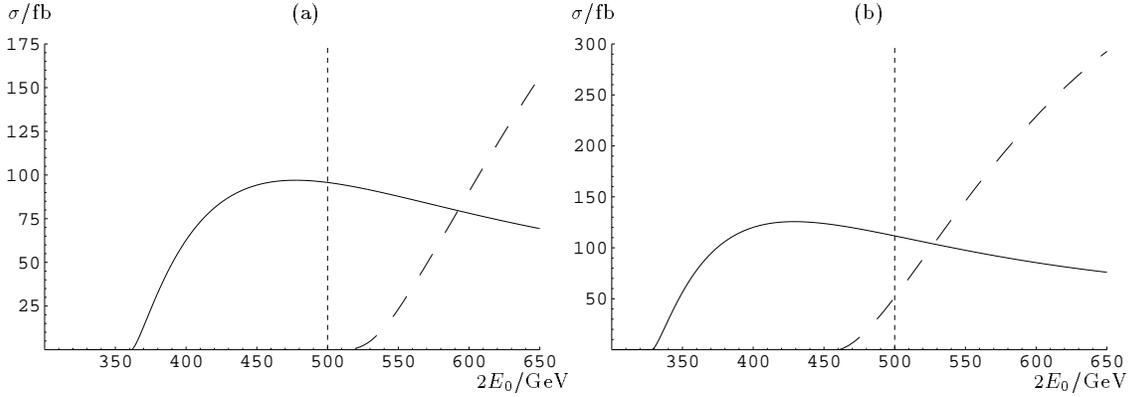}
\caption{Total cross sections for $\GE \rightarrow
  \tilde{\chi}_1^0\tilde{\EL}_{L/R}^-\rightarrow
  \tilde{\chi}_1^0\tilde{\chi}_1^0\EL^-$ (solid curves) for
  longitudinal polarisation $P_{\EL^-}=0.8$ and longitudinal (circular)
  polarisation $P_{\EL_c}=0.8$ ($\lambda_L=-1$) of the converted
  electrons (laser photons) compared to
  $\EPEM\rightarrow\tilde{\EL}_{L/R}^+\tilde{\EL}^-_{L/R}\rightarrow
  \tilde{\chi}_1^0\tilde{\chi}_1^0\EPEM$ (dashed curves) with
  longitudinally polarised electrons, $P_{\EL^-}=0.8$, and unpolarised
  positrons. MSSM parameters: $\mass{2}=152\GEVI$, $\mu=316\GEVI$,
  $\tan\beta=3$. (a) $\mass{{\tilde{\EL}_{R}}}=260\GEVI$,
  $\mass{{\tilde{\EL}_{L}}}=290\GEVI$. (b) $\mass{{\tilde{\EL}_{R}}}=230\GEVI$,
  $\mass{{\tilde{\EL}_{L}}}=270\GEVI$.} 
\label{egamma1}
\end{figure}

In Fig.~\ref{egamma1} the cross section  of the
process $\GE \rightarrow
\tilde{\chi}_1^0\tilde{\EL}_{L/R}^-\rightarrow
\tilde{\chi}_1^0\tilde{\chi}_1^0e^-$ is compared to the cross
section  of the process
$\EPEM\rightarrow\tilde{\EL}_{L/R}^+\tilde{\EL}^-_{L/R}
\rightarrow\tilde{\chi}_1^0\tilde{\chi}_1^0\EPEM$
for  the \MSSM\ parameters $\mass{2}=152\GEV$, $\mu=316\GEV$, $\tan\beta=3$ and
$\mass{{\tilde{\EL}_{R}}}=260\GEV$, $\mass{{\tilde{\EL}_{L}}}=290\GEV$ 
(Fig.~\ref{egamma1}a) and $\mass{{\tilde{\EL}_{R}}}=230\GEV$,
$\mass{{\tilde{\EL}_{L}}}=270\GEV$ (Fig.~\ref{egamma1}b)~\cite{bloechi1,bloechi2}. The
$\tilde{\chi}_1^0$ mass in this case is about $70\GEV$.  For
higher selectron masses pair production in \EPEM\ annihilation at
$2E_0=500\GEV$ is kinematically forbidden, whereas in
\GE\ collisions the cross section at $2E_0=500\GEV$ is
$96\fb$. According to~(\ref{eq:sebound}) the highest accessible selectron mass
for $2E_0 = 500\GEV$ is $\mass{{\tilde{\EL}}}<380\GEV$ in this scenario.

In some scenarios of supersymmetric extensions of the Standard Model
the stoponium bound states $\tilde{t}\bar{\tilde{t}}$ is formed.  A
photon collider would be the ideal machine for the discovery and study
of these new narrow strong resonances~\cite{Ilyin}.  About ten
thousand stoponium resonances for $\mass{S}=200\GEV$ will be produced for an
integrated luminosity in the high energy 
peak of $100\fb^{-1}$.  Thus precise measurements of the stoponium
effective couplings, mass and width should be possible.  At \EPEM\ 
colliders the counting rate will be much lower and in some scenarios
the stoponium cannot be detected due to the large background~\cite{Ilyin}.

\subsection{Extra dimensions}\label{s2.6}

\begin{figure*}[htb]
\centerline{
\psfig{figure=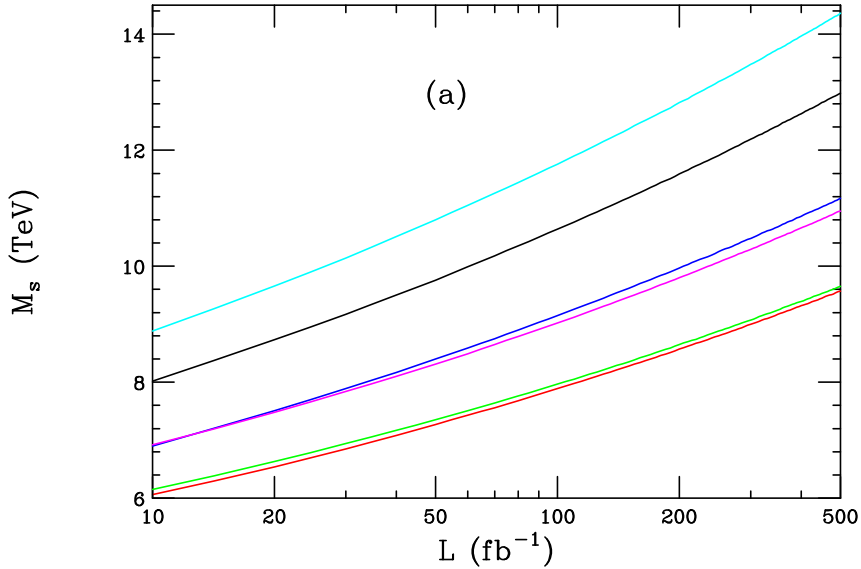,width=7.5cm,angle=90}}
\caption[*]{$M_s$ discovery reach for the process $\GG\to \WP\WM$ at a
  $2E_0=1\TEVI$ LC  as a function of the 
  integrated luminosity for the different initial state
  polarizations assuming $\lambda=1$.  From top to bottom on the
  right hand side of the figure the polarisations are $(-++-)$,
  $(+---)$, $(++--)$, $(+-+-)$, $(+---)$, and $(++++)$.}
\label{fig:extra6}
\end{figure*}

New ideas have recently been proposed to explain the weakness of the
gravitational force~\cite{add1,add2,add3}. The Minkowski world is extended by extra
space dimensions which are curled up at small dimensions $R$. While the gauge
and matter fields are confined in the (3+1) dimensional world, gravity
propagates through the extended 4+n dimensional world. While the effective
gravity scale, the Planck scale, in four dimensions is very large, the
fundamental Planck scale in 4+n dimensions may be as low as a few TeV so that
gravity may become strong already at energies of the present or next
generation of colliders.

Towers of Kaluza--Klein graviton excitations will be realised on the
compactified 4+n dimensional space. Exchanging these KK excitations between
\SM\ particles in high--energy scattering experiments will generate effective
contact interactions, carrying spin=2 and characterised by a scale $M_s$ of
order few TeV. They will give rise to substantial deviations from the
predictions of the Standard Model for the cross sections and angular
distributions for various beam polarisations~\cite{pheno1,pheno2,pheno3,pheno4,pheno5,rizzo}.

Of the many processes examined so far, $\GG\to\WW$ provides
the largest reach for $M_s$ for a given 
centre of mass energy of the LC~\cite{rizzo1,rizzo}.  
The main reasons are that the $\WW$ final
state offers many observables which are particularly sensitive to the
initial electron and laser polarisations and the very high
statistics due to the $80\pb$ cross section.

By performing a combined fit to the total cross sections and angular
distributions
for various initial state polarisation choices and the
polarisation asymmetries, the discovery
reach for $M_s$ can be estimated as a function of the total $\GG$
integrated luminosity. This is  shown in Fig.~\ref{fig:extra6}~\cite{rizzo}. The reach is in the range of
$M_s\sim (11$--$13)\cdot 2E_0$, which is larger than that obtained
from all other processes examined so far. By comparison, a
combined analysis of the processes $\EPEM\to \ffbar$ with the
same integrated luminosity leads to a reach of only $(6$--$7)\cdot 2E_0$.

Other $\GG$ final states are also
sensitive to graviton exchanges, two examples being the
$\GG$~\cite{Cheung,Davoudiasl} and $\ZZ$~\cite{rizzo1} final
states, which however result in smaller search reaches. 


\subsection{Gauge bosons}\label{s2.4}

New strong interactions
that might be responsible for the electroweak symmetry breaking can
affect the triple and quartic couplings of the weak vector bosons. Hence, the
precision measurements of these couplings, as well as corresponding
effects on the top quark couplings, can
provide clues to the mechanism of the electroweak symmetry breaking.

Due to the large cross sections of the
order of $10^2\pb$ well above the thresholds, the $\GG\to\WP\WM$ and
$\GE \to\nu \W$ processes seem 
to be ideal reactions to study such anomalous gauge interactions~\cite{GKPS1,GKPS2}.

\subsubsection{Anomalous gauge boson couplings}

The relevant process at the \EPEM\ collider is $\EPEM\to\WP\WM$.  This
reaction is 
dominated by the large t--channel neutrino exchange term which
however can be suppressed using electron beam polarisation. The cross
section of $\WP\WM$ pair production in $\EPEM$ collisions with
right--handed electron beams, for which the neutrino exchange is negligible,
has a maximum of about $2\pb$ at LEP2 and decreases at higher energy.

The two main processes at the Photon Collider are $\GG\to\WP\WM$ and 
$\GE\to\W\nu$.
Their total cross sections for centre--of--mass energies above $200\GEV$
are about $80\pb$ and $40\pb$, respectively, and they do not decrease with
energy. Hence the $\W$ production cross sections at the Photon Collider
are at least $20$--$40$ times larger than the cross section at the \EPEM\
collider. This enhancement makes event rates at 
the Photon Collider  one order of magnitude larger than at an
\EPEM\ collider, even when the lower \GG, \GE\ 
luminosities are taken into account. Specifically
for the integrated \GG\ luminosity of $100\fb^{-1}$, about
$8\times 10^6$ $\WP\WM$ pairs are produced at the Photon Collider.
Note that while $\GE \to \W\nu$ and $\GG \to \WW$ isolate
the anomalous photon couplings to the $\W$, $\EPEM\to\WW$ involves
potentially anomalous $\Z$ couplings so that the two LC modes are complementary
with each other.

The analysis of $\GG\ \to \WW$ has been performed in~\cite{JLCgg,TakahashiWW}
with the detector simulation. The \W\ boson by
photon colliders is compared to that from \EPEM\ colliders. The
results have been obtained only from analyses of the total cross
section. With the $\W$ decay properties taken into account 
further improvements can be expected. The resulting accuracy on 
$\lambda_{\gamma}$ is comparable with \EPEM\ analyses, while a similar
accuracy on $\delta \kappa_{\gamma}$ can be achieved at $1/20$--th of the
\EPEM\ luminosity.
In addition, the process $\GE\to\W\nu$, which has a large cross section, is
very sensitive to the admixture of right--handed currents in the $\W$
couplings with fermions: $\sigma_{\GE\to\W\nu} \propto (1$--$2 \lambda_{\EL})$.

Many processes of 3rd and 4th order have quite large cross sections~\cite{GinLCWS,3order,GinW,ginzburg} at the Photon
Collider:

\begin{center}
\begin{tabular}{l}
$\GE\to \EL\WW$ \\
$\GE\to \nu\W\Z$ \\
\end{tabular}\begin{tabular}{c}
\qquad\qquad
\end{tabular}\begin{tabular}{l}
$\GG\to \Z\WW$ \\
$\GG\to \WW\WW$ \\
$\GG\to \WW\ZZ$ \\
\end{tabular}
\end{center}

It should also be noted, that in $\GG$ collisions the
anomalous $\GG\WP\WM$ quartic couplings can be probed.
However, the higher event rate does not
necessarily provide better bounds on anomalous couplings.  In some
models electroweak symmetry breaking leads to large deviations mainly
in longitudinal $\W_L \W_L$ pair production~\cite{BBB}. On the other hand
the large cross section of the reaction $\GG\to\WP\WM$ is due to
transverse $\W_T \W_T$ pair production. In such a case transverse
$\W_T \W_T$ pair production would represent a background for the
longitudinal $\W_L \W_L$ production.  The relative yield of $\W_L \W_L$ can
be considerably improved after a cut on the $\W$ scattering angle.
Asymptotically for $s_{\GG} \gg \mass{\W}^2$ the production of $\W_L \W_L$ is
as much  as 5 times larger than at a \EPEM\ LC. 

However, if anomalous couplings manifest themselves in transverse
$\W_T \W_T$ pair production, e.g. in theories with large
extra dimensions, then the interference with the large
\SM\ transverse contribution is of big advantage in the Photon
Collider.

\subsubsection[Strong WW $\to$ WW, WW $\to$ ZZ scattering]
{Strong WW $\bto$ WW, WW $\bto$ ZZ scattering}
 
If the strong electroweak symmetry breaking scenario is realised in Nature,
$\W$ and $\Z$ bosons will interact strongly at high energies. If no
Higgs boson exists with a mass below $1\TEV$, the longitudinal
components of the electroweak gauge bosons must become strongly
interacting at energies above $1\TEV$. In such scenarios novel
resonances can be formed in $\W_L \W_L$ collisions at energies $\lsim 3\TEV$. 
If the energy of the $\GG$ collisions
is sufficiently high, the effective $\W$ luminosities in $\GG$
collisions  allow  the study of
$\WP\WM\to\WP\WM$, $\ZZ$ scattering in the reactions 
$$ 
\GG\to \WW\WW,\;\; \WW\ZZ 
$$ 
for energies in the threshold region of the new strong interactions. Each
incoming photon turns into a virtual $\WW$ 
pair, followed by the scattering of one $\W$ from each such pair to
form $\WW$ or $\ZZ$~\cite{boudjema1,boudjema2,brodsky,JikiaWWWW1,JikiaWWWW2,CheungWWWW1,CheungWWWW2}.  The
same reactions can be used to study quartic anomalous $\WW\WW$, $\WW\ZZ$
couplings.

\subsection{Top quark}\label{s2.5}

The top quark is heavy and up to now point--like at the same time.  The
top Yukawa coupling $\lambda_{t} = 2^{3/4}G_F^{1/2}\mass{t}$ is
numerically very close to unity, and it is not clear whether or not
this is related to a deep physics reason.  Hence one might expect
deviations from \SM\ predictions to be most pronounced in the top
sector~\cite{peccei1,peccei2}.  Besides, top quarks decay before 
forming a bound state with any other quark. Top quark physics
will be a very important part of research programs for all future
hadron and lepton colliders.  The $\GG$ collider is of
special interest because of the clean production mechanism and the
high rate (review~\cite{hewett}).
Moreover, the $S$ and $P$ partial waves of the final state top
quark--antiquark pair produced in $\GG$ collisions can be
separated by choosing the same or opposite helicities of the colliding
photons.

\subsubsection{Probe for anomalous couplings in t$\bbar{\rm \bf t}$ pair production} 

There is a difference for the case of \GG\ and \EPEM collisions with respect
to the couplings: the $\G\ttbar$ coupling is separated from $\Z \ttbar$
coupling in $\GG$ collisions while in \EPEM\ 
collisions both couplings contribute.

The effective Lagrangian contains four parameters $f_i^{\alpha}$ for the
electric and magnetic type couplings~\cite{Boos}, where $i=1$--$4$ and $\alpha = \gamma,\Z$ but only couplings with
$\alpha = \gamma$ occur in $\gamma\gamma$ collisions.  It was demonstrated~\cite{djouadi} that if the cross section can be measured with 2\%
accuracy, scale parameter for new physics
$\Lambda$ up to $10\TEV$ for $2E_0 = 500\GEV$ can be probed for form factors
taken in the form $f_i^{\alpha}=(f_i^{\alpha})^{\SM}(1+s/\Lambda^2)$.  The
sensitivity to the anomalous 
magnetic moment $f_2^{\gamma}$ is of similar size in \GG\ and
\EPEM\ collisions.  The $f_4^{\alpha}$ term describes the \CP\
violation. The best limit on the imaginary part of the electric dipole
moment $\Im(f_4^{\gamma}) \sim 2.3\times 10^{-17}e\CM$~\cite{rindani}
by measuring the forward--backward asymmetry $A_{fb}$ with initial--beam
helicities of electron and laser beams $\lambda_e^1=\lambda_e^2$ and
$\lambda_l^1= -\lambda_l^2$. The achievable limit for the real part of
the dipole moment is also of the order of $10^{-17}e\CM$ and is obtained
from the linear polarisation asymmetries~\cite{choi1,choi2}. 

\begin{figure}[htbp]
\centering{
\epsfig{file=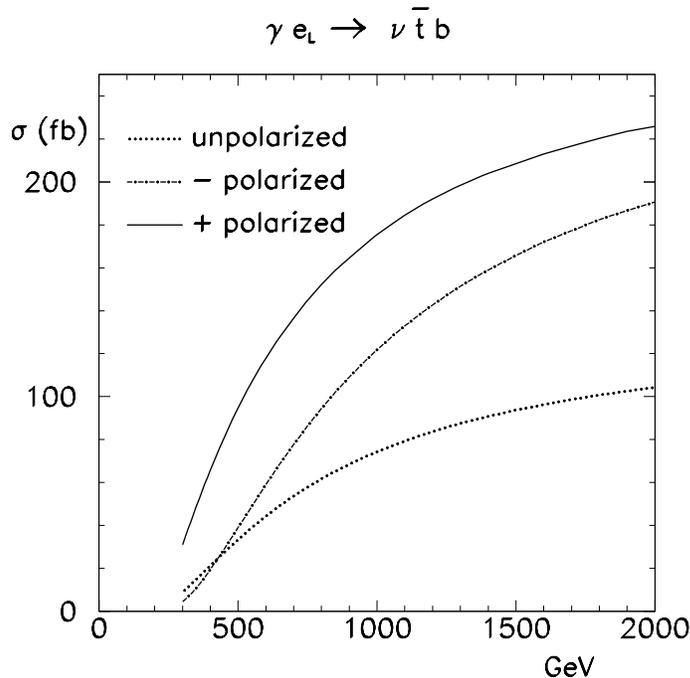,width=10cm
}}
\caption{Single top quark production cross section in $\GE$
collisions as a function of $2E_0$.}
\label{fig:f9}
\end{figure}

\subsubsection[Single top production in $\gamma\gamma$ and $\gamma$e
  Collisions]
{Single top production in $\bgamma\bgamma$ and $\bgamma$e
  Collisions}

Single top production in $\GG$ collisions results in the same  
final state as top quark pair production~\cite{boos-gg} and 
invariant mass cuts are required to suppress direct $\ttbar$  
contributions.  Single top production is preferentially realised in $\GE$
collisions~\cite{singletop1,singletop2,singletop3,singletop4,singletop5}. In contrast to the top 
pair production rate, the single top rate is directly proportional to 
the $\W tb$ coupling and the process is very sensitive to its 
structure. The anomalous part of the effective Lagrangian~\cite{Boos} contains
terms $f_{2L(R)} \propto 1/\Lambda$, where 
$\Lambda$ is the scale of a new physics. 

\begin{table}
 
 \begin{center}
    \begin{tabular}{|l||c|c|}\hline
        &{$f_{2L}$}
        &{$f_{2R}$} \\ \hline\hline
      TEVATRON ($\Delta_{sys.}\approx10\%$)
               & $-0.18 \div +0.55$ & $-0.24 \div +0.25$ \\
      LHC ($\Delta_{sys.}\approx5\%$)
               & $-0.052 \div +0.097$ & $-0.12 \div +0.13$ \\
      \hline
      \EPEM\ ($2E_0=0.5\TEV$) 
               & $-0.025\div +0.025$ & $-0.2\div +0.2$ \\  
      \GE\ ($2E_0 =0.5\TEV$)
               & $-0.045 \div +0.045$ & $-0.045$ $\div +0.045$ \\
      \GE\ ($2E_0=2.0\TEV$) 
               & $-0.008 \div +0.035$ & $-0.016 \div +0.016$\\
      \hline
    \end{tabular}%
  \end{center}
\caption{Expected sensitivity for the $\W tb$ anomalous couplings.
 The total integrated luminosity was assumed to be
$500fb^{-1}$ for \EPEM\ collisions and $250fb^{-1}$ and
$500~fb^{-1}$ for \GE\ collisions at $500\GEVI$ and $2\TEVI$,
respectively.
\label{tb:par}}
\end{table}

In  Table~\ref{tb:par}~\cite{boos-dudko-ohl,vtb-ee} 
limits on anomalous couplings from measurements at different
accelerators are collected. The best limits can be reached at 
very high energy \GE\ colliders, even in the case of 
unpolarised collisions.  In the case of polarised collisions, 
the production rate increases significantly as shown in 
Fig.~\ref{fig:f9}~\cite{boos-gg} and more stringent  
bounds on anomalous couplings may be achieved.

\subsection{QCD and hadron physics}\label{s2.7} 

Photon colliders offer a unique possibility to probe QCD in a new
unexplored regime. The very high luminosity, the (relatively) sharp
spectrum of the backscattered laser photons and their polarisation are of
great advantage.  At the Photon Collider the following measurements can be
performed, for example:
\begin{enumerate}
\item
  The total cross section for \GG\ fusion to hadrons~\cite{Godbole}.
\item  
  Deep inelastic \GE\ \NC\ and \CC\ scattering, and  measurement of the
  quark distributions in the photon at large $Q^2$.
\item 
  Measurement of the gluon distribution in the photon.
\item 
  Measurement of the spin dependent structure 
  function $g_1^{\G}(x,Q^2)$ of the photon.
\item 
  $J/\Psi$ production in \GG\ collisions as a probe of the hard QCD
  pomeron~\cite{GPS1,GPS2,GPS3}.
\end{enumerate}

\vspace{.3cm}
{\bf $\bgamma\bgamma$ fusion to hadrons} 
\vspace{.3cm}

The total cross section for hadron production in \GG\ collisions is a
fundamental observable. It provides us with a picture of hadronic fluctuations
in photons of high energy which reflect the strong--interaction dynamics as
described by quarks and gluons in QCD. Since these dynamical processes involve
large distances, predictions, due to the theoretical complexity, cannot be
based yet on first principles. Instead, phenomenological models have been
developed which involve elements of ideas which have successfully been applied
to the analysis of hadron--hadron scattering, but also elements transferred
from perturbative QCD in eikonalised mini--jet models. Differences between
hadron--type models and mini--jet models are dramatic in the TESLA energy
range. \GG\ scattering experiments are therefore extremely valuable in
clarifying the dynamics in complex hadronic quantum fluctuations of the
simplest gauge particle in Nature.

\vspace{.3cm}
{\bf Deep inelastic $\bgamma$e scattering (DIS)}
\vspace{.3cm}

The large c.m. energy in the \GE\ system and the possibility of precise
measurement of the kinematical variables 
$x,Q^2$ in DIS provide exciting opportunities at a photon collider.
In particular it allows precise measurements of
the photon structure function(s) with much better accuracy than in the
single tagged \EPEM\ collisions.  The \GE\ collider offers a
unique opportunity to probe the photon  at low values of $x$ ($ x\sim 10^{-4}$)
for reasonably large values of $Q^2 \sim 10\GEV^2$~\cite{PC}. At very large
values of $Q^2$ the virtual $\G$ exchange in deep inelastic \GE\ 
scattering is supplemented by
significant contributions from  $\Z$ exchange.  Moreover, at
very large values of $Q^2$ charged--current exchange becomes effective in deep
inelastic scattering,
$\GE \to \nu X$,
which is mediated by virtual $\W$
exchange.  The study of this process can in particular give
information on the flavour
decomposition of the quark distributions in the photon~\cite{ZERDR}.

\vspace{0.3cm}
{\bf Gluon distribution in the photon}
\vspace{.3cm}

The gluon distribution in the photon can be studied in dedicated
measurements of the hadronic final state in \GG\ collisions.
The following two processes are of particular interest:
 \begin{enumerate}
\item Dijet production~\cite{gamgtojets1,gamgtojets2}, generated by the subprocess 
  $\G g \to \qqbar$. 
\item  Charm production~\cite{gamgtocharm}, which is sensitive to the mechanism 
  $\G g \to \ccbar $ 
\end{enumerate}

Both these  processes, which are at least in certain kinematical
regions dominated by the photon--gluon fusion  mechanisms,  are sensitive to
the gluon distribution in 
the photon. The detailed discussion of these processes have been
presented in~\cite{WENGLER,ALBERT}.

\vspace{0.3cm}
{\bf Measurement of the spin dependent structure function
  g$_\bone^{\bgamma}$(x,Q$^\btwo$) of the Photon} 
\vspace{.3cm}

Using polarised beams,
photon colliders offer the possibility to measure the spin dependent
structure function $g_1^{\G}(x,Q^2)$ of the photon~\cite{GG11,GG12,GG13}.
This  quantity is completely unknown and its measurement in
polarised \GE\ DIS would be extremely interesting for testing QCD
predictions in a broad region of $x$ and $Q^2$.  The high--energy
photon colliders allow to probe this quantity for  very small
values of $x$
\cite{JKBZGG11,JKBZGG12}.  

\vspace{0.3cm}
{\bf Probing the QCD pomeron by  J/$\bPsi$ production 
in $\bgamma\bgamma$ Collisions}
\vspace{.3cm}

The exchange of the hard QCD (or BFKL) pomeron is presumably the
dominant mechanism of the process $\GG\to J/\psi\, J/\psi$.  Theoretical
estimates of the cross--section presented in~\cite{KMJPSI1,KMJPSI2} have demonstrated that measurement of the reaction 
$\GG\to J/\psi\, J/\psi$ at the Photon Collider should be
feasible.


\subsection{Table of gold--plated processes}\label{s2.8}

A short list of processes which we think are the most important ones
for the physics program of the Photon Collider option of the LC is
presented in Table~\ref{processes}.

\begin{table}[!hbtp]
{\renewcommand{\arraystretch}{1.2}
\begin{center}
\begin{tabular}{ l  c } 
\hline
$\quad$ {\bf Reaction} & {\bf Remarks} \\
\hline\hline
$\GG\to \hZ \to \bbbar$ & \SM\ (or \MSSM\ ) Higgs, 
$\mass{\hZ}<160\GEV$  \\
$\GG\to \hZ \to \WW(\WW^*)$    & \SM\ Higgs,
$140\GEV<\mass{\hZ}<190\GEV$ \\
$\GG\to \hZ \to \ZZ(\ZZ^*)$      & \SM\ 
Higgs, $180\GEV<\mass{\hZ}<350\GEV$ \\
\hline
$\GG\to \H,\A\to \bbbar$  &
 \MSSM\ heavy Higgs, for intermediate $\tan\beta$\\
$\GG\to \tilde{f}\bar{\tilde{f}},\
\tilde{\chi}^+_i\tilde{\chi}^-_i,\ \HP\HM$ & large cross sections,
possible observations of FCNC \\ 
$\GG\to S[\tilde{t}\bar{\tilde{t}}]$ & 
$\tilde{t}\bar{\tilde{t}}$ stoponium  \\
$\GE \to \tilde{\EL}^- \tilde{\chi}_1^0$ & $\mass{{\tilde{\EL}^-}} < 
0.9\times 2E_0 - \mass{{\tilde{\chi}_1^0}}$\\
\hline
$\GG\to \WP\WM$ & anomalous $\W$ interactions, extra dimensions \\
$\GE^-\to \WM\nu_{\EL}$ & anomalous $\W$ couplings \\
$\GG\to \WW\WW$,$\WW\ZZ$& strong $\WW$ scatt., 
quartic anomalous $\W$, $\Z$  couplings\\
\hline
$\GG\to \ttbar$ & anomalous top quark interactions \\
$\GE^-\to \bar t b \nu_e$ & anomalous $\W tb$ coupling \\
\hline
$\GG\to$ hadrons & total \GG\ cross section \\
$\GE^-\to \EL^- X$ and $\nu_{\EL}X$ & \NC\ and \CC\ structure functions
(polarised and unpolarised) \\ 
$\G g\to \qqbar,\ \ccbar$ & gluon distribution in the photon \\
$\GG\to J/\psi\, J/\psi $ & QCD Pomeron \\
\hline
\end{tabular}
\end{center}
}
\caption{Gold--plated processes at photon colliders
\label{processes}}
\end{table}

Of course there exist many other possible manifestations of new
physics in $\GG$ and $\GE$ collisions which we have not
discussed here. The study of resonant production of
excited electrons  $\GE\to \EL^*$, the production of
excited fermions $\GG\to f^* f$, leptoquark production 
$\GE\to (eQ)\bar{Q}$~\cite{leptoquark1,leptoquark2}, a magnetic monopole signal in the
reaction of $\GG$ elastic scattering~\cite{monopole1,monopole2} etc. may be mentioned in
this context. 

To summarise, the Photon Collider will allow us to study the physics
of the EWSB in both the weak--coupling and the strong--coupling
scenarios.
Measurements of the two--photon Higgs width of the
$\h$, $\H$ and $\A$ Higgs states provide a strong physics motivation for
developing the technology of the $\GG$ collider option.  Polarised photon beams, large cross sections and
sufficiently large luminosities allow to significantly enhance the
discovery limits of many new particles in SUSY and other extensions of the
Standard Model. Moreover, they will   
substantially improve the accuracy of the precision measurements of
anomalous $\W$ boson and top quark couplings, thereby complementing and
improving the measurements at the \EPEM\ mode of TESLA. Photon
colliders offer a unique possibility for probing the photon structure
and the QCD Pomeron. 
\clearpage

\section{Electron to Photon Conversion}\label{s3}
\subsection{Processes in the conversion region}\label{s3.1}
\subsubsection{Compton scattering}\label{s3.1.1}

Compton scattering is the basic process for the production of high
energy photons at photon colliders. The fact that a high energy
electron loses a large fraction of its energy in collisions with an
optical photon was realized a long time ago in astrophysics~\cite{finberg}.
The method of generation of high energy \G--quanta by Compton
scattering of the laser light on relativistic electrons has been
proposed soon after  lasers were invented~\cite{ARUT1,ARUT2} and has already been used
in many laboratories  for more than 35 years~\cite{kulikov,ballam}.
In first experiments the conversion efficiency of electron to photons
$k=N_{\G}/N_{\EL}$ was very small, only about $10^{-7}$~\cite{ballam}.
At linear colliders, due to small bunch sizes one can focus the laser
 to the electron beam and get $k\approx 1$ at rather
moderate laser flash energy, about $1$--$5\J$. Twenty years ago when
photon colliders were proposed~\cite{GKST81,GKST83} such flash
energies could already be obtained but with a low rate \footnote{The
  proposed linear collider VLEPP (Novosibirsk) had initially only $10\Hz$ rep.
  rate with one bunch per ``train'', in present projects the collision 
  rate is about $10\kHz$ which is much more difficult.} 
and a pulse duration
longer than  is necessary.  Progress in laser
technology since that time now presents  a real possibility for the
construction of a laser system for a photon collider.

\vspace{.3cm}
{\bf Kinematics, photon spectrum}
\vspace{.3cm}

Let us consider the most important characteristics of Compton scattering.
In the conversion region a laser photon with  energy $\omega_0$
scatters at a small collision angle $\alpha_0$ off a high energy
electron with  energy $ E_0$.  The energy of the scattered photon
$\omega$ depends on the photon scattering angle as follows~\cite{GKST83}:

\begin{equation}
\omega = \frac{\omega_m}{1+(\vartheta/\vartheta_0)^2},\;\;\;\;
\omega_m=\frac{x}{x+1}E_0, \;\;\;\;
\vartheta_0= \frac{mc^2}{E_0} \sqrt{x+1},
\end{equation}
where

\begin{equation}
x=\frac{4E_0 \omega_0 }{m^2c^4}cos^2{\alpha_0/2}
 \simeq 15.3\left[\frac{E_0}{\TEV}\right]
\left[\frac{\omega_0}{\eV}\right] = 
 19\left[\frac{E_0}{\TEV}\right]
\left[\frac{\mu m}{\lambda}\right],
\label{e3:x}
\end{equation}
$\omega_m$ is the maximum energy of scattered photons (in the direction of the
electron, Compton ``backscattering'').

For example: $E_0=250\GEV$, $\omega_0 =1.17\eV$ ($\lambda=1.06\MKM$) (region of
most powerful solid--state lasers) $\Rightarrow x=4.5$ 
and $\omega_m/E_0 = 0.82$.

The energy spectrum of the scattered photons is defined by the
Compton cross section 

\begin{equation}
\frac{1}{\sigma_c} {d\sigma _{c}\over d y}   = {2\sigma_{0}\over x\sigma_{c}}
\left[{1\over 1-y} + 1-y - 4r(1-r) +
2\lambda_{\EL}P_c r x(1-2r)(2-y)\right]\,,
\label{e3:spect}
\end{equation}
$$ 
y = \omega /E_0, \; \;\;\;\;    r=\frac{y}{(1-y)x}, \;\;\;
\sigma_{0}=
\pi r_e^2 =
\pi\left({e^2\over mc^2}\right)^2 = 2.5\cdot
10^{-25}\CM ^{2}\, , 
$$
where $\lambda_{\EL} $ is the mean electron helicity
($|\lambda_{\EL}| \leq 1/2$) and $P_c $ is that of the laser photon
($|P_c | \leq 1$). It is useful to note that $r\to 1$ for 
$y\to y_m$.

The total Compton cross section is

$$
\sigma_{c}\;=\;\sigma_{c}^{0}\;+\;2\lambda_{\EL}P_c \,
\sigma_c^{1}\, ,
$$
\begin{equation}
\sigma_{c}^{0}\;=\; {2\sigma_{0}\over x}
\left[\left(1-{4\over x} - {8\over x^{2}}\right)\ln (x+1)
+{1\over 2} +{8\over x} - {1\over 2(x+1)^{2}}\right] \, ,
\label{e3:cross}
\end{equation}
$$
\sigma_c^{1}\;=\; {2\sigma_{0}\over x}
\left[\left(1+{2\over x}\right) \ln
(x+1)-{5\over 2}+{1\over x+1}-{1\over 2(x+1)^{2}}\right]\,.
$$
Polarisations of initial beams influence the differential and the
total cross section only if both their helicities are nonzero,
i.e. at $\lambda_{\EL} P_c \neq 0$.
In the region of interest

\begin{equation}
x = 1 \div  5,\;\;\;\sigma_c^{0}=(1.5\div 0.7)\, \sigma_0\,,\;\;\;
|\sigma_c^{1}|/ \sigma_{c} < 0.1\, ,
\label{e3:scsc1}
\end{equation}
i.e. the  total  cross section  only  depends slightly on  the
polarisation.

On the contrary, the energy spectrum  strongly depends on
the value of $\lambda_{\EL}P_c$. The ``quality" of the photon beam,
i.e. the relative number of hard photons, is improved when one
uses beams with a negative value of $\lambda_{\EL} P_c$. For
$2\lambda_{\EL}P_c=-1$ the peak at $\omega =\omega_{m}$ nearly
doubles, significantly improving  the
energy spread of the $\G$ beam

$$
{d \sigma_c (y_m,2\lambda_{\EL}P_c=-1)/d y\over d\sigma_c
(y_m,2\lambda_{\EL}P_c=0)/d y}  = {2\over 1+ (x+1)^{-2} }\,.
$$
The full width of the spectrum at
the half of maximum is $\Delta \omega_{1/2} \approx \omega_m/(x+2)$
for unpolarised beams, and even smaller at $\lambda_{\EL}P_c < 0$.
 Photons in this high energy peak have the characteristic angle
$\theta _{{\rm char}} = 1/\gamma = mc^2/E = 0.51 /E_0 [\TEV]\;
\mu \mbox {rad}$.   

To increase the maximum photon energy, one should use a laser with a
higher energy. This also increases the fraction of hard photons.
Unfortunately, at large $x > 4.8$, a new phenomenon takes place: the
high energy photons disappear from the beam, producing \EPEM\ pairs in
collisions with laser photons (see Section~\ref{s3.1.3}).   
Therefore, the value $x\approx 4.8$ is the most preferable.

\begin{figure}[!hbp]
\centering
\includegraphics[width=12cm,angle=0]{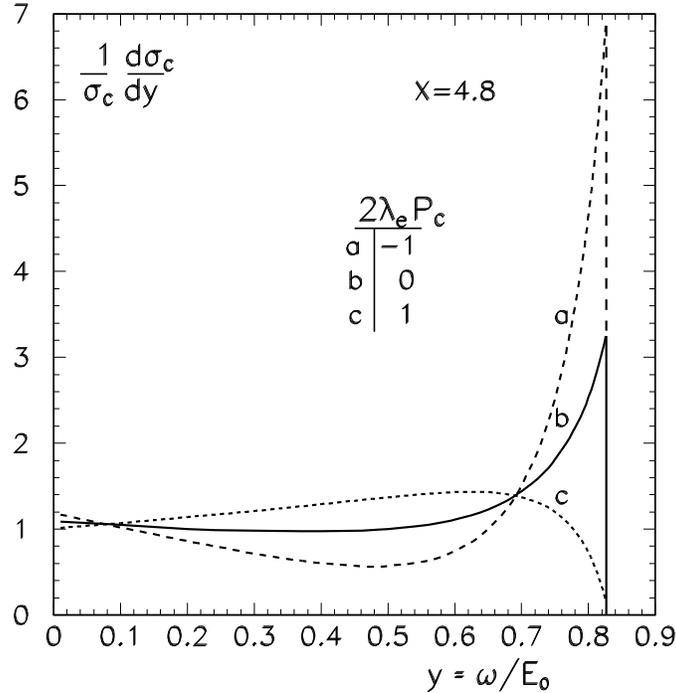}
\caption{Spectrum of the Compton scattered photons for different
polarisations of the laser and electron beams.}
\label{f3:fig4}
\end{figure}

The energy spectrum of the scattered photons for $x=4.8$ is shown in
Fig.~\ref{f3:fig4} for various helicities of electron and laser beams.
As was mentioned before, with the polarised beams at $2\lambda_{\EL}P_c =
-1$, that the number of high energy photons nearly doubles and the
luminosity in collisions of these photons is larger by a factor of 4.
This is one of the important advantages of polarised electron beams.

The photon energy spectrum presented in Fig.~\ref{f3:fig4} corresponds
to the case of a small conversion coefficient.  In the realistic case
when the thickness of the laser target is about one collision length
each electron may undergo multiple Compton scattering~\cite{TEL95}.
This probability is not small because, after a large energy loss in the
first collision, the Compton cross section increases and approaches 
the Thomson cross section $\sigma_T = (8/3)\sigma_0$.  The secondary
photons are softer  and populate the low energy part of the
spectrum. Multiple Compton scattering leads also to a low energy tail in
the energy spectrum of the electron beam after the $\EL \to \G$ conversion.
This creates a problem for the removal of the beams (see Section~\ref{s4.2}).

\vspace{.3cm}
{\bf Polarisation of scattered photons}
\vspace{.3cm}

The averaged helicity of photons after Compton scattering is~\cite{GKST84}

\begin{equation}
\langle\lambda_{\gamma}\rangle =
{-P_c(2r-1)[(1-y)^{-1}+1-y]+2\lambda_{\EL} xr[1+(1-y)(2r-1)^2]
\over (1-y)^{-1}+1-y-4r(1-r)-2\lambda_{\EL} P_c xr(2-y)(2r-1)}\; .
\label{e3:helicity}
\end{equation}
The final photons have an averaged helicity  
$\langle \lambda_\gamma \rangle \neq 0$ if either the laser light has circular
polarisation $P_c \neq 0$ or the electrons have mean helicity
$\lambda_{\EL} \neq 0$.  Moreover, 
$\langle \lambda_\gamma (\omega=\omega_m) \rangle = -P_c$
at $P_c= \pm 1$ or $\lambda_e=0$.

\begin{figure}[!hbp]
\centering
\includegraphics[width=12cm,angle=0,]{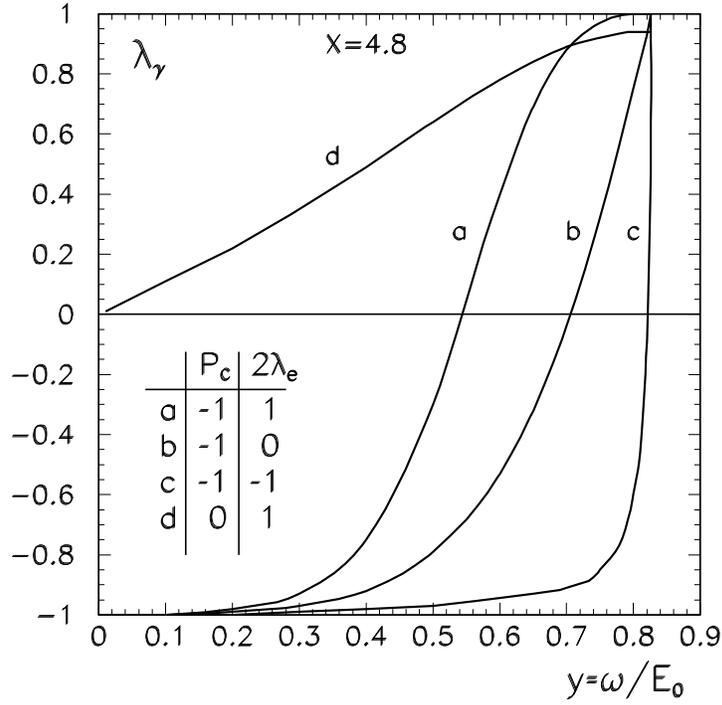}
\caption{Mean helicity of the scattered photons.}
\label{f3:fig6}
\end{figure}
\begin{figure}[!bp]
\centering
\includegraphics[width=12cm,angle=0,]{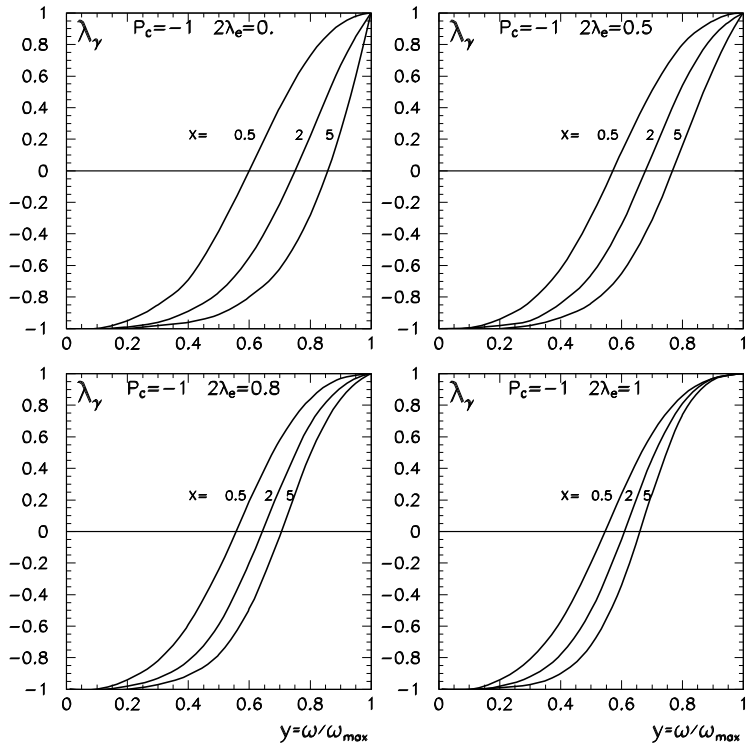}
\caption{Mean helicity of the scattered photons for various $x$ and 
degree of the longitudinal electron polarisation.}
\label{f3:polx}
\end{figure}
The mean helicity of the scattered photons at $x=4.8$ is shown in
Fig.~\ref{f3:fig6} for various helicities of the electron and laser
beams~\cite{TEL95}.  For $ 2P_c\lambda_e = -1$ (the case with minimum
energy spread) all photons in the high energy peak have a high
degree of like--sign polarisation.  This is the most valuable region
for experiments.  If the electron polarisation is not 100\% and
$|P_c|= 1$,  the helicity of the photon with the maximum energy 
is still 100\%  but the energy region with a high helicity is reduced,
 see~\ref{f3:polx}.

Low energy photons are also polarised (especially in the case
$2\lambda _{\EL} P_c = +1$ which corresponds to the broad spectrum), but
due to contribution of multiple Compton scattering and beamstrahlung photons
produced during the beam collisions the low energy region is not
attractive for polarisation experiments. 

A high degree of longitudinal photon polarisation is essential for
the suppression of the QED background in the study of the intermediate
Higgs boson (Section~\ref{s2}).  Note that at a $0.5\TEV$ linear collider the
region 
of the intermediate Higgs can be studied with rather small $x$.  In
this case the helicity of scattered photons is almost independent of
the polarisation of the electrons, and, if $P_c=1$, the high energy
photons have very high circular polarisation over a wide range near the
maximum energy, even with $\lambda_{\EL}=0$. Nevertheless,  electron
polarisation is very desirable even for rather low $x$ because, as was
mentioned before, it  increases the relative number of high energy
photons.

\begin{figure}[htbp]
\centering
\includegraphics[width=12cm,angle=0,]{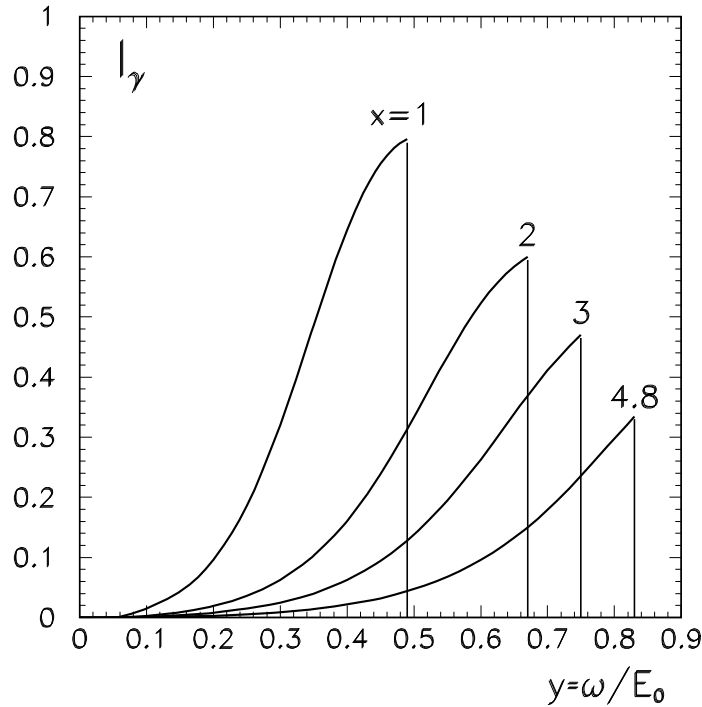}
\caption{Linear polarisation of the scattered photons for 
various $x$ for unpolarised electrons and $P_l = 1$.}
\label{f3:linp}
\end{figure}
The averaged degree of the linear polarisation of the final
photons is~\cite{GKST84}

\begin{equation}
\langle l_{\gamma } \rangle ={2r^2\; P_l \over
(1-y)^{-1}+1-y-4r(1-r)-2\lambda_{\EL} P_c xr(2-y)(2r-1)}\,.
\label{e3:linear}
\end{equation}
If the laser light has a linear polarisation, then the high-energy photons
are polarised in the same direction.  The degree of this polarisation
$\langle l_\gamma \rangle$ depends on the linear polarisation of laser
photons $P_l$ and $2\lambda_{\EL}\,P_c$. For $P_l=1$ (in this case 
$P_c=0$) the linear polarisation is maximum for the photons with the
maximum energy.  At $y=y_m$ the degree of linear polarisation for the
unpolarised electrons

\begin{equation}
l_\gamma=\frac{2}{1+x+(1+x)^{-1}}
\label{e3:lmax}
\end{equation}
is 0.334, 0.6, 0.8 for  $x=4.8,2,1$ respectively.
The dependence of the linear polarisation on the photon energy for
unpolarised electron beams and 100\% linear polarisation of laser 
photons is shown in Fig.~\ref{f3:linp}

It is of interest that varying polarisations of laser and electron
beams one can get larger $\langle l_\gamma \rangle$, up to 
$\langle l_\gamma \rangle =1$. For example, at $P_t=2(x+1)/(x^2+2x+2)$ and
$2\lambda_{\EL} \, P_c=x(x+2)/(x^2+2x+2)$ the quantity 
$\langle l_\gamma \rangle $ at $y=y_m$ can reach 1. Unfortunately, in this case
$2\lambda _{\EL}\, P_c \approx +1$, which corresponds to curve $c$ in Fig.~\ref{f3:fig4}, when the number of photons with the energy $\omega$ near
$\omega _m$ is small.

Linear polarisation is necessary for the measurement of the \CP--parity of
the Higgs boson in \GG\ collisions (Section~\ref{s2}).  Polarisation asymmetries
are proportional to $l_{\gamma,1}l_{\gamma,2}$, therefore low $x$
values are preferable.

\subsubsection{Nonlinear effects}\label{s3.1.2}

For the calculation of the $\EL \to \G$ conversion efficiency,
beside the geometrical properties of the laser beam and the 
Compton effect, one has to consider also {\it nonlinear effects} in the
  Compton scattering.  The field in the laser wave at the conversion
region is very strong, so that the electron (or the high--energy photon)
can interact simultaneously with several laser photons (so called
nonlinear QED effects).  These nonlinear effects are characterised by
the parameter~\cite{Berestetskii,GKP1,GKP2,GKP3}

\begin{equation}
\xi^2 = \frac{e^2\bar{F^2}\hbar^2}{m^2c^2\omega_0^2} = 
\frac{2 n_{\gamma} r_e^2 \lambda}{\alpha},
\label{xi2}
\end{equation}
where $\bar{F}$ is the r.m.s. strength of the electric (magnetic)
field in the laser wave, $n_{\gamma}$ is the density of laser photons.
At $\xi^2 \ll 1$ the electron is scattered on one laser photon, while
at $\xi^2 \gg 1$  on several (like synchrotron radiation in a wiggler).
Nonlinear effects in Compton scattering at photon colliders are
considered in detail in~\cite{Galynskii} and references therein.

\begin{figure}[t!hbp]
 \hspace{-1cm}  \includegraphics[width=9cm,angle=0,]{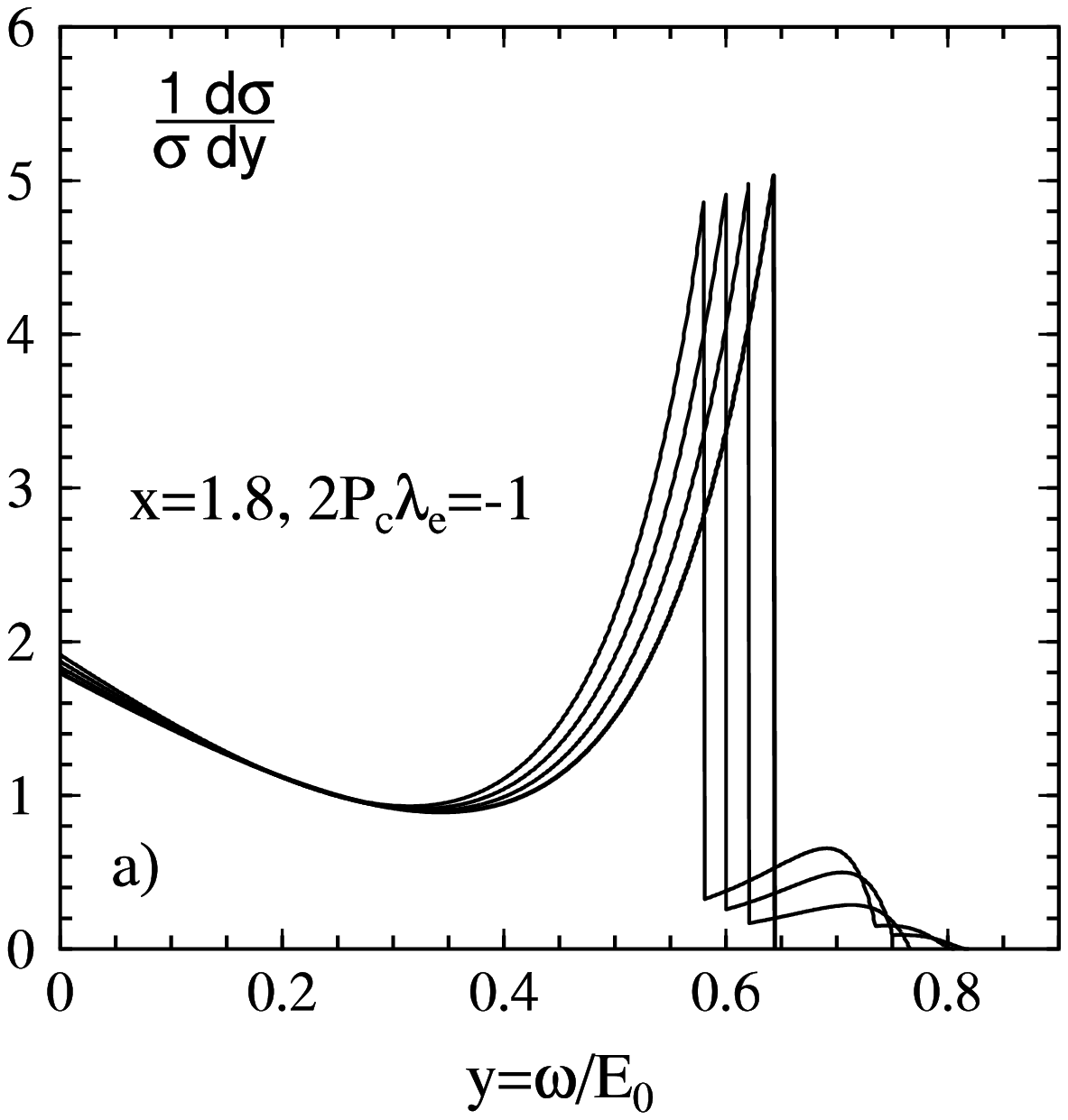}
\hspace{-1.7cm} \includegraphics[width=9cm,angle=0,]{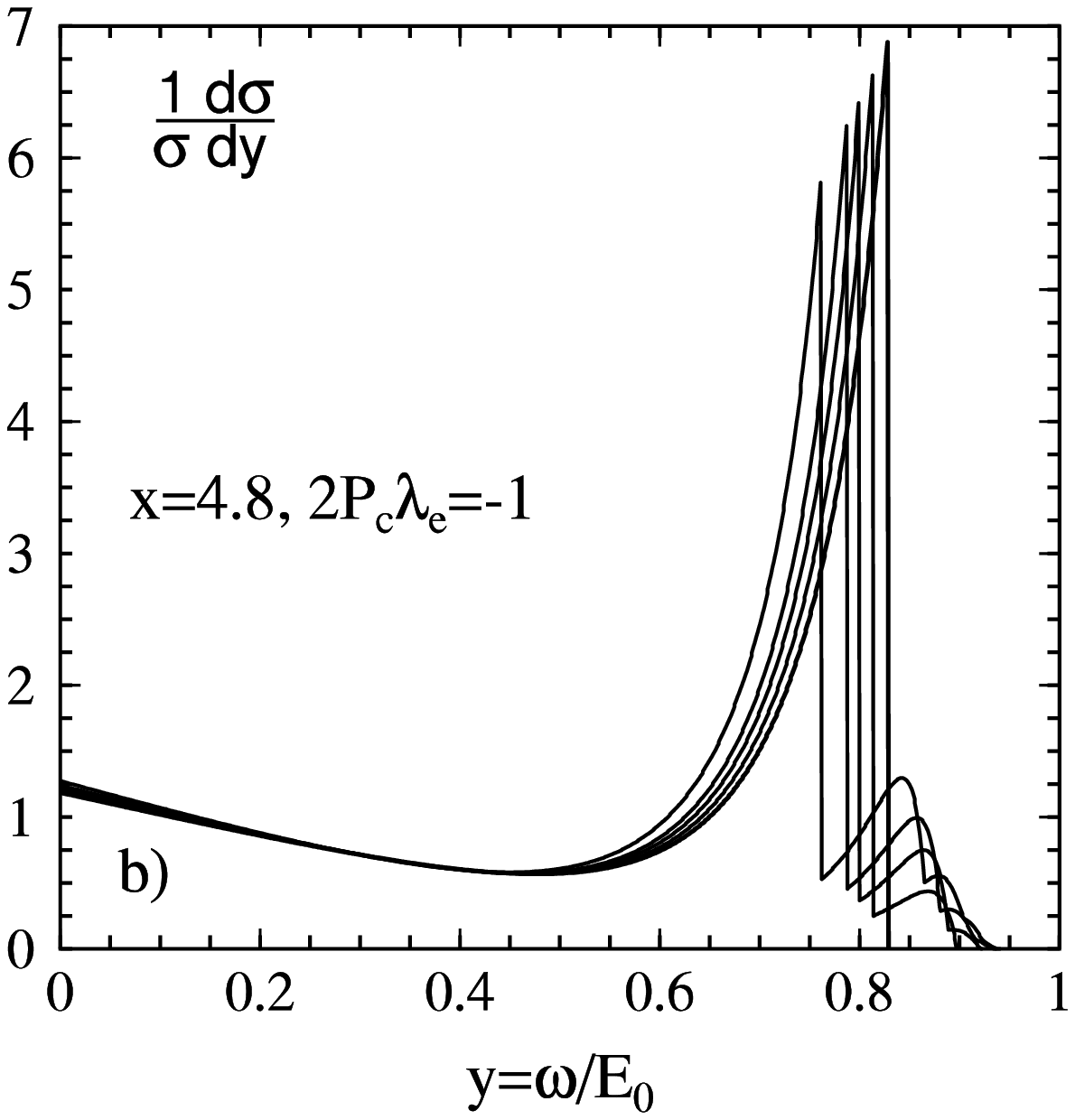}
\caption{Compton spectra for various values of the parameter $\xi^2$. 
Left figure is for $x=1.8$, right for $x=4.8$. Curves from right to left
correspond to $\xi^2 = 0, 0.1, 0.2, 0.3, 0.5$ (the last for $x=4.8$, only).}
\label{f3:gal12}
\end{figure}
\begin{figure}[h!btp]
 \hspace{-1cm} \includegraphics[width=9.5cm,angle=0,]{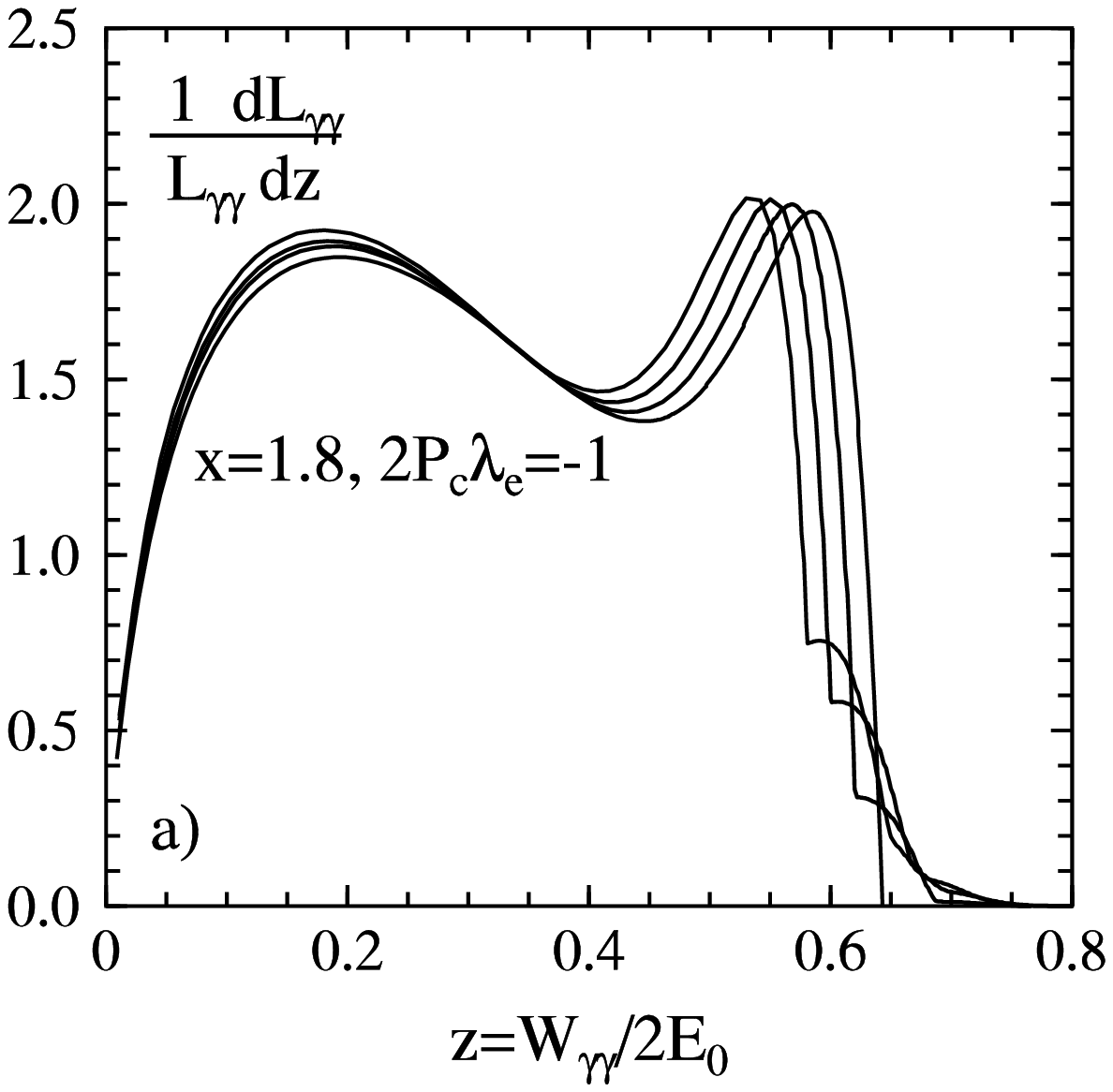}
\hspace{-1.7cm} \includegraphics[width=9.5cm,angle=0,]{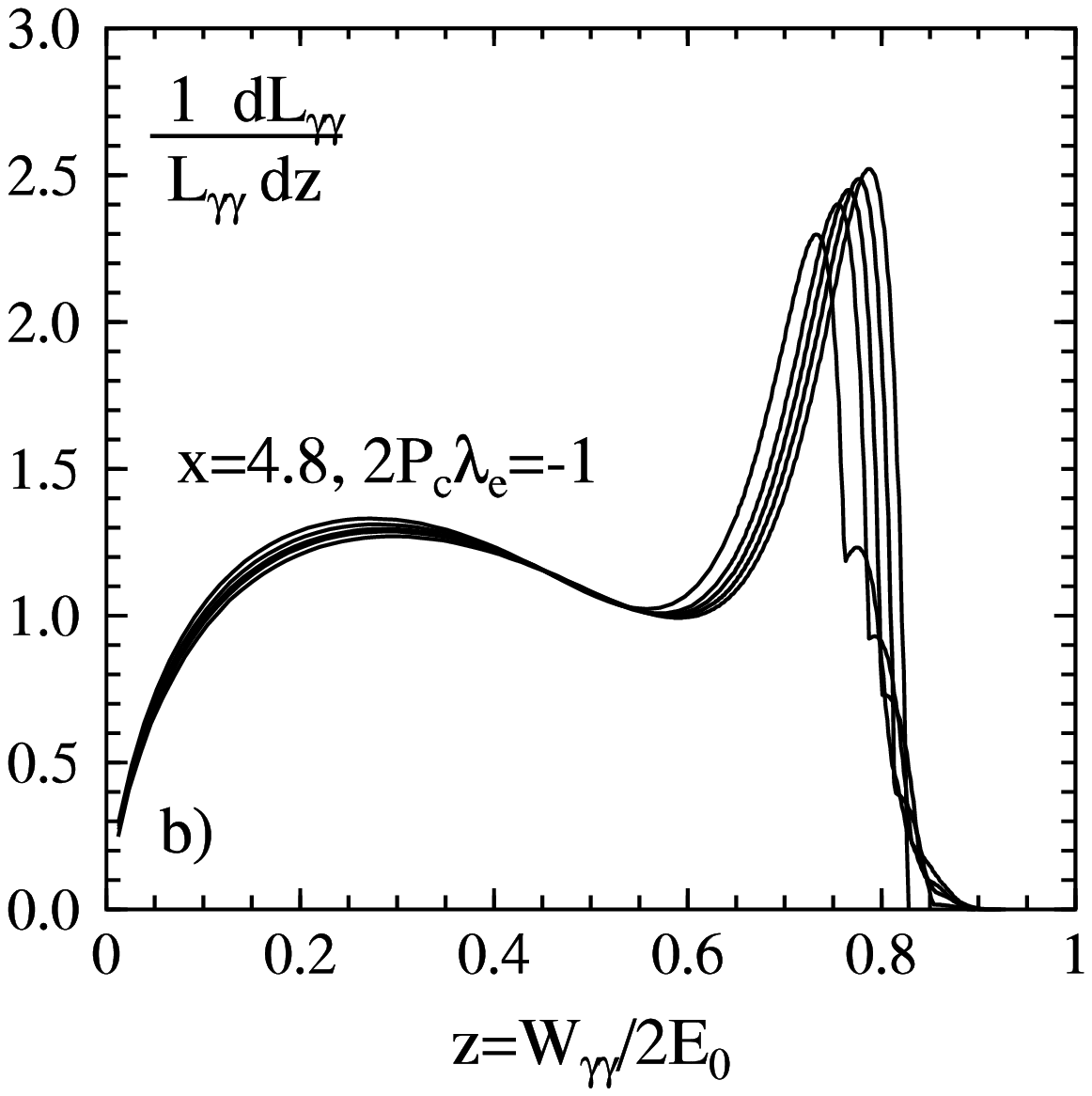}
\caption{Idealised (see the text) \GG\ luminosity distributions for various 
  values of the parameter $\xi^2$. Left figure is for $x=1.8$,
  right for $x=4.8$. Curves from right to left
correspond to $\xi^2 = 0, 0.1, 0.2, 0.3, 0.5$.}
\label{f3:gal34}
\end{figure}  
The transverse motion of an electron in the electromagnetic wave leads
to an effective increase of the electron mass: $m^2 \rightarrow
m^2(1+\xi^2)$, and the maximum energy of the scattered photons
decreases: $\omega_m/E_0 = x/(1+x+\xi^2)$. The relative shift 
$\Delta \omega_m /\omega_m \approx \xi^2/(x+1)$. At $x=4.8$ the value of
$\omega_m/E_0$ decreases by 5\% at $\xi^2=0.3$~\cite{TEL95}. This
value of $\xi^2$ can be taken as the limit. For smaller $x$ it should
be even lower.

The evolution of the Compton spectra as a function of $\xi^2$ for $x=$ 4.8
and 1.8 (the latter case is important for the Higgs study) is shown in
Fig.~\ref{f3:gal12}~\cite{Galynskii}. One can see that with increasing 
$\xi^2$ the Compton spectrum becomes broader, is shifted to lower
energies and higher harmonics appear. These effects are clearly seen
also in the \GG\ luminosity distributions (Fig.~\ref{f3:gal34}) which,
under certain conditions (Section~\ref{s5}), are  a simple convolution of the
photon  spectra.

For many experiments (such as scanning of the Higgs) it is very
advantageous to have a sharp edge of the luminosity spectrum. This
requirement restricts the maximum values of $\xi^2$  to $0.1$--$0.3$,
depending on $x$.

\subsubsection[e$^+$e$^-$ Pair creation and choice of the laser
wavelength]
{e$^{\bp}$e$^{\bm}$ Pair creation and choice of the laser wavelength}
\label{s3.1.3}

As it was mentioned with increasing  $x$, the energy of the
back--scattered photons increases and the energy spectrum becomes narrower.
However, at high $x$, photons may be lost due to
creation of \EPEM\ pairs in the collisions with laser
photons~\cite{GKST83,TEL90,TEL95}. The threshold of this reaction is
$\omega_m \omega_0 = m^2c^4$, which gives $x=2(1+\sqrt{2})\approx 4.83$.

The cross section for \EPEM\ production in a photon-photon collision is
given by~\cite{Ispirian1,Ispirian2,BGMS}
\newcommand{\XG}{\ensuremath{x_{\gamma}}}

\begin{equation}
\sigma_{\GG \to \EPEM} = \sigma_{np} + \lambda_1\lambda_2 \sigma_1,
\label{e3:sigmae+e-}
\end{equation}
$$
 \sigma_{np} = \frac{4\sigma_0}{\XG}\left[2 \left(1 + \frac{4}{\XG} -
\frac{8}{\XG{^2}}\right)\ln{\frac{\sqrt{\XG} + \sqrt{\XG\ -4}}{2}} -
\left(1+ \frac{4}{\XG}\right)\sqrt{1-\frac{4}{\XG}}\;\right],
$$
\begin{equation}
 \sigma_1 = \frac{4\sigma_0}{\XG}\left[2\ln{\frac{\sqrt{\XG} + 
\sqrt{\XG\ -4}}{2}} - 3\sqrt{1-\frac{4}{\XG}}\;\right]
\label{e3:sigmae+e-1},
\end{equation}
where $\XG\ = 4\omega_m \omega_o/ m^2c^4  = x^2/(x+1)$, 
$\lambda_1, \lambda_2$ are photon helicities.

\begin{figure}[!htb]
\centering
\includegraphics[width=10cm,angle=0,]{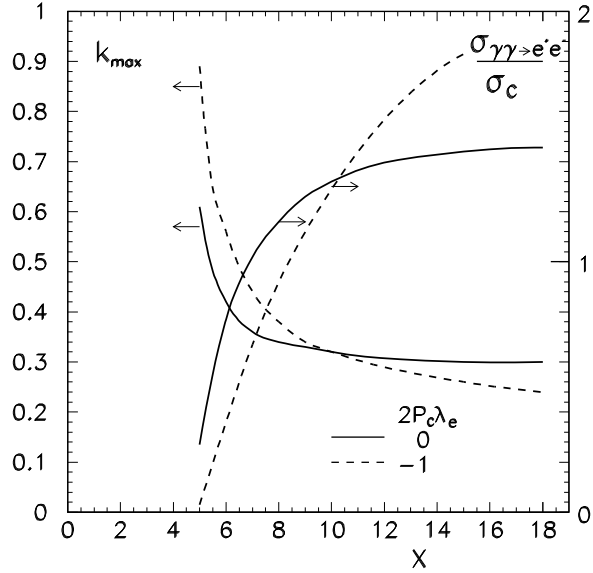}
\caption{The ratio of cross sections for \EPEM\ pair creation in the collision
  of laser and high energy photons and for Compton scattering; and
  the corresponding dependence of the maximum conversion efficiency on $x$
  assuming $\omega=\omega_m$.}
\label{f3:ratio}
\end{figure}

The ratio $\sigma_{\GG \to \EPEM}/\sigma_c$ and the maximum conversion
efficiency is shown in Fig.~\ref{f3:ratio}~\cite{TEL90,TEL95}.

One can see that above the threshold, ($x \approx$ 8--20) the \EPEM\ 
cross section is larger by a factor of $1.5-2$, the maximum conversion
coefficient is limited to $25$--$30\%$. Therefore, the value of $k^2$
which is proportional to the \GG\ luminosity  is only $0.06$--$0.09$. For
these reasons it is preferable to work at $x \leq 4.8$ where 
$k^2 \approx 0.4$ (one collision length) or even higher values are possible.

The wavelength of the laser photons corresponding to  $x=4.8$ is

\begin{equation}
 \lambda= 4.2 E_0 [\mathrm{TeV}]\MKM.
\label{e3:lamopt}
\end{equation}
For $2E_0 = 500\GEV$ it is about $1\MKM$, which is exactly the
region of the most powerful solid state lasers.  This value of $x \approx 4.8$ is
preferable for most measurements. However, for experiments 
with linear photon polarisation (see above) lower values of $x$ are
preferable.  Larger values of $x$ may be useful, for example, for reaching
somewhat higher energy.

The nonlinear effects, considered in the previous section for Compton
scattering are important for the \EPEM\ pair creation as well. First
of all, due to the high photon density \EPEM\ pairs can be produced in
collisions of a high energy photon with several laser photons. This
process is possible even at $x<4.8$. For the considered values of
$\xi^2$ such effect is not important for conversion, but the presence of
positrons may be important for the beam removal.

It is even more important that the threshold for \EPEM\ collision in
the collision with one laser photon increases because the effective
electron mass in the strong laser field increases: 
$m^2 \to m^2(1+\xi^2)$ (see previous section). This means that the threshold
value of $x$ is shifted from $x=4.8$ to 

\begin{equation}
x_{eff}=4.8(1+\xi^2). 
\end{equation}
For example, for the maximum TESLA energy $2E_0=800\GEV$ and 
$\lambda = 1.06\MKM$ from~(\ref{e3:x})  $x=7.17$. For estimation of
the \EPEM\ production one can use  Fig.~\ref{f3:ratio} where
all $x$ values are multiplied by a factor of $1 + \xi^2$. Equivalently
one can take the conversion probability in Fig.~\ref{f3:ratio}(dashed
lines) for $7.17/(1+\xi^2)$. For $\xi^2 =0.4$ (which is acceptable for
such $x$ values) we get $7.17/1.4=5.12$. One can see that the \EPEM\ creation
probability for such $x$ is negligible. To be more accurate, the
values of $\xi^2$ vary in the laser beam, but the main contribution
to the \EPEM\ probability comes from regions with values of $\xi^2$ close to
maximum.  Thus a laser with $\lambda=1.06\MKM$ can be used at all TESLA
  energies. This is confirmed by simulation (Section~\ref{s4.5})

\subsubsection{Low energy electrons in multiple compton
  scattering}\label{s3.1.4} 

For the removal of the disrupted electrons it is important to know the values
of the maximum disruption angle and minimum energy of the  electrons.

The disruption angles are created during beam collisions at the IP.
Electrons with lower energies have larger disruption angles.
The simulation code (to be described in the next section) deals with
about 5000 (initial) macro--particles and can not describe the tails of
distributions. But, provided that the minimum energy and the energy
dependence of the disruption angle are known, we can correct the value
of maximum disruption angle obtained by the simulation.

\begin{figure}[htbp]
\centering
\includegraphics[width=7.cm,angle=0,trim=30 30 30 80]{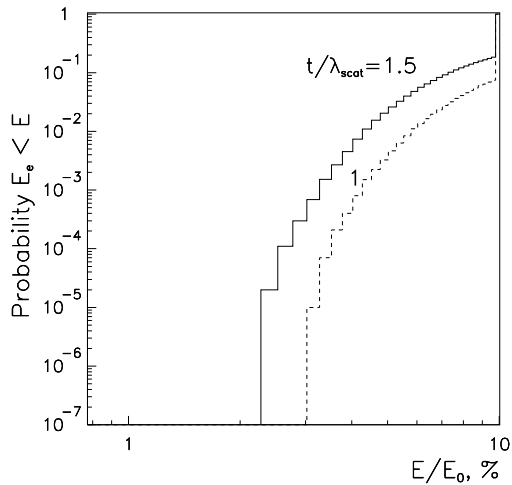}
\caption{Probability for
  an electron to have an energy below $E/E_0$ after the conversion
  region.}  
\label{f3:emin}
\end{figure}
Low energy electrons are produced at the conversion region due to
multiple Compton scattering~\cite{TEL90}.
Fig.~\ref{f3:emin}~\cite{TESLAgg} shows the probability that an
electron which has passed the conversion region has an energy below
$E/E_0$. The two curves were obtained by simulation of $10^5$ electrons
passing the conversion region with a laser target thickness of 1 and
1.5 of the Compton collision length (at $x=4.8$). Extrapolating these
curves (by tangent line) to the probability $10^{-7}$ we can obtain
the minimum electron energy corresponding to this probability: 2.5\%
and 1.7\% of $E_0$ for $t/\lambda_{scat}=1$ and 1.5 respectively.
The ratio of the total energy of all these electrons to the beam
energy is about $2\cdot10^{-9}.$ This is a sufficiently low fraction
compared with other backgrounds (see Section~\ref{s5}).  We conclude
that the minimum energy of electrons after the conversion region is
about 2\% of the initial energy, in agreement with the analytical
estimate~\cite{TEL90}.

The minimum energy of electrons after $n$  Compton collisions
$E_{min} = E_0/(nx+1) \approx E_0/nx$~\cite{GKST83}. The last approximation is
done  because the tails correspond to $n > 10$~\cite{TEL90}.  
After $1$--$2$ collisions the Compton cross
section approaches the Thompson one. This, together with the
simulation result gives the scaling for the minimum energy as a
function of the $x$ and the thickness of the laser target in units of
the collision length (for electrons with the initial energy)

\begin{equation}
E_{min} \approx 6 \frac{\sigma_c(x)/\sigma_c(4.8)}
{(\omega_0[\EV]/1.25)(t/\lambda_{scat})} \GEV.
\end{equation}
The results of this section will be used for calculation of the
disruption angle (Section~\ref{s4.2.5}).

\subsubsection{Other processes in the conversion region}\label{s3.1.5}

Let us enumerate some other processes in the conversion region which
are not dominant but nevertheless should be taken into account.

\begin{enumerate}
\item 
  {\it Nonlinear \EPEM\ pair creation} $\gamma + n\gamma_0 \to \EPEM$
  below the single photon threshold $x=4.8$ (see~\cite{GKP1,GKP2,GKP3} and references therein).
  The probability of this process is not small and should be taken into account
  when the beam removal is considered.

\item 
  {\it Variation of the high energy photon polarisation in the laser wave}~\cite{kotser}.  It
  is well known that an electromagnetic 
  field can be regarded as an anisotropic medium~\cite{Berestetskii}. Strong
  laser fields also have such properties. As a result, the polarisation
  of high energy photons produced in the Compton scattering may be
  changed during  the propagation through the polarised laser
  target. This effect is large only at $x \approx 4.8$ (the threshold for
  \EPEM\ production).  Note, that in the most important case,
  $2P_c\lambda_e= - 1$, the polarisation of high energy circularly
  polarised photons propagating in the circularly polarised laser wave
  does not change. It also does not change  for linearly polarised high
  energy photons propagating in a linearly polarised laser wave because
  they have the same direction.

  In principle, using two adjacent conversion regions one can first
  produce circularly polarised photons (using a circularly polarised
  laser) and then change the circular polarisation to the linear one
  using a linearly polarised laser~\cite{TELvav2,TELvav3}.  However, it does
  not appear to be technically feasible and moreover the quality will
  be worse than in the ideal case due to a strong dependence of the
  rotation angle on the photon energy and  the additional 
  $\EL \to\G$ conversions on the second laser bunch.
  
  A similar effect also exists  at the interaction region of photon
  colliders (Section~\ref{s4.2}), the beam field influences the photon
  polarisation~\cite{TELvav2,TELvav3}.

\item 
  {\it Variation of polarisation of unscattered electron}~\cite{KPerltS}.  
  Compton scattering chan\-ges the electron polarisation.  Complete
  formulae for the polarisation of the final electrons in the case of linear
  Compton scattering have been obtained in~\cite{KPS1997}, for the
  nonlinear case in~\cite{Galynskii1992,Galynskii}.  However,
  additional effects have to be taken into account when simulating 
  multiple Compton scattering.

  Let us first consider a simple example: an unpolarised
  electron beam collides with a circularly polarised laser
  pulse. Some electrons pass this target without  Compton scattering.
  Their polarisation is changed, since the cross
  section of the Compton scattering depends on the product
  $P_c\lambda_{\EL}$ and the unscattered electron beam already contains
  unequal 
  number of electrons with forward and backward helicities. When considering
  the multiple Compton scattering, this effect should be taken into
  account.
  
  General formulae for this effect have been obtained in~\cite{KPerltS}, where the variation in polarisation of the
  unscattered electrons was considered to be the result of the interference of
  the incoming electron wave with the wave scattered at  zero angle.
\end{enumerate}

\subsection{The choice of laser parameters}\label{s3.2}

For the $\EL\to\G$ conversion the following laser characteristics
are important: wavelength, flash energy, duration, optimum focusing.
The problem of optimum wavelength was considered in Section~\ref{s3.1.3}.
The other items are considered below.

\subsubsection{Conversion probability, laser flash energy}\label{s3.2.1}

For the calculation of the conversion efficiency  it is
useful to remember the correspondence between the parameters of the
electron and laser beams. The emittance of the Gaussian laser
beam with diffraction limited divergence is $\epsilon_{x,y} = \lambda/4\pi$.
The ``beta--function'' at a laser focus $\beta\equiv Z_R$, where $Z_R$ is
known as the Rayleigh length in optics literature. 

The r.m.s. transverse radius of a laser near the conversion region
depends on the distance $z$ to the focus (along the beam) as~\cite{GKST83}

\begin{equation}
\sigma_{L,r}(z)= \sigma_{L,r}(0) \sqrt{1+z^2/Z_R^2},
\label{sLrz}
\end{equation}
where the r.m.s. radius  at the focus

\begin{equation}
a_{\gamma} \equiv \sigma_{L,r}(0) = \sqrt{\frac{\lambda Z_R}{2\pi}}.
\label{a_g}
\end{equation}
We see that the effective length of the conversion region is about
$2Z_R$.  The r.m.s. beam sizes on $x,y$ projections
$\sigma_{L,i}(z)=\sigma_{L,r}(z)/\sqrt{2}$.

The r.m.s. angular divergence of the laser light in the focal point

\begin{equation}
\sigma_{L,x^{\prime}} = \frac{\lambda}{4\pi\sigma_{L,x}} = 
\sqrt{\frac{\lambda}{4 \pi Z_R}}\; .
\label{xprime}
\end{equation}
The density of laser photons in a Gaussian laser beam

\begin{equation}
n_{\gamma} = \frac{A}{\pi\sigma^2_{L,r}(z)\omega_0} exp({-r^2/\sigma^2_{L,r}(z)})
\; F_L(z+ct),
\label{ngam}
\end{equation}
$$
\int F_L(z) dz =1,
$$
where $A$ is the laser flash energy and the function $F_L(z)$
describes the longitudinal distribution (can be  Gaussian as well).

Neglecting multiple scattering, the dependence of the conversion
coefficient on the laser flash energy  $A$ can be written as

\begin{equation}
     k  = N_\gamma /N_e \approx 1-\exp (-A/A_0 ),
\label{kdef}
\end{equation}
where $A_0$ is the laser flash energy for which the thickness of the 
laser target is equal to one Compton collision length. The value of $A_0$
can be roughly estimated from  the collision probability 
$p \approx n_{\gamma}\sigma_{c}l = 1$, where
$ n_\gamma \approx A_0/(\pi \omega_0 a_{\gamma}^{2} l_\gamma )$,
$\sigma_c$ is the Compton cross section ($\sigma_c =
1.8\cdot10^{-25}\;$ cm$^2$ at $x=4.8$), $l$ is the length of the region
with a high photon density,
which is equal to $2Z_R$ at $Z_R \ll \sigma_{L,z}\approx\sigma_z$
($\sigma_z$ is the r.m.s. electron bunch length). This gives

\begin{equation}
  A_0 \approx \frac{\pi\hbar c\sigma_z}{\sigma_c} \approx 5 \sigma_z [\MM]\; \J
  \;\;\; \mbox{for}\;\; x=4.8.
\label{A0estimate}
\end{equation}
Note that the required flash energy decreases when the
Rayleigh length is reduced to $\sigma_z$, and it  hardly  changes with
further decreasing of $Z_R$. This is because the density of photons
grows but the length having a high density decreases and as a result
the Compton scattering probability is almost constant.  It is not
helpful to make the radius of the laser beam at the focus smaller than
$\sigma_{L,x} \approx \sqrt{\lambda\sigma_z/4\pi}$, which may be much
larger than the transverse electron bunch size in the conversion
region.

From~(\ref{A0estimate}) one can see that the flash energy $A_0$ is
proportional to the electron bunch length and for  TESLA 
($\sigma_z = 0.3$ mm) it is about 1.5\J.  

\begin{figure}[!htb]
\centering
\epsfig{file=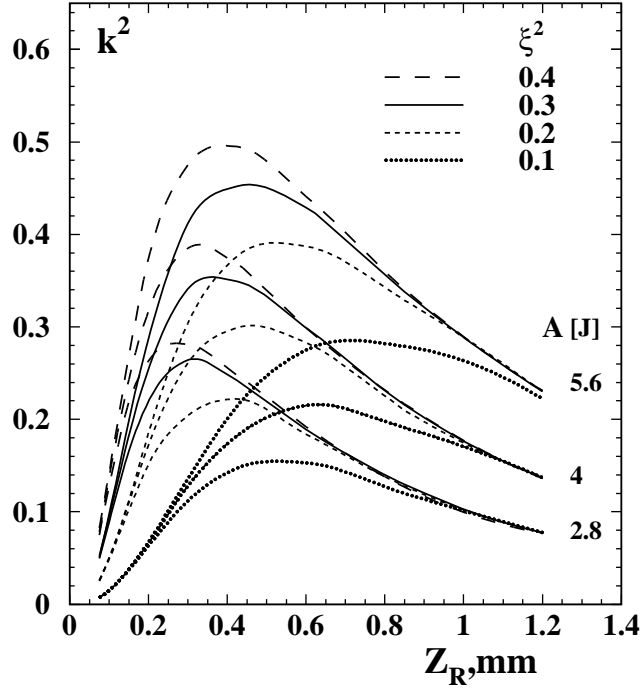,width=10cm,angle=0}
\caption{Square of the conversion probability (proportional to the \GG\ 
  luminosity) as a function of the Rayleigh length for various
  parameters $\xi^2$ and laser flash energies; $x=4.8, \lambda =1.06\MKM$
  are assumed. The mirror system is situated outside the
  electron beam trajectories (collision angle 
  $\theta = 2\sigma_{L,x^{\prime}}$). The crab crossing angle $30\MRADI$ is taken
  into account. See also the text.}
\label{k2}
\end{figure}

More precise calculations of the conversion probability in  head-on
collision of an electron with a Gaussian laser beam can be found
elsewhere~\cite{GKST83,TEL90,TEL95}. However, this is not a complete
picture, one should also take into account the following effects:

\begin{itemize}
\item 
  {\it Nonlinear effects in Compton scattering}.
  In the laser focus the value of the  parameter $\xi^2$
  (Section~\ref{s3.1.2}) is given by 
  \begin{equation}
    \xi^2 = \frac{4r_e\lambda A}{(2\pi)^{3/2}\sigma_{L,z}mc^2 Z_R},
    \label{xi2a}
  \end{equation}
  this follows from eqs~(\ref{xi2},\ref{ngam}). For example, for $A=2\J$,
  $\lambda=1.06\MKM$  and $\sigma_{L,z} = Z_R = \sigma_z = 0.3\MM$, we get 
  $\xi^2 \approx 0.2$. This is still acceptable, but for shorter bunches nonlinear
  effects will determine the laser flash energy.
  
\item 
  {\it Collision angle}.  A maximum conversion probability for a
  fixed laser flash energy can be obtained in a head-on collision of
  the laser light with the electron beam. This variant was considered
  in the TESLA Conceptual Design~\cite{TESLAgg}. In this case
  focusing mirrors should have holes for the incoming and outgoing
  electron beams. From the technical point of view it is  easier to
  put all laser optics outside  the electron beams. In this case, the
  required laser flash energy is larger by a factor of $2-2.5$, but
  on the other hand it is much simpler and this opens a way for a
  multi--pass laser system, such as an external optical cavity
  (Section~\ref{s5.1}).
  Below we assume that the laser optics is situated outside the electron beams.
  
\item 
  {\it Transverse size of the electron beam}.  For the removal of
  disrupted beams at photon colliders it is necessary to use a
  crab--crossing beam collision scheme (see Fig.~\ref{ggcol} and
  Section~\ref{s4.1}). In this scheme the electron beam is tilted
  relative to its direction of motion by an angle
  $\alpha_c/2\approx 15\MRAD$. Such a method allows to collide beams at
  some collision angle (to make easier the beam removal) without
  decrease of the luminosity. 

  Due to the tilt the electron beam at
  the laser focus has an effective size $\sigma_x=\sigma_z\alpha_c /2$ which
  is $4.5\MKM$ for TESLA. This should be compared with the laser spot
  size (eq.\ref{a_g}), for $Z_R = \sigma_z = 0.3\MM$ and 
  $\lambda = 1.06\MKM$ of 
  $\sigma_{L,x}= \sqrt{\lambda Z_R/4\pi} \approx 5\MKM$. The sizes are
  comparable, which leads to some increase of the laser flash energy.

\end{itemize}
The result of the simulation~\cite{telnov} of $k^2$ ($k$ is the
conversion coefficient) for the electron bunch length $\sigma_z= 0.3\MM$
(TESLA project), $\lambda=1.06\MKM$, $x=4.8$ as a function of the
Rayleigh length $Z_R$ for various flash energies and values of the
parameter $\xi^2$ are shown in Fig.~\ref{k2}.

It was assumed that the angle between the laser optical axis and the
electron beam line is $\theta = 2\sigma_{L,x^{\prime}}$, where
$\sigma_{L,x^{\prime}}$ is the angular divergence of the laser beam in
the conversion region (eq.~\ref{xprime}), and the mirror system is
situated outside the electron beam trajectories. One conversion length
corresponds to $k^{2} = (1- e^{-1})^2 \approx 0.4$. One can see that
$k^2=0.4$ at $\xi^2=0.3$ can be achieved with the minimum flash energy
$A=5\J$. The optimum value of $Z_R$ is about 0.35\MM.

The r.m.s. duration of the laser pulse can be found from~(\ref{xi2a}),
for the considered case $\sigma_{L,z}=0.44\MM$ or 1.5\ps.

Above we have considered the requirements for the laser at
$\lambda=1.06$  $x \approx 4.8$, which is the case for a $2E_0 =500\GEV$ 
collider.  The required flash energy as about 5\J\ for  
$\xi^2 = 0.3$. Next we discuss what changes when the electron beam energy is
decreased or increased?

When we decrease the energy to $E_0 = 100\GEV$, keeping the laser
wavelength constant,  the Compton cross section increases from
$\sigma_C/\sigma_0 =0.7$ ($x=4.8$) to 1.24 ($x=1.8$). This case
corresponds to $W_{\GG,m}\approx 130\GEV$. Calculations similar to the one
presented in 
Fig.~\ref{k2} show that for this case $k^2=0.4$ can be obtained with
$A \approx 3.8\J$ at $\xi^2 =0.1$ (and $Z_r \approx 0.6\MM$) or with $A \approx
2.5\J$ at $\xi^2 =0.3$ (and $Z_r \approx 0.3\MM$). So, for the study of the low
mass Higgs one needs a laser with somewhat lower flash energy and
values of $\xi^2$ can be lower than that at $x \approx 4.8$.
 
Another variant for study of $W_{\GG,m}\approx 130\GEV$ involves decreasing the
electron beam energy keeping $x = const = 4.8$. This requires 
$\lambda = 1.06/3\MKM$. Calculations show that using a 5\J\ laser flash one can
obtain only $k^2=0.35$ at $\xi^2 =0.3$.  The conversion coefficient is
lower than that for $x=4.8$ and $\lambda = 1.06$.  This
result is quite surprising, because for the shorter wavelength the
nonlinear effects are less important and according to~(\ref{A0estimate}) the minimum flash energy does not depend on the
wavelength.  Such behaviour is connected with the effective transverse
electron bunch size due to the crab--crossing (see above) which
restricts the minimum laser spot size, and to the fact that for shorter
wavelength the energy of each photon is larger.
  
Comparing the two methods of reaching the low mass Higgs region
we come to the conclusion that it is easier to use a 
$\lambda = const = 1.06\MKM$ laser due to the lower flash energy, lower
$\xi^2$ and the fact that this is the region of powerful solid state lasers
(production of the second or third harmonics require $2$--$3$ times larger
initial flash energy). There are also some advantages for physics, namely, a
high degree of linear polarisation.

In Section~\ref{s3.1.3} it was shown that it is possible to work with a
$\lambda=1.06\MKM$ laser even at the maximum TESLA energy of 
$2E_0 = 800\GEV$, in spite of a value of $x=7.17$. This is due to the nonlinear
effects 
which increase the threshold for \EPEM\ pair production from $x=4.8$
to $x=4.8(1+\xi^2)$.  The Compton cross section for the value of 
$x = 7.17$ is lower than at $x=4.8$ by a factor of 1.32. Nevertheless,
with 5\J\ flash energy and $\xi^2 =0.4$, one can obtain $k^2 \approx 0.35$.

So, we can conclude that a laser with $\lambda \approx 1\MKM$ is
suitable for all TESLA energies.

\subsubsection{Summary of requirements to the laser}\label{s3.2.2}

From the above considerations it follows that to obtain a conversion
probability of $k\approx 63\%$  at all TESLA energies a laser with the
following parameters is required:

\begin{center}
\begin{tabular}{ll}
Flash energy & $\approx 5\J$ \\
Duration & $\tau(rms) \approx 1.5\ps$ \\
Repetition rate & TESLA collision rate, $\approx 14\kHz$ \\
Average power & $\approx 140\kW$ (for one pass collision) \\
Wavelength &   $\approx 1\MKM$ (for all energies).
\end{tabular}
\end{center}

\section{The Interaction Region} \label{s4}

\subsection{The collision scheme, crab--crossing} \label{s4.1}

The basic scheme for photon colliders is shown in Fig.~\ref{ggcol}
(Section~\ref{s1}).  The distance between the conversion point (CP) and the IP,
$b$, is chosen from the relation $b \approx \gamma \sigma_y$, so that the
size of the photon beam at the IP has equal contributions from the
electron beam size and the angular spread from Compton scattering. At
TESLA $\sigma_y \approx 4\NM$ gives $b \approx 2\MM$ at 
$2E_0=500\GEV$. Larger $b$ values lead to a decrease of the \GG\ luminosity,
for  
smaller $b$ values the low--energy photons give a larger contribution to the
luminosity (which is not useful for the experiment but causes additional
backgrounds).

In the TESLA Conceptual Design four years ago two schemes were
considered: with magnetic deflection and without. At that time
$\sigma_y$ was assumed to be about $16\NM$, and the distance 
$b \approx 1\CM$ was sufficient for deflection of the electron beam from the IP
using a small magnet with $B \approx 5\kG$. With the new TESLA
parameters with $b$ about 5 times smaller  this option is practically
impossible (may be only for a special experiment with reduced
luminosity). We now  consider only one scheme: without magnetic
deflection, when all particles after the conversion region travel to
the IP producing a mixture of \GG, \GE, \EM\EM\ collisions. The beam
repulsion leads to some reduction of the \GE\ luminosity and a considerable
suppression of the \EMEM\ luminosity.

There are two additional constraints on the CP--IP distance. It should
be larger than the half-length of the conversion region (which is
about $Z_R \approx 0.35\MM$ (Section~\ref{s3})), and larger than about $2$--$3$
$\sigma_z$ ($\sigma_z$ is the electron bunch length) because the
$\EL\to\gamma$ conversion should take place before the beginning of
electron beam repulsion. So, the minimum distance $b$ for the TESLA is
about $1\MM$.

The removal of the disrupted beams can best be done using the
crab-crossing scheme~\cite{PALMER}, Fig.~\ref{ggcol}, which is foreseen
in the NLC and JLC projects for \EPEM\ collisions.  In this scheme the
electron bunches are tilted (using an RF cavity) with respect to the
direction of the beam motion, and the luminosity is then the same as
for head--on collisions. Due to the collision angle the outgoing
disrupted beams travel outside the final quads.  The value of the
crab--crossing angle is determined by the disruption angles (see the next
section) and by the final quad design (diameter of the quad and its
distance from the IP).  In the present TESLA design 
$\alpha_c = 34\MRAD$.

\subsection[Collision effects in $\gamma\gamma$, $\gamma$e collisions]
{Collision effects in $\bgamma\bgamma$, $\bgamma$e collisions} \label{s4.2}

 The luminosity in \GG,\GE\ collisions may be limited by several factors:
\begin{itemize}
\item 
  geometric luminosity of the electron beams;
\item 
  collision effects (coherent pair creation, beamstrahlung, beam
  displacement);
\item 
  beam collision induced background (large disruption angles of
  soft particles);
\item 
  luminosity induced background (hadron production, \EPEM\ pair
  production).
\end{itemize}

For optimisation of a photon collider it is useful to know
qualitatively the main dependences. In this section we will consider
collision effects which restrict the \GG, \GE\ luminosity.

Naively, at first sight, one may think that there are no collision
effects in \GG\ and \GE\ collisions because at least one of the beams
is neutral. This is not correct because during the beam collision
electrons and photons are influenced by the field of the opposite
electron beam, which leads to the following effects~\cite{TEL90,TEL95}:


\vspace{2mm}

 \GG\ collisions: conversion of photons into \EPEM\ pairs 
         (coherent pair creation).

\vspace{2mm}

\GE\ collisions: coherent pair creation; beamstrahlung;
   beam displacement.

\vspace{2mm}

Below we consider the general features of these phenomena and then present the
results of simulations where all main effects are included.

\subsubsection{Coherent pair creation} \label{s4.2.1}

The probability of pair creation per unit length  by  a photon
with the energy $\omega$ in the magnetic field $B$ ($|B|+|E|$  for our
case) is~\cite{TEL90,CHTEL}

\begin{equation}
\mu (\kappa) = \frac{\alpha^2}{r_e} \frac{B}{B_0} T(\kappa), \;\;\;
\kappa = \frac{\omega}{mc^2}\frac{B}{B_0},\;\; B_0 = \frac{\alpha e}{r_e^2} 
= 4.4 \cdot 10^{13}\;
\mbox{G},
\end{equation}
where $B_0$ is the the critical field, the function $T(\kappa) \approx
0.16 \kappa^{-1} K^{2}_{1/3}(4/3\kappa)$. At $ \kappa < 1 $, it is
small, $T \approx 0.23 \exp (-8/3\kappa)$, and $ T \approx 0.1$ at $ \kappa
= 3$--$10$. 

In our case, $\omega \approx  0.8 E_0$ ,  therefore one can put
$\kappa \approx 0.8\Upsilon \equiv \gamma B/B_0$.

Coherent pair creation is exponentially suppressed for $\Upsilon < 1$,
but for $\Upsilon > 1$ most  high energy photons can convert to
\EPEM\ pairs during the beam collision. The detailed analyses of these
phenomena at photon colliders are presented in~\cite{TEL90,TEL95,TSB2}.

Without disruption the beam field $B \approx eN/(\sigma_x \sigma_z)$ (we
assume that $\sigma_x > \sigma_y$). Therefore,   coherent
\EPEM\ creation restricts  the minimum  
horizontal beam size.

For example, for $N=2\times 10^{10}$, $\sigma_x=50\NM$, 
$\sigma_z =0.3\MM$, $E_0 = 500\GEV$, we obtain $\kappa_{av} \approx 1.2$, 
$T \approx 0.01$
and the $\gamma \to\EPEM$ conversion probability $p \approx \mu \sigma_z= 0.06$
(rather small).  For $\sigma_x = 10\NM$ it would be about 0.5 (40\% 
loss of the \GG\ luminosity).

However, it turns out that at TESLA energies and beam parameters 
$N, \sigma_z$ the coherent pair creation is further suppressed due to
the repulsion of the electron beams~\cite{TELSH,TSB2}.  Due to the
repulsion, the characteristic size of the disrupted beam 
$r \approx \sqrt{\sigma_z r_e N/8\gamma}$, would be about $45\NM$ for the previous
example.  Therefore, with decreasing $\sigma_x$ the field at the IP
increases  to a maximum value $B \approx 2eN/(r \sigma_z)$.  The
corresponding parameter $\Upsilon \propto (E_0/\sigma_z)^{3/2}N^{1/2}$.
As a result, at a sufficiently low beam energy and long beams the
field may be below the threshold  for coherent pair creation even
for zero initial transverse beam sizes. This fact allows,
in principle, very high \GG\ luminosity to be reached. This interesting
effect is confirmed by the simulation~\cite{TSB2} (Section~\ref{s4.4}).

One comment on the previous paragraph: although the beam disruption helps to
suppress the coherent pair creation and to keep the \GG\ luminosity  close
to the geometric one,   there is,  nevertheless, some restriction on the
field strength due to background caused by coherent pair creation.
One can show that the minimum energy of electrons (at the level of
probability of $W \approx 10^{-7}$) in coherent pair creation is about
$E_{min}/\omega \approx 0.05/\kappa$. Therefore at $\kappa >2$ this
energy is lower than the minimum energy of electrons after 
multiple Compton scattering and the resulting  disruption angles
will be determined by the coherent pair creation. 

Electrons of similarly low energies are also produced in hard
beamstrahlung with approximately similar probability.  However, in the
TESLA case, beamstrahlung is less important because electrons radiate
inside the disrupted beam, while in the case of coherent pair creation
the head of the Compton photon bunch travels in the field of the
undisturbed oncoming electron beam and passes the region with the
maximum (undisturbed) beam field.  Simulation results for luminosity
and disruption angles taking of all these effects into account are
presented in Section~\ref{s4.4}.

\subsubsection{Beamstrahlung} \label{s4.2.2}

The physics of beamstrahlung (radiation during beam collisions) at
linear \EPEM\ colliders is very well understood~\cite{Noble,Yokoya}.
Consequences of beamstrahlung for \GG, \GE\ colliders have been
considered in~\cite{TEL90,TEL95}.

For \GG\ collisions beamstrahlung is not important.  However,
beamstrahlung photons collide with opposing Compton and beamstrahlung
photons, increasing the total \GG\ luminosity by a significant factor 
(mainly in the the region of rather low invariant masses, below the
high energy luminosity peak.)

In the \GE\ collisions beamstrahlung leads to a decrease of the electron
energy and, as a result, the \GE\ luminosity in the high energy peak
also decreases.  In addition, the beamstrahlung photon contribution to the
\GE\ luminosity considerably worsens the \GE\ luminosity spectrum.

\subsubsection{Beam--beam repulsion} \label{s4.2.3}

During the collision opposing beams either attract or repulse each
other.  In \EPEM\ collisions this effect leads to some increase of the
luminosity (the pinch effect), while in \EMEM\ collisions the
attainable luminosity is reduced~\cite{Balakin1,Balakin2,Balakin3}.

Photon colliders are based on \EMEM\ beams. For \GG\ 
collisions the effects of the beam repulsion are only positive: the coherent
pair creation is suppressed; the beamstrahlung 
photons emitted by the deflected electrons have a smaller probability of
colliding with the Compton or beamstrahlung photons from the opposite
electron beam; \GE\ background is smaller due to the relative shift of the
electron beams.

For \GE\ collisions the effect of beam repulsion is  negative. It leads to a
displacement of the electron beam, and hence to a decrease of 
the \GE\ luminosity.

The beam repulsion also leads to a considerable decrease of the \EMEM\ 
``background'' luminosity. 

Beam--beam deflection is very useful for the diagnostics of beam
collisions and for the stabilisation of the luminosity both at \EPEM\ and
photon photon colliders.

\subsubsection{Depolarisation}  \label{s4.2.4}

Depolarisation effects are not included in our simulation code, therefore
we give an estimation of these effects~\cite{TEL90}.

\subsubsubsection{Depolarisation of electrons}

When an electron is bent by the angle $\theta$, its spin rotates,
relative to its trajectory, by the angle~\cite{Berestetskii}
\begin{equation}
\theta^{\prime} = \frac{\mu^{\prime}}{\mu_0}\gamma\theta \approx 
\frac{\alpha\gamma}{2\pi}\theta,
\label{thetap}
\end{equation}
where $\mu_0$ and $\mu^{\prime}$ are the normal and the anomalous magnetic
moments of the electron, $\alpha = e^2/\hbar c = 1/137$.
  
  In the absence of disruption, the beam field
\begin{equation}
B \approx \frac{eN}{\sigma_z \sigma_x}.
\label{bfield}
\end{equation}
The bending angle during  beam collisions (on the length $\sigma_z$) is
$\theta \approx eB\sigma_z/E_0 = r_e N/(\sigma_x \gamma)$. This gives
\begin{equation}
\theta^{\prime} \approx \frac{\alpha r_e N}{2\pi \sigma_x}.
\end{equation}

For example, for TESLA with $N=2\times 10^{10},\; \sigma_x \approx 100\NM$, 
we get $\theta^{\prime} = 0.65$. The corresponding polarisation (for
$\lambda_{e,0} =1$) is $\lambda_e \approx \cos {\; \theta^{\prime}} \approx 0.8$. The
effect is not small. 

Let us now consider the same case with beam repulsion taken into
account. In  \GE\ collisions, the electrons collide with the high
energy photons until their vertical displacement is smaller then
$\sigma_y$ (this is the case with the high energy photons for
$b=\gamma \sigma_z$ (see Section~\ref{s4.1})).  The deflection angles are derived
from $\rho\theta^2/2\approx \sigma_y$ and $\rho \approx \gamma mc^2/eB$.
This gives
\begin{equation}
\theta^{\prime} \approx \frac{\alpha\gamma}{2\pi}
\sqrt{\frac{2\sigma_y r_e N}{\sigma_x \sigma_z \gamma}}.
\end{equation}
For the previous set of parameters and $\sigma_y = 4\NM$, $2E_0 = 500\GEV$, we
obtain $\theta^{\prime} = 0.1$ and 
$\lambda_e \approx \cos {\; \theta^{\prime}} \approx 0.995$.

Although this estimate is rough, one can see that a factor of $2$--$3$ will not
change the conclusion that the Depolarisation of electrons in \GE\ 
collisions is negligible.

\subsubsubsection{Depolarisation of photons}

It is well known that a strong electromagnetic field can be treated as
an anisotropic medium with some refraction index $n$~\cite{Berestetskii}.
In fact, the conversion of photons to \EPEM\ pairs (absorption)
considered above is the manifestation of the imaginary part of the
refraction index.  The values $n$ are different for photons with
linear polarisation parallel and perpendicular to the field direction.
As a result, the polarisation of photons travelling in this field can change. In
Section~\ref{s3.1.4} we mentioned already one such effect in the conversion
region.  Here we will consider the influence of the beam field 
on the polarisation of the high energy photons.
\begin{figure}[!htb]
\centering
\epsfig{file=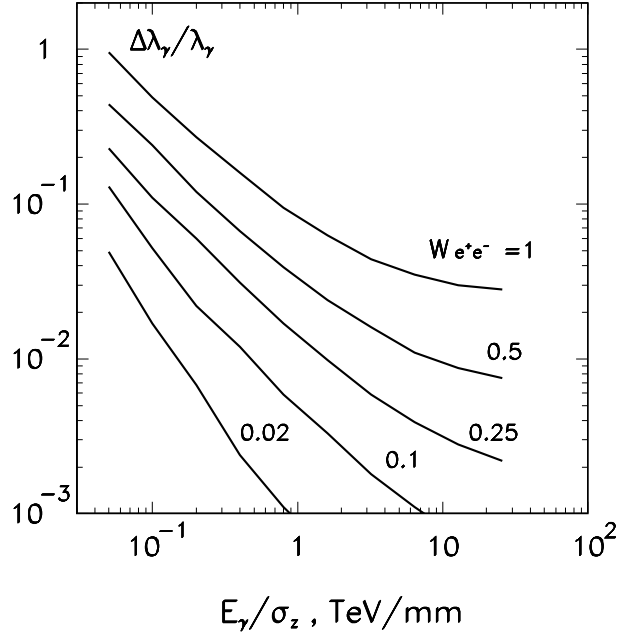,width=10cm}
\caption{ Decrease in photon helicity during beam collisions for
  various beam parameters and probabilities of coherent pair
  creation $W_{\EPEM}$~\protect\cite{TELvav2,TELvav3}. See comments in the text.}
\label{depol1}
\end{figure}

This problem was considered in detail in~\cite{TELvav2,TELvav3}.  The beam
field can transform the circular photon polarisation into a
linear polarisation and vice versa. The degree of Depolarisation as a
function of $E_{\gamma}/\sigma_z$ is shown in Fig.~\ref{depol1}.
Instead of the field strength each curve corresponds to a certain
value of the coherent pair creation probability $W_{\EPEM}$ which is defined
in units of collision lengths.  In this case,
consideration of the beam disruption is not necessary, as it is
included in the \EPEM\ conversion probability which is kept under
control at photon colliders.

For example, for TESLA beams $E_{\gamma}/\sigma_z \approx 10\TEV/\mathrm{cm}$.  We
see that even for 50\% \EPEM\ conversion probability the decrease of
the photon polarisation is only about 1\%. Moreover, as was mentioned
before, due to the beam repulsion the coherent pair creation
probability at TESLA is small, therefore the Depolarisation will be
even smaller. Hence, the Depolarisation of photons is negligibly
small.

\subsubsection{Disruption angle}   \label{s4.2.5}

The maximum disruption angle is an important issue for photon
colliders, it determines the value of the crab--crossing angle.

One source of  large angle particles are low energy electrons from
the conversion region. The minimum energy is about
$0.02E_0$ (section \ref{s3.1.4}). The second source of soft particles is hard
beamstrahlung and coherent pair creation with the minimum energy of about
$0.05/\Upsilon$. Particles from these sources can carry very large
energies, therefore the crab--crossing angle should be sufficient for
removal of all these particles from the detector without hitting the
quads or detector components.

Another source of even lower energy particles are \EPEM\ pairs
produced incoherently in collisions of individual particles at the IP.
This unavoidable background is proportional to the luminosity. A large
fraction of these particles (with large energy and small angles) can
also escape from the detector through the exit hole for disrupted
beams.  This source of background carries much less power than
enumerated in the previous paragraph and can be handled without 
crab--crossing, as  in the \EPEM\ TESLA option.

The deflection angle for soft electrons in the field of the opposite beam 
is given approximately by~\cite{TEL90,TESLAgg}
\begin{equation}
\vartheta_d \approx 0.7\left(\frac{4\pi r_e N}
{\sigma_z \gamma_{min}} \right)^{1/2} \approx
 9\left(\frac{N/10^{10}} {\sigma_z [\mathrm{mm}] E_{min} [\mathrm{GeV}]}
\right)^{1/2}\MRAD.
\label{thdisr}
\end{equation} 
In the first approximation the deflection angle for very soft
electrons does not depend on the transverse beam size. The coefficient
0.7 here was found by tracking particles in the field of the beam with
a Gaussian longitudinal distribution for the TESLA range of
parameters.  For example: at $2E_0 = 500 \GEV$, $E_{min}/E_0 = 0.02$
(Compton, $x=4.8$), $N= 2\times 10^{10}, \sigma_z = 0.3\MM$ we
get $\vartheta_d \approx 10.4\MRAD$. This estimate will help us to
understand   results of the simulation.

The coefficient 0.7 in (\ref{thdisr}) corresponds to the collision of
a low energy electron with the electron beam. If a low energy electron
is produced near the centre of the opposing beam then it is more
accurate to use the coefficient 1.2 instead of 0.7.

\subsection{The simulation code}\label{s4.3}

As we have seen, the picture of beam collisions at photon colliders is 
complicated and the best way to obtain final results is a
simulation. In the present study we used the code described in~\cite{TEL95}.

It serves for  simulation of \EPEM, \EMEM, \GE, \GG\ beam
collisions in linear colliders and the present version takes into
account the following processes:

\begin{enumerate} 
\item 
  {\it Compton scattering in the conversion region}. At present we use the
  formulae for linear Compton scattering, including all polarisation effects.
  Nonlinear effects are considered approximately by smearing $x$ 
  ($x\to x/(1+\xi^2)$) according to the variable density of laser photons
  in the conversion region.

\item 
  {\it \EPEM\ pair creation in the conversion region} for $x>4.8$. 

\item 
  {\it Deflection by  magnetic fields and synchrotron radiation} in the region
  between the CP and IP, due to special magnets or the solenoidal detector
  field (it has an effect due to the crab--crossing angle).

\item 
  {\it Electromagnetic forces, coherent pair creation and beamstrahlung}
  during beam collisions at the IP.   

\item 
  {\it Incoherent \EPEM\ creation} in \GG, \GE, \EPEM\ collisions.
\end{enumerate}

The initial electron beams are described by about 3000 macro--particles
(m.p.) which have a shape of flat rectangular bars with the horizontal
size equal to 0.4$\sigma_x$ and zero vertical size. In
the longitudinal direction the electron bunch has a Gaussian shape
($\pm 3\sigma$) and is cut into about 150 slices. It is assumed that the
macro--particles 
have only a transverse field and influence macro--particles of the
opposite bunch which have the same z--coordinate (this coordinate
changes by steps).  At initial positions macro--particles move
to the collision region according to the beam emittances and beta
functions.  During the simulation new macro--particles (photons,
electrons and positrons) are produced which are included in the
calculation in the same way as the initial macro-particles.

Low energy particles can get too large a deflection during one step
(because the step is too large). This problem is solved by artificial
restriction of the deflection angle (and the corresponding transverse
displacement) for one step.  The resulting angles will be simulated
correctly because the repulsion length for the soft electron is much
shorter than the bunch length and the charge distribution (the beam
field) in the next steps is approximately the same.

The code was used for simulation of photon colliders  in NLC
Zero Design and the TESLA Conceptual Design. The results are in agreement~\cite{TAK} with the code CAIN~\cite{YOK} written later for the same purpose.

\subsection{Luminosity limitations due to beam collision effects} \label{s4.4}

Beam collision effects in \EPEM\ and \GG, \GE\ collisions are
different.  In particular, in \GG\ collisions there are no
beamstrahlung or beam instabilities.  Therefore, it was of interest
to study limitations of the luminosity at the TESLA photon collider
due to beam collision effects.  The simulation~\cite{Tfrei,telnov} was
done for the TESLA beams and the horizontal size of the electron beams was
varied.

\subsubsection{Ultimate luminosities} \label{s4.4.1}

\begin{figure}[!htb]
\centering
\hspace*{-0.7cm} \epsfig{file=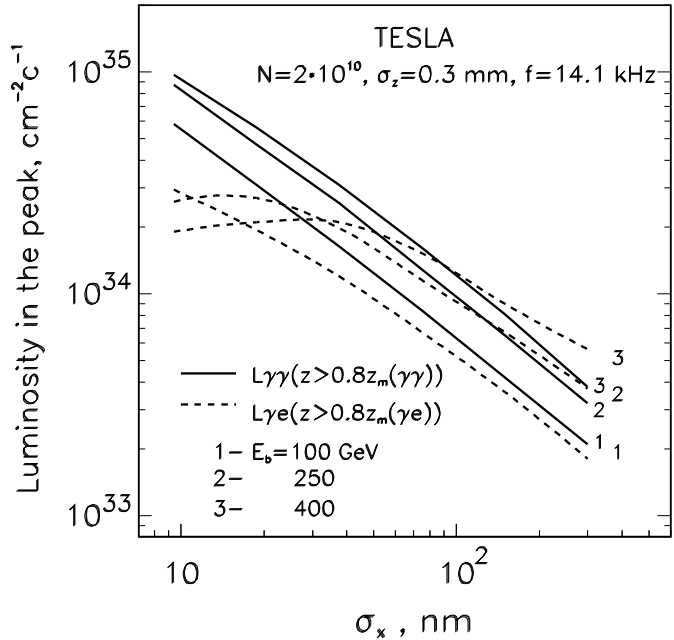,width=11.5cm,angle=0} 
\caption{Dependence of \GG\ and \GE\ luminosities in the high energy
  peak on the horizontal beam size for TESLA  at various
  energies. See also comments in the text.}
\label{sigmax}
\end{figure}

Fig.~\ref{sigmax} shows the dependence of the
\GG\ (solid curves) and the \GE\ (dashed curves) luminosities on the
horizontal beam size for several energies.
  The horizontal beam size was varied by changing  the horizontal 
beam emittance keeping the horizontal beta function at the IP 
constant and equal to $1.5\MM$. 

One can see that all curves for the \GG\ luminosity follow their natural
behaviour: $L\propto 1/\sigma_x$ (values of $\sigma_x < 10\NM$ are not
considered because too small horizontal sizes may introduce problems
with the crab--crossing scheme).  Note that while in \EPEM\ collisions
$\sigma_x\approx 500\NM$,  in \GG\ collisions the attainable
$\sigma_x$ with the planned injector (damping ring) is about $100\NM$
(Section~\ref{s4.5}).

In \GE\ collisions the luminosity at small $\sigma_x$ is lower than
follows from the geometric scaling due to beamstrahlung and
displacement of the electron beam during the beam collision.  So, we
can conclude that for \GG\ collisions at TESLA one can use beams with
a horizontal beam size down to $10\NM$ (maybe even smaller) which is
much smaller than that in \EPEM\ collisions.  Note, that the vertical
beam size could also be additionally decreased by a factor of two (for
even smaller electron beam size the effective photon beam size will be
determined by the Compton scattering contribution). As a result, the
\GG\ luminosity in the high energy peak can be, in principle, several
times higher than the \EPEM\ luminosity (Table~\ref{tabtel}).

Production of the polarised electron beams with emittances lower than
those possible with damping rings is a challenging problem.  There is
one method, laser cooling~\cite{TSB1,Monter,Tlasv1} which allows, in
principle, the required emittances to be reached.  However this method
requires a laser power  one order of magnitude higher than is
needed for $\EL \to \gamma$ conversion. This is not excluded, but since
many years of R\&D would be required, it should be considered as a
second stage of the photon collider, maybe for a Higgs factory.

\subsubsection{Disruption angles}\label{s4.4.2}

As it was mentioned before, for small beam sizes one can expect
the production of low energy particles in the processes of coherent pair
creation and beamstrahlung. The luminosity may not be
affected, but there is the problem with background due to the deflection of
the low energy particles by the opposing electron beam.
\begin{figure}[!htb]
\hspace*{+0.0cm} \epsfig{file=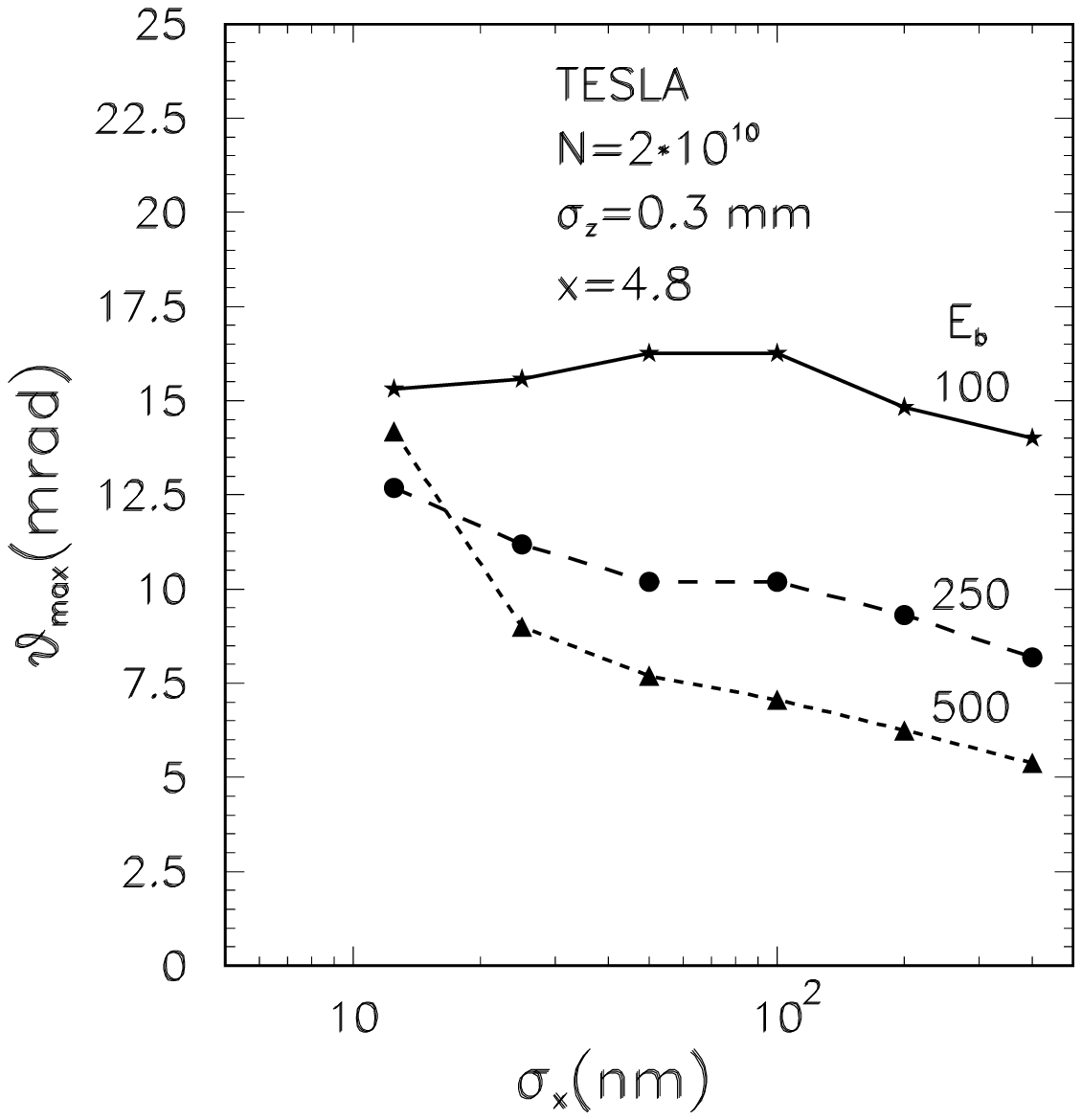,width=8.4cm,angle=0} 
 \hspace*{-1.5cm}           \epsfig{file=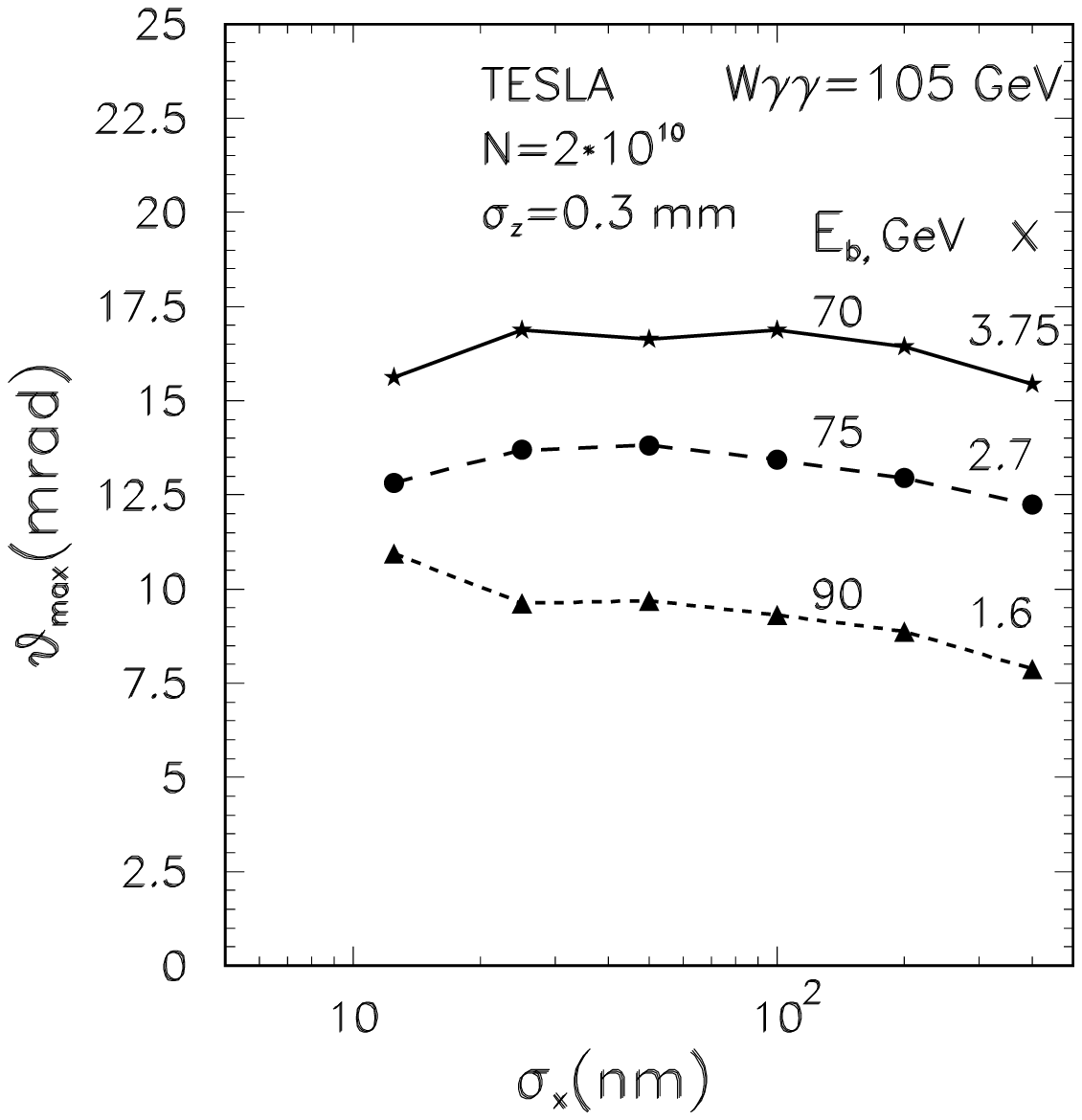,width=8.5cm,angle=0}
\caption{Dependence of the maximum disruption angle on the horizontal 
  beam size for TESLA at various energies. Left figure for $x=4.8$ and
  several beam energies. Right figure corresponds to the invariant
  mass $W_{\GG} = 105\GEVI$, $x$ values $1.6$, $2.7$, $3.75$ correspond to
  the laser wave lengths $1.06$, $1.06/2$, $1.06/3 \MKM$, respectively.}
\label{tmax}
\end{figure} 
Fig.~\ref{tmax} shows the dependence of the maximum disruption angle on
the horizontal beam size. In the left figure the parameter $x = 4.8$,
the right figure corresponds to the c.m.s. energy of the \GG\ collider
equal to $105\GEV$.
The total statistics in the simulation is about $10^5$ particles, so the
tails which can lead to background are not simulated. However, we know
the scaling and therefore can make corrections. From the simulation
we have found the angle corresponding to the probability $10^{-4}$
and multiplied it by a factor of 1.25. The angle shown in 
Fig.~\ref{tmax} is the angle above which the energy of background particles
is less than about $10\TEV$, that is less than the energy of the incoherent
\EPEM\ pairs (Section~\ref{s4.7}) which have larger angles and represent
an unavoidable background.

In  Fig.~\ref{tmax} (left) we see  that at large
$\sigma_x$ the angle is smaller for higher beam energies, in agreement with
(\ref{thdisr}). With decreasing $\sigma_x$ the contribution of the low
energy particles from coherent pair creation and beamstrahlung is
seen.

Fig.~\ref{tmax} (right) shows that at the fixed \GG\ center--of--mass energy
$\WGG$ the disruption 
angle is larger for larger $x$. It is easy to show that
\begin{equation}
\vartheta \propto \frac{x}{\sqrt{(x+1)\sigma_c(x)}},
\end{equation}
where the Compton cross section $\sigma_c(x)$ decreases with increasing
$x$. This gives a factor of two difference between $x=1.6$ and $3.75$. We
think that one can study the low mass Higgs with $\lambda \approx 1.06\MKM$, i.e.
with the same laser at all energies below $2E_0 = 500\GEV$. 
Lower $x$ have the advantage of a higher degree of  linear
polarisation (Section~\ref{s3.1}).  As higher $x$ values also have also 
some advantages (sharper edge) we can foresee the possibility of 
a frequency  doubled laser. With these assumptions we conclude that
the maximum disruption angle is about $14\MRAD$. 
For the laser with $\lambda \approx 1\MKM$ $12\MRAD$ will be sufficient. In the
present 
design the crab--crossing angle in the second IP is $34\MRAD$.  These
values put restrictions on possible quadrupole designs.

\subsection[$\gamma\gamma$ and $\gamma$e luminosities at TESLA]
{$\bgamma\bgamma$ and $\bgamma$e luminosities at TESLA} \label{s4.5}

\subsubsection{Parameters of the electron beams}\label{s4.5.1}

In this section we discuss what luminosities can be obtained with the
technology presently available. It depends strongly on the emittances of the
electron beams. There are two methods of 
production, low--emittance electron beams: damping rings and
low--emittance RF--photo--guns (without damping rings).  The second
option is promising, but at the moment there are no such photo--guns
producing polarised electron beams~\cite{Ferrario}.  Polarisation of
electron beams is very desirable for photon colliders (sect~\ref{s2}).
So, there is only one choice now --- damping rings.

Especially for a photon collider the possibility of decreasing the beam
emittances at the TESLA damping ring has been studied~\cite{Decking}
and it was found that the horizontal emittance can be reduced by a
factor of 4 compared to the previous design. Now the normalised
horizontal emittance is $\ENX\ = 2.5\times 10^{-6}\M$. 

\begin{figure}[!htb]
\hspace*{2.cm} \epsfig{file=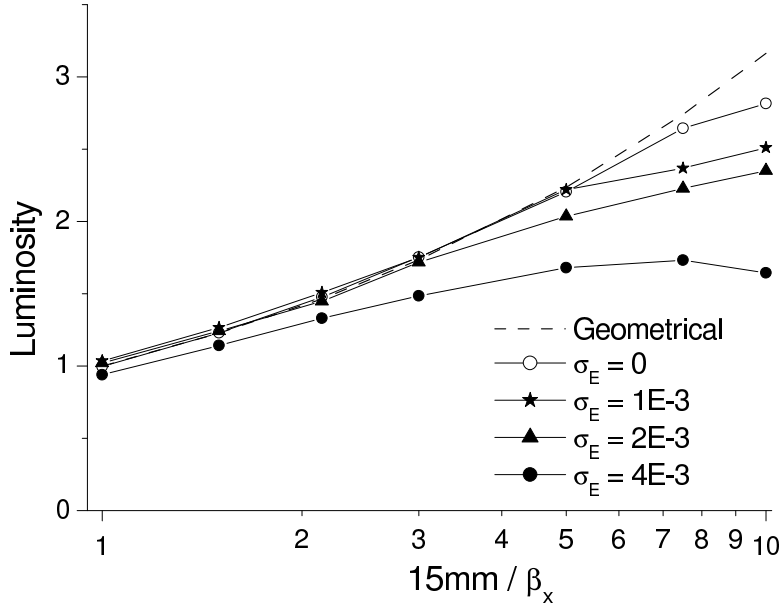,width=8.cm,angle=-90} 
\caption{Dependence of the geometric \EMEM\ luminosity on the 
  horizontal $\beta$--function (SLAC design). For TESLA the relative
  energy spread ($\sigma_E$ in the figure) is $10^{-3}$. } 
\label{seryi}
\end{figure} 

The luminosity also depends on the $\beta$--functions at the interaction
points: $L \propto 1/\sqrt{\beta_x \beta_y}$.  The vertical $\beta_y$ is
usually chosen close to the bunch length $\sigma_z$ (this is the design  for
\EPEM\ collisions and can also be realized for \GG\ collisions).  Some
questions remain about the minimum horizontal $\beta$--function. For
\EPEM\ collisions, $\beta_x \approx 15\MM$ which is larger than the bunch
length $\sigma_z = 0.3\MM$, because beams in \EPEM\ collisions 
must be
flat to reduce beamstrahlung. In \GG\ collisions, $\beta_x$ could be
about $1\MM$ (or even somewhat smaller).  There are two fundamental
limitations: the beam length and the Oide effects~\cite{Oide}
(radiation in final quads). The latter is not important for the beam
parameters considered. There is also a certain problem with the angular
spread of the synchrotron radiation emitted in the final quads. But,
for the photon collider the crab--crossing scheme will be used
and in this case there is sufficient clearance  for the removal of the
disrupted beams and synchrotron radiation.

Very preliminary studies of the existing scheme for the TESLA final
focus have shown~\cite{Walker} that chromo--geometric aberrations
dominate at $\beta \leq 6\MM$. However, this  is not a fundamental limitation
and it is very likely that after further study and optimisation a
better solution will be found.  At SLAC  a new
scheme for the final focus system has recently been proposed~\cite{Sery}.  The
first check without optimisation has shown~\cite{Sery1} that, with the
new scheme, one can obtain $\beta_x \approx 1.5\MM$ with small
aberrations, see Fig.~\ref{seryi}, and further optimisation is
possible.  For the present study we assume $\beta_x=1.5\MM$.

Some uncertainties remain for the operation of TESLA at low energies.  For
the low mass Higgs the minimum required energy is about $75\GEV$.  In
this case TESLA should work either at reduced accelerating gradient
or  a bypass after about $100\GEV$ should be used. In the case of a bypass one
can consider that the luminosity is approximately proportional to the
beam energy (due to the adiabatic change of the beam emittances).

In principle, the loss of luminosity at low energies could be
compensated by an increase of the repetition rate as 
$f \propto 1/E_0$. In this case the RF power (for the linac) is constant.
However, for the present design of the TESLA damping ring, the
repetition rate may be increased at most by a factor of 2. Further
decrease of the damping time is possible but at additional cost
(wigglers, RF--power).  The factor of 2 is almost sufficient, but,
unfortunately, at low gradients beam loading (RF efficiency) may be
problem.  Its adjustment requires the change of the coupler position, which
for  TESLA is technically very difficult or even impossible.

For the present study we assume the bypass solution and use the same beam
parameters ($N,\sigma_z$, normalised emittances, collision rate) for
all energies, that gives $L \propto E_0$.

\subsubsection[$\gamma\gamma$, $\gamma$e luminosities, summary table]
{$\bgamma\bgamma$, $\bgamma$e luminosities, summary table}  \label{s4.5.2}

The resulting parameters of the photon collider at TESLA for $2E_0=$
$200$, $500$ and $800\GEV$ are presented in Table~\ref{sumtable}. It is
assumed that the electron beams have 85\% longitudinal polarisation
and that the laser photons have 100\% circular polarisation. The
thickness of the laser target is one scattering length for $2E_0=500$
and $800\GEV$ and 1.35 scattering length for $2E_0=200\GEV$ (the
Compton cross section is larger), so that $k^2 \approx$ 0.4 and 0.55,
respectively. The parameter $\xi^2 = 0.15, 0.3, 0.4$ for $2E_0=200,$
$500$, $800 \GEV$, as explained in Section~\ref{s3.2}.  The laser wave
length is $1.06\MKM$ for all energies.  The conversion point is
situated at a distance $b=\gamma\sigma_y$ from the interaction point.

\begin{table}[!hbtp]
{\renewcommand{\arraystretch}{1.3}
\begin{center}
\begin{tabular}{l c c c} \hline
$2E_0$ [GeV] & 200 & 500 & 800   \\ \hline
$\lambda_L$ [\MKM]/$x $& 1.06/1.8 & 1.06/4.5 & 1.06/7.2 \\
$t_{L}$ $[\lambda_{scat}]$& 1.35 & 1 &1 \\
$N/10^{10}$& 2 & 2 & 2  \\  
$\sigma_{z}$ [mm]& 0.3 & 0.3 & 0.3  \\  
$f_{rep}\times n_b$ [kHz]& 14.1 & 14.1 & 14.1  \\
$\gamma \epsilon_{x/y}/10^{-6}$ [m$\cdot$rad] & 2.5/0.03 & 2.5/0.03 & 
2.5/0.03 \\
$\beta_{x/y}$ [mm] at IP& 1.5/0.3 & 1.5/0.3 & 1.5/0.3 \\
$\sigma_{x/y}$ [nm] & 140/6.8 & 88/4.3 & 69/3.4  \\  
b [mm] & 2.6 & 2.1 & 2.7 \\
$L_{\EL\EL}\, (geom)$ [$10^{34}$\, \CMS] & 4.8 & 12 &  19 \\  
$\LGG (z>0.8z_{m,\GG\ }) [10^{34}\CMS] $ & 0.43 & 1.1 &  1.7  \\
$\LGE (z>0.8z_{m,\GE\ }) [10^{34}\CMS] $ & 0.36 & 0.94 & 1.3 \\
$\LEE (z>0.65) [10^{34}\CMS] $ & 0.03 & 0.07 & 0.095 \\
\end{tabular}
\end{center}
}
\caption{Parameters of  the \GG\ collider based on TESLA. 
\label{sumtable}}
\end{table}

\begin{figure}[!htb]
\centering
\hspace*{-0.7cm} \epsfig{file=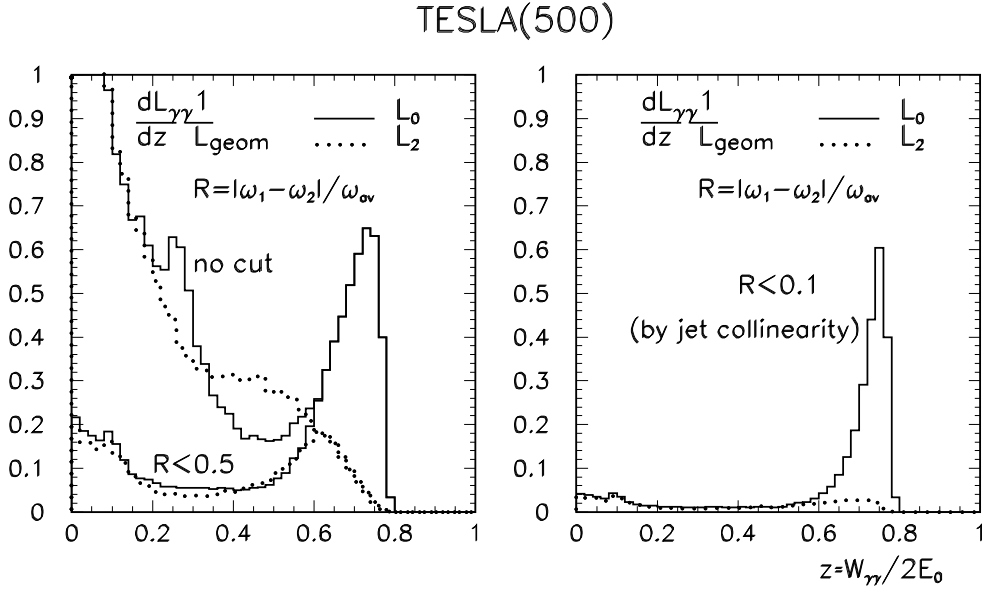,width=17cm,angle=0}
\caption{\GG\ luminosity spectra at TESLA(500) with various cuts 
  on longitudinal momentum. Solid line for total helicity of the two
  photons 0 and dotted line for total helicity 2. See also 
  Table~\protect\ref{sumtable}.}
\label{Ldist250}
\end{figure} 

As it was already mentioned in the introduction, 
for the same energy
\begin{equation}
\LGG(z>0.8z_m) \approx \frac{1}{3} L_{\EPEM}. 
\label{lgge+e-a}
\end{equation}

The relation (\ref{lgge+e-a}) is valid only for the beam parameters
considered. A more universal relation is (for $k^2=0.4$)
\begin{equation}
\LGG(z>0.8z_m) \approx 0.09 L_{\EL\EL}(geom).   
\label{lgge-e-}
\end{equation}

The normalised \GG\ luminosity spectra for 2E$_0 = 500 \GEV$ are
shown in Fig.~\ref{Ldist250}~\cite{telnov}.

The luminosity spectrum is decomposed into two parts with the total
helicity of the two photons 0 and 2. We see that in the high energy part
of the luminosity spectra the photons have a high degree of polarisation.
In addition to the high energy peak, there is a factor $5$--$8$ higher 
luminosity at low energy. It is produced mainly by photons after multiple
Compton scattering and beamstrahlung photons. These events have a
large boost and can be easily distinguished from the central high
energy events.  Fig.~\ref{Ldist250}  shows the same
spectrum with an additional cut on the longitudinal momentum of the
produced system, which suppresses the low energy luminosity to a low level.
For two jet events ($H\to b\bar b,\; \tau\tau$, for example) one can
restrict the longitudinal momentum using the acollinearity angle
between the jets. The resulting energy spread of collisions can be
about 7.5\%, see Fig.~\ref{Ldist250} (right).
\begin{figure}[hbt]
\hspace{-0.5cm}   \epsfig{file=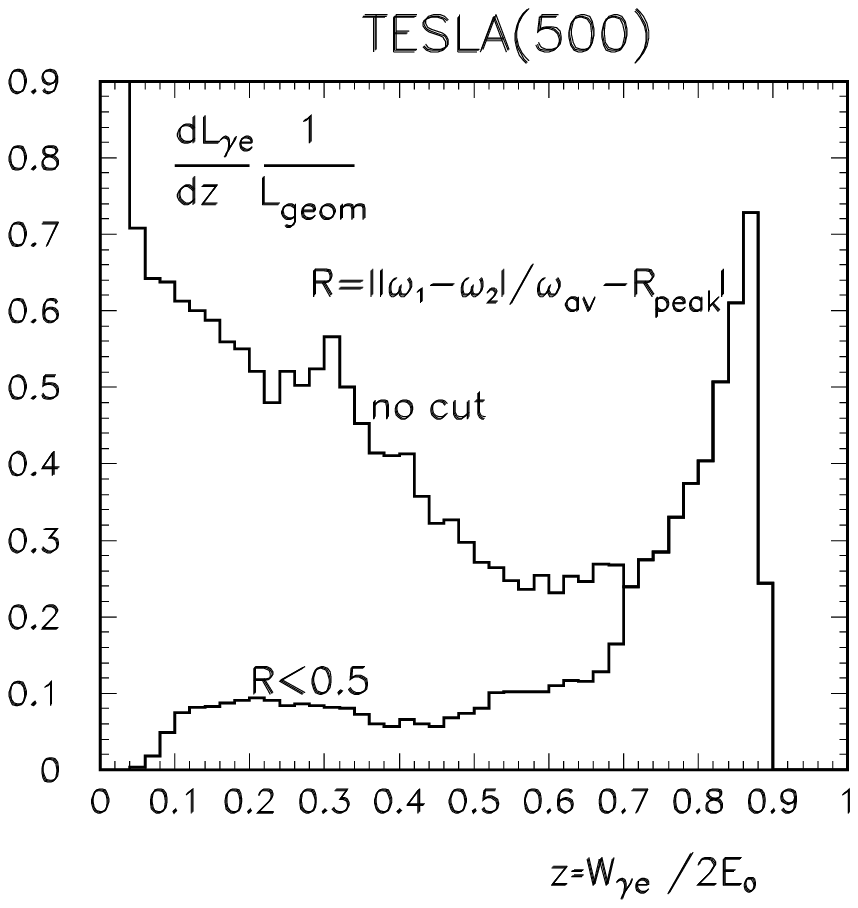,width=9.cm,angle=0} 
\hspace{-2cm} \epsfig{file=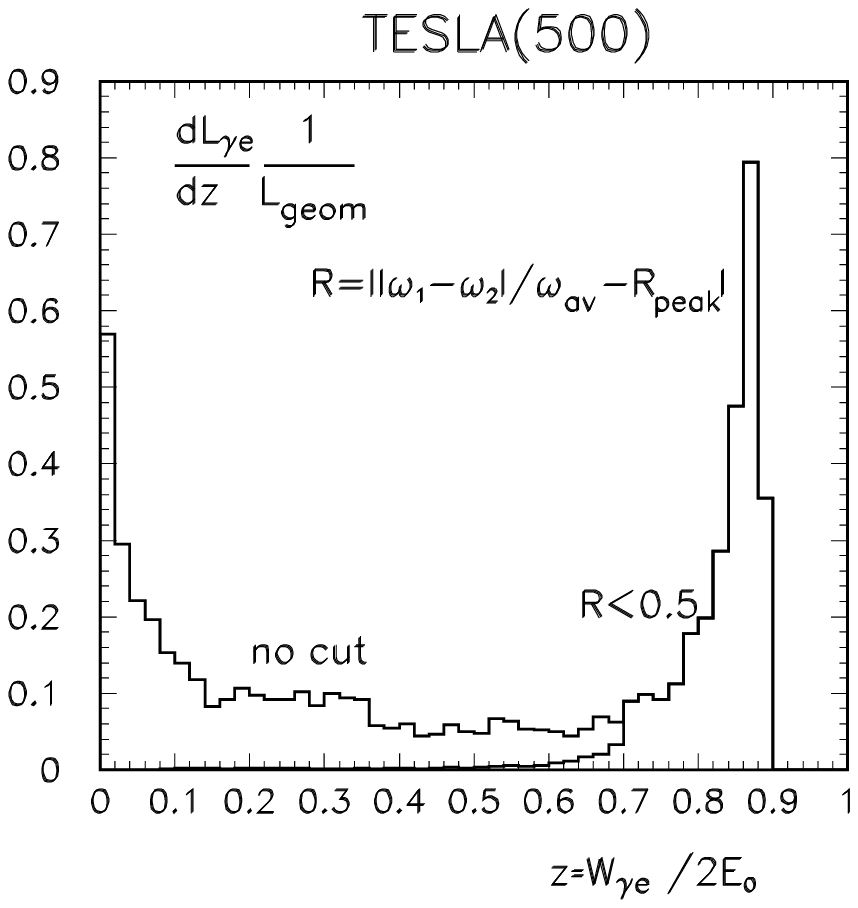,width=9.cm,angle=0} 
\caption{Left: normalised \GE\ luminosity spectra at TESLA(500) 
  when the photon collider is optimised for \GG\ collisions and there
  is $\gamma \to \EL$ conversion for both electron beams, parameters
  are given in table~\protect\ref{sumtable}. Right figure: there is
  $\gamma \to \EL$ conversion only for one electron beam and the
  distance between interaction and conversion point is 1.7\CM.  See
  comments in the text.}
\label{Ldist250e}
\end{figure} 
\begin{figure}[hbt]
\centering
\hspace*{-0.5cm} \epsfig{file=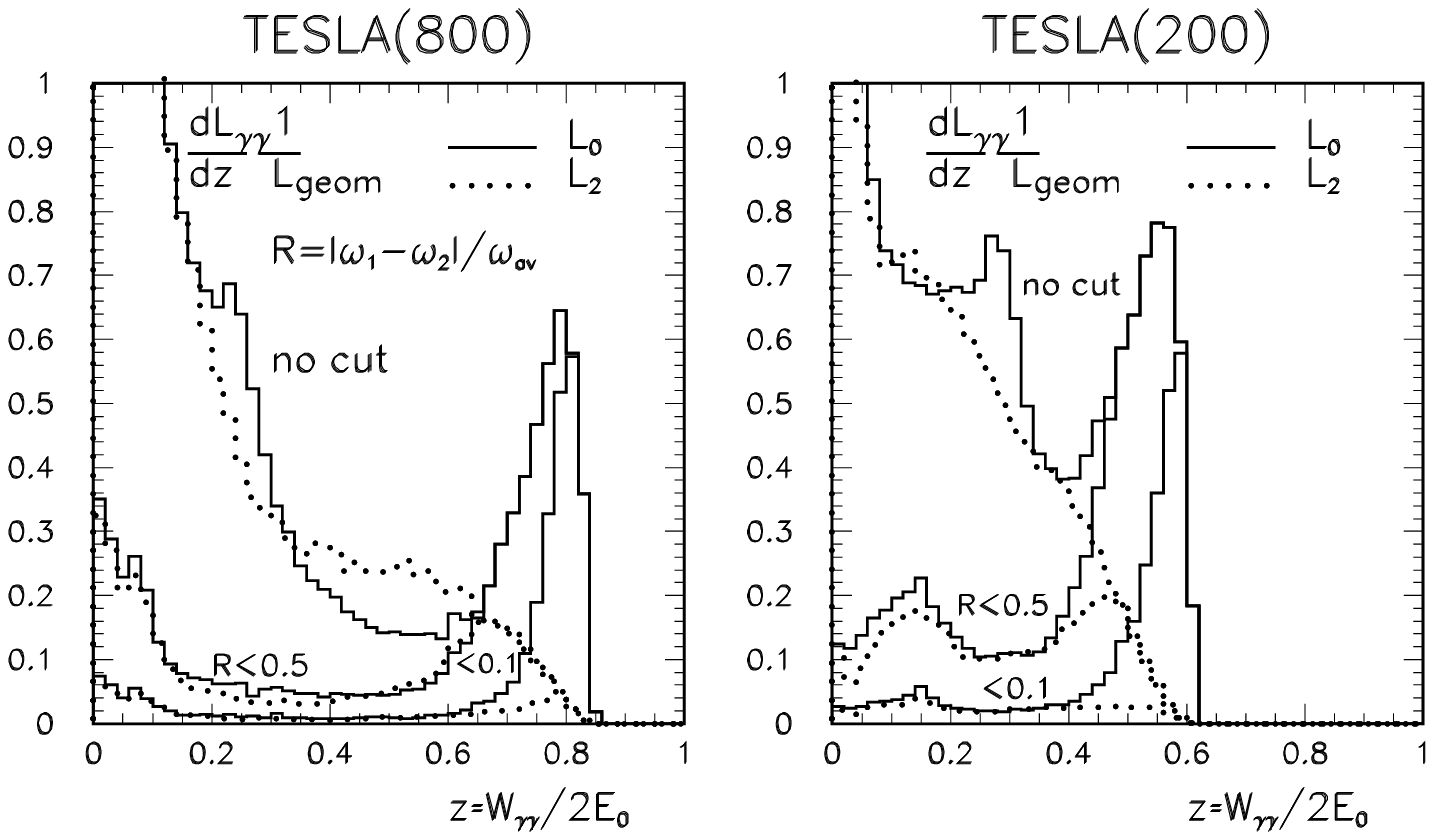,width=16.5cm,angle=0}
\caption{The \GG\ luminosity spectra at TESLA for $2E_0 = 800$ and $200\GEV$ 
  (for Higgs(120)) with various cuts on longitudinal momentum (the
  case of $2E_0= 500\GEVI$ is shown in Fig.~\protect\ref{Ldist250}).  The solid
  line is for the total helicity of the two photons 0 and the dotted
  line for the total helicity 2. See also Table~\protect\ref{sumtable}.}
\label{Higgs130}
\end{figure}

The high energy part of the \GG\ luminosity spectrum is almost
independent of collision effects at the IP (beamstrahlung and multiple
Compton scattering). For theoretical studies one can calculate the
high energy part of the luminosity spectrum with sufficient accuracy
by convolution of the Compton function~\cite{GKST84}. Recently, a
simple analytical formula for the Compton spectrum has been
obtained~\cite{Galynskii} which takes into account nonlinear effects
in the conversion region for sufficiently small  values of $\xi^2$.

The normalised \GE\ luminosity spectra for $2E_0=500\GEV$ are shown in
Fig.~\ref{Ldist250e}\-(left). Again, besides the high energy peak there is a
several times higher \GE\ luminosity at low invariant masses.  Note,
that the \GE\ luminosity in the high energy peak is not a simple
geometric characteristic of the Compton scattering process (as it is in \GG\ 
collisions). For the case considered it is suppressed by a factor of
$2$--$3$, mainly due to the repulsion of the electron beams and
beamstrahlung.  The suppression factor depends strongly on the
electron beam parameters. 

For dedicated $\GE$ experiments one can convert only one electron beam, 
increase the distance between the conversion and the interaction
points and obtain a much more monochromatic \GE\ luminosity spectrum.
One of such examples is shown in Fig.~\ref{Ldist250e}\-(right).

The luminosity distributions for $2E_0=800\GEV$ is presented in
Fig.~\ref{Higgs130}~(left), and for $2E_0=200\GEV$ on
Fig.~\ref{Higgs130}~(right). The latter case corresponds to 
$W_{\GG,m} \approx 120\GEV$.  At $2E_0=800\GEV$ the value $x \approx 7.2 > 4.8$,
however, due to nonlinear effects in the conversion region there is no
suppression of the luminosity which might be due to \EPEM\ creation (Section~\ref{s3.1.3}).

For the Higgs the production rate is proportional to $dL_0/dW_{\GG}$
at $W_{\GG}=\mass{\H}$.  For the case considered, $\mass{\H} \approx 120\GEV$,
and 
$x=1.8$, $dL_0/dW_{\GG} = 1.87\times 10^{32}\CMS/\GEV,$ so that the
coefficient in Fig.~\ref{hcross} characterising the width of the peak
is about 5.3 (instead of 7).

Several other important accelerator aspects of the photon collider
at TESLA are discussed in~\cite{Walker}.

\subsection[Monitoring and measurement of the $\gamma\gamma$ and $\gamma$e 
  luminosities]
{Monitoring and measurement of the $\bgamma\bgamma$ and $\bgamma$e 
  luminosities} \label{s4.6}

\subsubsection[Luminosity measurement in $\gamma\gamma$ collisions]
{Luminosity measurement in $\bgamma\bgamma$ collisions}\label{s4.6.1}

At photon colliders the luminosity spectrum is broad, photons and
electrons may have various polarisations. One should have method  to
measure all luminosity characteristics. Let us start from $\GG$ 
collisions.

We consider the head-on collisions of photons with 4--momenta
$k_{1,2}$ and  energies $\omega_{1,2}$. The
    $z$--axis is chosen along the momentum of the first photon, all the
azimuthal angles are referred to one fixed orthogonal $x$--axis.
The polarisation properties of the {\it i}--th photon are
described by three parameters:  $\lambda_{i}$ the mean
helicity (or degree of the circular polarisation), $l_i$ and
$\gamma_i$ the mean degree of the linear polarisation and the
azimuthal angle of its direction. The total cross section
$\sigma$ for the $\gamma\gamma$ collisions after summing over
polarisations of final particles has the form~\cite{BGMS}
\begin{equation}
\sigma =\sigma^{np}+ \lambda_{1}\lambda _{2}\; \tau^{c} + l_{1}
l_{2} \; \tau^l \; \cos {2(\gamma_1- \gamma_2)}
\label{1.4.13}
\end{equation}
where $\sigma^{np}$ is the total cross section for unpolarised
photons and $\tau^{c}$ ($\tau^{l}$) is the asymmetry related to
the circularly (linearly) polarised photons. Besides, we use the
notations $\sigma_0= \sigma^{np} + \tau^{c}$ and
$\sigma_2= \sigma^{np} - \tau^{c}$ where $0$ and $2$ denote
values of $|\lambda_1-\lambda_2|$ --- the total helicity of the
produced system. The system produced in a $\gamma \gamma$
collision is characterised by its invariant mass $W_{\gamma
\gamma}=\sqrt{4\omega_1 \omega_2}$ and rapidity $\eta = 0.5\,
\ln(\omega_1/\omega_2)$.

Let us fist consider the important case when both photons are
circularly polarised. In this case we should have a method
to measure a spectral luminosity $dL/dW_{\gamma \gamma} d\eta$
and the product of helicities $\lambda_1 \lambda_2$ or, in other
words, the spectral luminosities $dL_0/dW_{\gamma \gamma} d\eta$
and $dL_2/dW_{\gamma \gamma} d\eta$ with the total helicity $0$
and $2$.

These luminosities can be measured using the process $\gamma \gamma
\to l^+l^-$, where $l=\EL$ or
$\mu$~\cite{GKST83,TEL93,TEL95,MILLER,YASUI}. The cross section of
this process for colliding photons with total helicity $0$ and $2$ and
for $W^2_{\gamma \gamma} \gg m^2$ is ($\hbar = c = 1$)
$$
\sigma_0 (|\cos \vartheta| < a) \approx \frac{4\pi\alpha^2}{W^2_{\gamma\gamma}}
 {8m^2\over W^2_{\gamma\gamma}}
\left[\frac{1}{2} \ln \left(\frac{1+a}{1-a}\right) + \frac{a}{1-a^2}\right]
$$
\begin{equation}
\sigma_2 (|\cos \vartheta| < a) \approx \frac{4\pi\alpha^2}{W^2_{\gamma\gamma}}
\left[2 \ln \left(\frac{1+a}{1-a}\right) -2a \right].
\end{equation}

One can see that $\sigma_0 /\sigma_2 \sim m^2/W^2_{\gamma
\gamma} \ll 1$ (excluding the region of small angles). For
photons with arbitrary circular polarisations the cross section
is
\begin{equation}
\sigma_{\GG \to \EPEM} = \frac{1+\lambda_1 
\lambda_2}{2} \sigma_0
+ \frac{1-\lambda_1 \lambda_2}{2} \sigma_2,
\label{sigma02}
\end{equation}
where $\sigma_2 \gg \sigma_0$.

Hence the number of events
\begin{equation}
dN_{\gamma \gamma \to \mu^+ \mu^-} \approx dL \,{1-\lambda_1
\lambda_2\over 2}\, \sigma_2 \equiv dL_2\, \sigma_2\,,
\label{1.4.12}
\end{equation}
and one can measure the luminosity $dL_2/dW_{\gamma \gamma} d\eta$.
Measurement of $dL_0/dW_{\gamma \gamma} d\eta$
 is done by inversion of the
helicity of  one photon beam 
simultaneously changing the signs of the helicities of the laser beam
used for the  $e \to \gamma$ conversion and that of the
electron beam~\cite{TEL93}.  In this case the
spectrum of scattered photons is not changed while the product
$\lambda_{1}\lambda_{2}$ changes its sign. In other
words, $L_0$ "becomes" now $L_2$, which is measurable.  The cross
section for this process is 
$\sigma(|\cos\vartheta|<0.9)\approx 10^{-36}/W^2_{\GG}[\mathrm{TeV}] \CM^2$.  This
process is very easy to select due to a zero coplanarity angle.

Linear photon polarisations can also be measured using the above
processes.  At large angles the cross section has a strong correlation
between the plane of the final state particles and the directions of the photon
polarisations. Let us consider  the general case in more detail.

The differential cross section can be written in the form~\cite{GKST84} 
\begin{equation}
d\sigma={\alpha^2 \,T\over W^2_{\GG}\,(m^2+{\bf p}_{-\bot}^2)^2}\,
d\Gamma\,,\;\;
d\Gamma = \delta(k_1+k_2-p_--p_+) {d^3 p_- d^3 p_+
\over E_- E_+} = {dt d\varphi_-\over W^2_{\GG}},
\label{3}
\end{equation}
where ${\bf p}_{-\perp}$ is the transverse momentum of the
electron, $t=(k_1-p_-)^2$ and $\varphi_-$ is the azimuthal
angle of the electron. The quantity $T$ is
\begin{equation}
T= T_{00} + \lambda_1 \lambda_2\,T_{22} - 2T_\varphi\,,
\label{4}
\end{equation}
with
$$
T_{00}=m^2(W^2_{\GG}-2m^2)+{\bf p}_{-\bot}^2(W^2_{\GG}-2{\bf p}_{-\bot}^2)\,,
$$
\begin{equation}
T_{22}=m^2(W^2_{\GG}-2m^2)-{\bf p}_{-\bot}^2(W^2_{\GG}-2{\bf p}_{-\bot}^2)\,,
\label{5}
\end{equation}
and
$$
T_\varphi=  l_1 l_2 \,[\,m^4\,\cos
{(2\phi_1-2\phi_2)} + ({\bf p}_{-\bot}^2)^2
\cos{(2\phi_1+2\phi_2)}\,] -
$$
\begin{equation}
-2m^2 {\bf p}_{-\bot}^2 \,
[\, l_1\,\cos{2\phi_1} +l_2\, \cos{2\phi_2}\, ] \, ,
\label{6}
\end{equation}
where $\phi_i = \varphi_- - \gamma_i$ is the (azimuthal) angle
between the vector ${\bf p}_{-\perp}$ and the direction of the
linear polarisation of {\it i}-th photon (therefore, the angle
$\phi_2-\phi_1 = \gamma_1-\gamma_2$).
From (\ref{4}), ignoring the azimuthal term, the
contribution of the total helicity 0 corresponds to the sum $T_{00}+T_{22}$
and the helicity 2 to the term $T_{00}-T_{22}$, which is smaller by a factor of
$m^2/p^2_\bot$, in agreement with our previous observation (see~\ref{sigma02}).

At high energy and not too small angles the cross section is
\begin{equation}
d\sigma={\alpha^2\over W^2_{\GG}} \, \left[(1-\lambda_1\lambda_2)
\left({W^2_{\GG}\over
{\bf p}_{-\bot}^2}-2\right)-2l_1\, l_2\, \cos(2\phi_1+
2\phi_2)\,\right] \, d\Gamma\,,
\label{7}
\end{equation}
$$
d\Gamma = {2\omega_1\omega_2 \over
\left[\omega_1(1-\cos{\theta_-})+ \omega_2(1+\cos{\theta_-})
\right]^2}\, d\Omega_-\,,\;\; W^2_{\GG} \gg m^2,\quad {\bf
p}_{-\bot}^2\gg m^2
$$
where $d\Omega_-$ is  the electron solid angle.
One sees that at large angles ($p_\bot \sim W_{\GG}/2$) the cross
  section depends strongly on the degrees of both the circular and the linear
  photon polarisations.

The cross section of the calibration processes 
$\GG\ \to \EPEM (\mu^+\mu^-)$ is larger than those for most processes to be
studied 
and only the processes $\GG \to$W$^+$W$^-$ and 
$\GG\ \to \mathrm{hadrons}$ have larger cross sections.  However, taking the
detection efficiency for WW into account, the counting rate of WW
pairs will be comparable with that of the calibration processes.  As
for hadrons, the expected number of calibration events is sufficient
to measure the properties of hadronic reactions with high accuracy.

Note that the momenta of electrons (muons) in the processes under
discussion can be measured with a high accuracy which is very
important for the determination of the luminosity distribution near the
high energy edge. 

Other processes with large cross sections which can be used for the
luminosity measurement are $\GG\  \to \WP\WM$~\cite{YASUI} and 
$\GG\ \to\mu^+\mu^-\mu^+\mu^- $~\cite{GKST83,Kapusta}. The first process has a
total cross section of $8\times10^{-35}\,\CM^2$ the second one
$1.6\times10^{-34}\,\CM^2$.  The first process depends on the photon
polarisations especially in the region of large angles~\cite{GKPS1,GKPS2}.
The second processes is sensitive only to the linear photon
polarisation.  These processes may be useful, for an
independent check and a fast monitoring of the luminosity.

\subsubsection[Luminosity measurement in $\gamma$e collisions]
{Luminosity measurement in $\bgamma$e collisions}\label{s4.6.2}
 
   For the absolute \GE\ luminosity measurement, one can use  the
process of Compton scattering, which is strongly polarisation
dependent.

Let us consider the polarisation properties of Compton scattering at
high energies.  For an \GE\ collider we consider the head--on
collision of an electron with 4--momentum $p$ and a photon with
4--momentum $k$, energies $E$ and $\omega$ of the same order and the
squared invariant mass of $\GE$ system 
$W^2_{\GE}= (p+k)^2 \approx 4E\omega$.  We choose the $z$--axis along the
momentum of the electron. 
The polarisation properties of the electron are described by its mean
helicity $\lambda_{\EL}$ ($|\lambda_{\EL}|\leq 1/2$), transverse polarisation
$\zeta_\perp$ ($\zeta_\perp \leq 1$), and the azimuthal angle $\beta$
of the direction of the transverse polarisation.  The polarisation
properties of the photon are described by three parameters:
$\lambda_{\gamma}$ the mean helicity (or degree of the circular
polarisation), $l_\gamma$ and $\gamma$ the mean degree and the
direction of the linear polarisation.

The total and differential cross sections for the process 
$\EL(p)+ \gamma(k)\to \EL(p')+ \gamma(k')$ and
their dependence on the polarisation of the initial particles are
discussed in~\cite{GKST84}. We consider here the case of high
energies $W^2_{\GE} \gg m^2$ only. In this case the total cross section
\begin{equation}
\sigma\approx (1+2\lambda_{\EL}\lambda_\gamma)\,{2\pi\alpha^2\over
W^2_{\GE}}\,\ln{W^2_{\GE}\over m^2}, \quad W^2_{\GE}\gg m^2
\label{10}
\end{equation}
depends strongly on the circular photon polarisation and on the longitudinal
electron polarisation only. Here the mean electron
helicity is defined as a projection of its spin and 100\% polarisation
corresponds to $\lambda_{\EL} = 1/2$. 

The differential cross section depends on the degrees of the
circular and linear polarisations of the photon and on its angle
$\gamma$ which determines the direction of the linear photon
polarisations as well as on the electron polarisation.  It can be
written in the the form
\begin{equation}
d\sigma={\alpha^2 \,F_0\over m^2x\,}\,d\Gamma\,,\;\;
d\Gamma = \delta(p+k-p'-p') {d^3 p' d^3 k'
\over E'\omega' } = {dy d\varphi_\gamma}
\label{11}
\end{equation}
where
$$
x= {2pk\over m^2}\approx {4E\omega\over m^2} \gg 1\,,\;\;
y=1- {pk'\over pk}\,,\;\; r={y\over (1-y)x}
$$
and $\varphi_\gamma$ is the azimuthal angle of the final photon.
The quantity $F_0$ is
\begin{equation}
F_0 = {1\over 1-y} + 1-y - 4r(1-r)\left[1+ l_\gamma\,
\cos{2(\varphi -\gamma)} \right]-
\label{12}
\end{equation}
$$
-y\lambda_\gamma \left[2\sqrt{r(1-r)} \, \zeta_\perp \, \cos{(\varphi
-\beta)}- {2-y\over 1-y}\,(1-2r)\, 2\lambda_{\EL} \right]\,.
$$

In the region of angles $\theta_\gamma \gg m/E$, we
have
\begin{equation}
1-y = {E(1-\cos{\theta_\gamma})\over E(1-\cos{\theta_\gamma})+
\omega(1+\cos{\theta_\gamma})}\,,\;\;
d\Gamma =
{2E\omega\over [E(1-\cos{\theta_\gamma})+
\omega(1+\cos{\theta_\gamma})]^2}\, d\Omega_\gamma\,.
\label{13}
\end{equation}
If the angle $\theta_\gamma \approx 1$ all terms in expression
(\ref{12}) have to be taken into account. Thus   by
detecting the final state particles at large angles, one can measure all
polarisation parameters of the colliding particles.

In the region ${m/E} \ll \theta_\gamma \ll 1$, which corresponds to a
large cross section, the expression for the differential cross section is
\begin{equation}
d\sigma = {\alpha^2 \over (E\theta_\gamma)^2}\,
\left(1+2\lambda_{\EL} \lambda_\gamma\right) \,d\Gamma\,,\;\;
d\Gamma = {E\over 2\omega}\, d\Omega_\gamma
\label{14}
\end{equation}
which depends strongly only on the circular photon polarisation and
longitudinal electron polarisation only.

For the luminosity tuning in \GG\ and \GE\ collisions one can use the
beam--beam deflection (same as for \EPEM) and ``background'' processes like
incoherent \EPEM\ and hadron production which are discussed in the
next section.

\subsection{Backgrounds} \label{s4.7}

Backgrounds cause problems for
recording data (complicating triggers) and data analysis
(underlying background processes, overlapping of ``interesting'' and
background events) and also damage of detectors. It is well
known that at \EPEM\ colliders background conditions are  much less severe
than at $pp$ or $p\bar{p}$ colliders because  the total $pp$/$p\bar p$ cross
section is much larger. 

The photon collider is based on electron--electron linear colliders and
therefore has a lot of common with \EPEM\ colliders as far as
backgrounds are concerned.  Like the electron, the photon interacts
electromagnetically and does not participate directly in strong
interactions.  Photon colliders produce a mixture of \EM\EM, \GE\ and
\GG\ collisions.  Electromagnetic interactions of these particles
between each other (incoherently) as well as with the beam field
(coherently) generate beamstrahlung photons, \EPEM\ pairs and other
reactions which are quite similar to those at \EPEM\ colliders.  These
QED backgrounds have small transverse momenta and cause problems
mainly for the vertex detector, the small angle calorimeter and the
luminosity monitor. Many of these particles hit the final quads
generating   showers for which some of these particles may
backscatter into the detector.  These backgrounds at photon
colliders are smaller than at \EPEM\ colliders because of the
crab--crossing collision scheme which provides a clear angle for
disrupted beams and for the most energetic part of the
luminosity--induced background.

On the other hand, due to virtual  $q\bar{q}$ pairs
the photon behaves as a hadron with the probability of about 1/200. The
corresponding cross section 
$\sigma(\GG\ \to \mathrm{hadrons}) \approx 5\times 10^{-31}\CM^2$ is smaller than the
total $pp$ cross section by 5 
orders of magnitude. However, the TESLA bunch crossing rate 
($\nu = 14$ kHz) is about 3000 times lower than that at the $pp$ collider
LHC. For the same luminosity the probability of accidental coincidence (or
the number of background events per bunch crossing) at the photon
collider will be smaller by a factor of 30.   At the \GG\ luminosity
planned at TESLA the average number of hadronic background events per
one bunch collision will be of the order of $1$--$3$ and we should expect
some problems with the analysis of certain physics processes. 

However, there is very big difference between $pp$ and \GG\ colliders
because the rate of hadronic events per second at photon colliders is
by 5 orders of magnitude smaller. Correspondingly there should be no problem
with the radiation damage of the detector, nor the
trigger.


In addition, photon colliders have several very specific background
problems.  Electrons after the Compton scattering have a very broad
energy spectrum, $E \approx (0.02$--$1)E_0$, and an angular spread of
about $5$--$10\MRAD$.  Removal of the disrupted beams requires the
crab--crossing beam collision.  This was discussed in Section~\ref{s4}.

Another specific problem is connected with the presence of the optical
mirrors very close to the beams. The mirrors are bombarded by the
large angle X--ray Compton scattered photons, by large angle
beamstrahlung photons and by synchrotron radiation from beam tails.
Also \EPEM\ pairs produced at the interaction point will hit the
mirrors. 

 Below the backgrounds are considered in the following order:

\begin{enumerate}
\item 
  Particles with large disruption angles hitting the final quads
  and mirrors.  The sources are multiple Compton scattering, hard
  beamstrahlung, Bremsstrahlung (in \EM\EM );

\item 
  \EPEM\ pairs created in the processes of $\EM\EM  \to  \EM\EM \EPEM$~
  (Landau--Lifshitz, LL), $\GE \to \EL\, \EPEM$~(Bethe--Heitler, BH),
  $\GG \to \EPEM$~(Breit--Wheeler, BW). This is the main source
  of low energy particles, which can cause problems in the vertex detector;
  
\item 
  $\GG \to \mathrm{hadrons}$;

\item 
  X--ray background (for optical mirrors).  
\end{enumerate}

\subsubsection{Low energy electrons} 

In Section~\ref{s4} we considered already the disruption angles of low
energy particles from  multiple Compton scattering, hard
beamstrahlung and coherent pair creation, and found that one can remove
these particle from the detector with low backgrounds using the
crab--crossing scheme with about 14\MRAD\ (radius) holes for the disrupted
beams. 
The low energy electrons after the hard bremsstrahlung may be
sufficiently deflected by the opposite beam and hit the quads. A
simple estimate shows that the total energy of these particles per
bunch collision is of the order of one TeV which is much smaller than
that of the \EPEM\ pairs discussed below.

\subsubsection[Incoherent e$^+$e$^-$ pairs]
{Incoherent e$^{\bp}$e$^{\bm}$ pairs} 
   
This source of background at the photon collider is  less
important than for the TESLA \EPEM\ collider because 1) one of the main
sources (LL) is almost absent; 2) many particles with almost 99\% of
the total energy escape through the hole for the disrupted beams, while
in \EPEM\ collisions at TESLA (without crab--crossing) they almost all 
 hit the quads. 

Nevertheless, we will consider here the main characteristics of \EPEM\ 
pairs which are important for designing the vacuum chamber near the IP and
for the vertex detector design.

This background was considered in detail in the CDR on the photon
collider at TESLA~\cite{TESLAgg}. Since that time the geometric design
luminosity has increased by one order of magnitude, but the \GE\ 
luminosity/per bunch collision has increased only 2 times, while
for \EMEM\ even decreased 3 times. So, with a good accuracy we can use
the previous numbers.
   
   Most of the $\EM$ and $\EP$ produced in LL, BH, BW processes travel
in the forward direction, but due to the kick in the field of the
opposing electron beam they get much larger angles and can cause 
problems in the detector.

In one bunch collisions about 50000 \EPEM\ pairs are produced with
a total energy of about $10^6\GEV$. A large fraction of these
particles escape the detector through the hole for the disrupted beams
(about $10-15\MRAD$) without interactions, and only particles with
$\vartheta>10\MRAD$ and $p\leq 1\GEV$ (the latter due to crab--crossing in the
solenoidal field) will hit the quads and mirrors.  The total energy of
these particles is much smaller: $2\times10^4\GEV$ (we use the CDR
number). We see that this energy is almost two orders of magnitude
lower than in the case of \EPEM\ collisions (without crab--crossing) where it
was found that the backgrounds are acceptable for 
the detector.  However, at the photon collider there are optical mirrors in
the way of the large angle particles which may lead to  differences in
the flux of back scattered particles. This has to be simulated more
accurately.

In the incoherent \EPEM\ background there are two classes of
particles: a) with large initial angles and b) with angles determined
by the beam--beam interaction.  The first class is an unavoidable
background (and rather small), the second class of particles, which carry
most of the total energy, can be suppressed by proper choice of the
beam pipe and vertex detector geometry.

The shape of the zone occupied by the deflected electrons with an energy
spectrum from 0 to $E_0$ is described by the formula~\cite{BATTEL,TESLAgg}
\begin{equation}
r^2_{max} \simeq \frac{25Ne}{\sigma_zB}z \approx 0.12\frac{N}{10^{10}}
\frac{z\, [\mathrm{cm}]}{\sigma_z\, [\mathrm{mm}]B\, [\mbox{T}]},
\end{equation} 

\noindent where $r_{max}$ is the radius of the envelope at a distance 
$z$ from the IP, $B$ is the longitudinal detector field.
For example, for TESLA with $N=2\times10^{10}, \sigma_z=0.3\MM$,
and $B=3\T$, $r=0.52 \sqrt{z\textrm{[cm]}}\CM$. This simple formula
can be used to define  the vertex detector radius and the shape 
of the vacuum chamber.

\begin{figure}[!hbt]
\centering
\includegraphics[width=5.5in,angle=0,trim=30 -5 30 30]{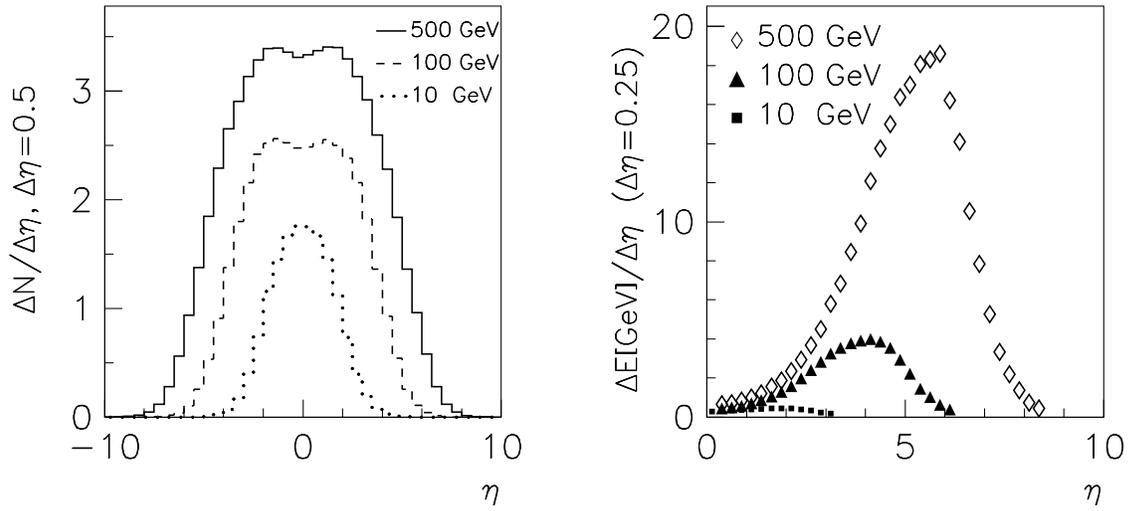}
\caption{Distribution of particle flow (left) and energy flow (right)
  in pseudo--rapidity in $\GG \to \mathrm{hadrons}$\ events for various values
  of $W_{\GG}$ assuming equal energies photons).}
\label{fig16}
\end{figure}

\begin{figure}[!hbt]
\centering
\includegraphics[width=1.2in,angle=0,trim=200 100 200 120]{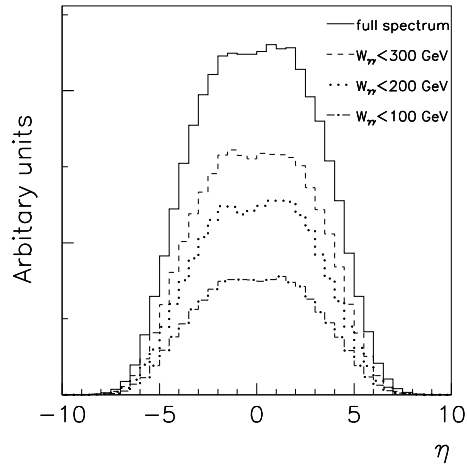}
\caption{Distribution of the number of particles 
  in pseudo--rapidity for different ranges of \GG\ 
  invariant mass for $2E_0=500\GEVI$.}
\label{fig17}
\end{figure}

\begin{figure}[!hbt]
\centering
\includegraphics[width=4.in,angle=0,trim=30 0 30 30]{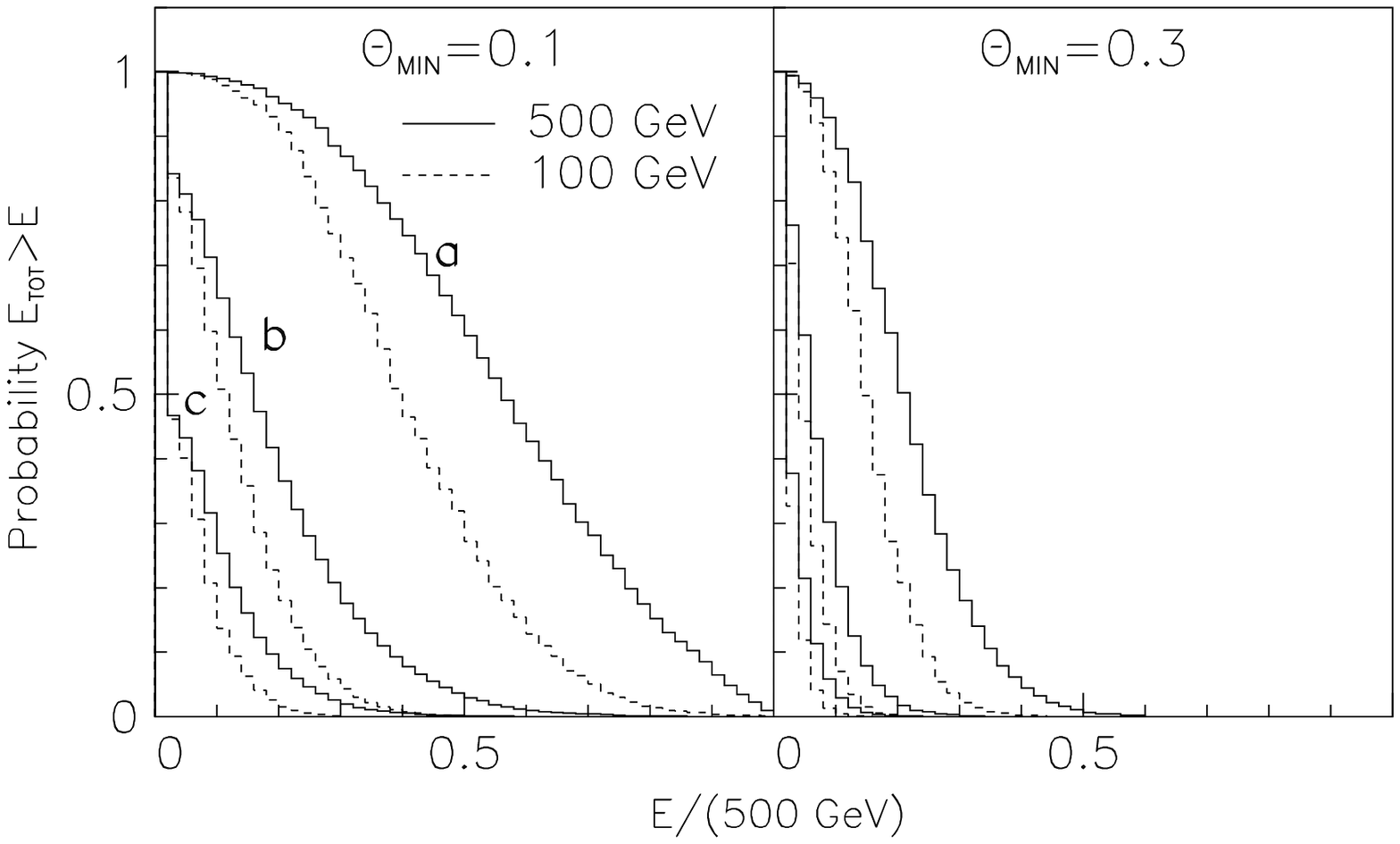}
\caption{The probability of an energy deposition in the detector above the
  value E due to the process $\GG \to \mathrm{hadrons}$.  The polar angle
  acceptance is $\vartheta > 0.1\RAD$ (left plot) and $\vartheta >
  0.3\RAD$ (right plot).  Curves a), b), c) correspond to 7, 2 and 0.7
  hadronic events on the average per beam collision respectively. The
  collision energy $W_{\GG}$ is 500\GEVI\ (solid line) and 100\GEVI\ 
  (dashed line); both photons have equal energies.}
\label{fig18}
\end{figure}

\begin{figure}[!htb]
\centering
\includegraphics[width=4.1in,angle=0,trim=30 40 30 40]{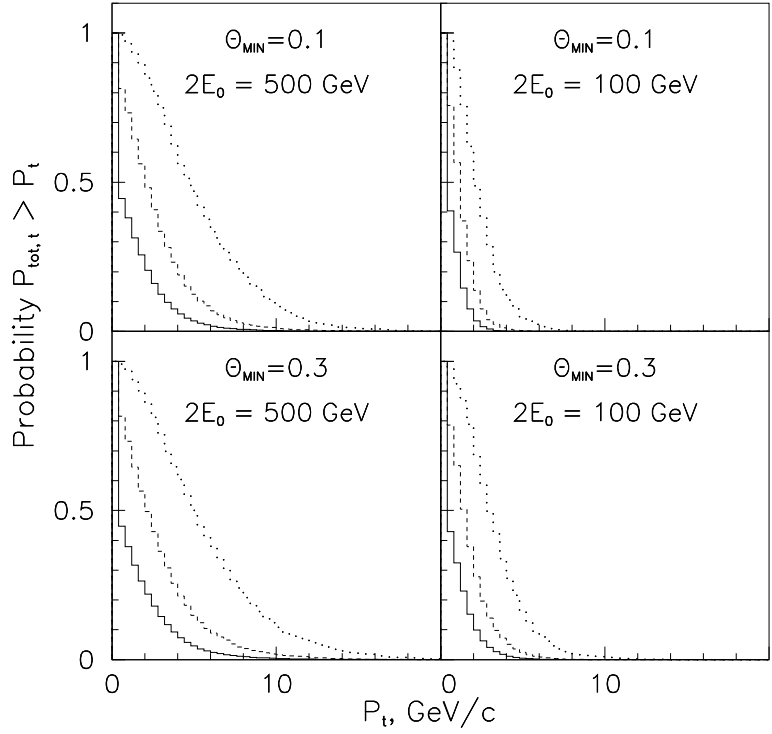}
\caption{The probability to find an unbalanced transverse 
  momentum above some $p_t$. Dotted, dashed and solid curves
  correspond to 7, 2, 0.7 $\GG \to \mathrm{hadrons}$ events on the average per
  beam collision. The polar angle acceptance is $\vartheta > 0.1\RAD$
  (upper plots) and $\vartheta > 0.3\RAD$ (lower plots).  The
  collision energy $W_{\GG}$ is 500\GEVI\ (left plots) and 100\GEVI\ (right
  plots), both photons have equal energies.}
\label{fig19}
\end{figure}

\begin{figure}[!htb]
\centering
\includegraphics[width=4.2in,angle=0,trim=30 20 30 20,clip]{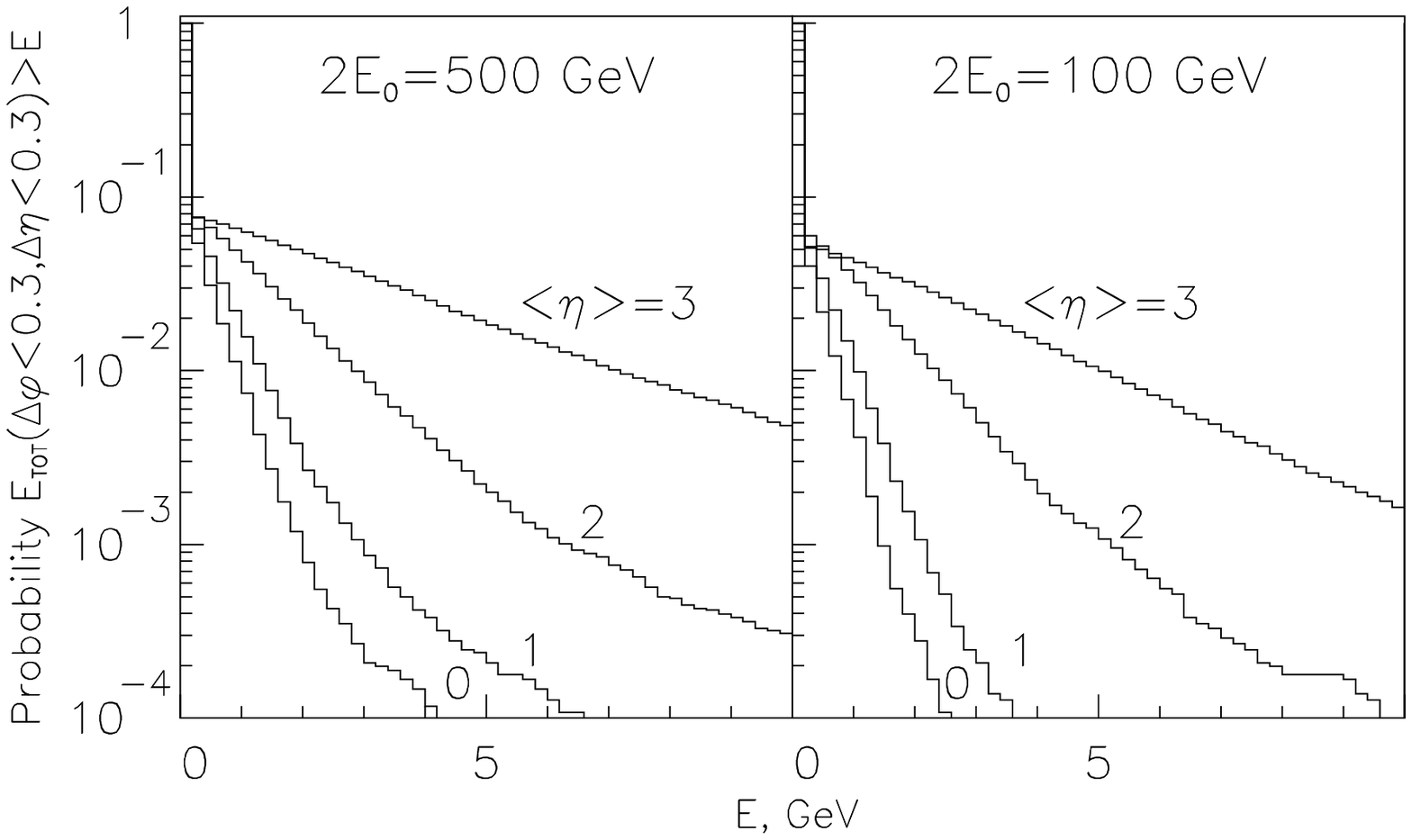}
\caption{The probability to have the energy flow into
$\Delta\phi \times \Delta\eta=0.3\times0.3$\ cell above
some threshold (abscissa value) for 4 pseudo--rapidity points: 
$\eta=$0, 1, 2, 3. $W_{\GG}$ is 500\GEVI\ (left)
and 100\GEVI\ (right).}
\label{fig20}
\end{figure}

\subsubsection[$\gamma \gamma$ $to$  hadrons]
{$\bgamma\bgamma$ $\bto$  hadrons}

The cross section of this process is about $400$--$600\nb$ at
$W_{\GG}=10$--$500\GEV$.  The \GG\ luminosity at the TESLA Photon Collider
(Table~\ref{sumtable}) is about $10^{35}\CMS$ in total, 
$5 \times 10^{34}$ with $z=W_{\GG}/2E_0 > 0.1$ and $1.2\times 10^{34}$ with 
$z > 0.65$.  The corresponding numbers of  hadronic events per bunch
crossing at $2E_0 = 500\GEV$ is about 3.5, 1.7 and 0.4, respectively.

We now discuss the consequences for the experiment and for the 
maximum luminosity. Detailed studies have been performed
for the TESLA CDR using the PYTHIA code 5.720~\cite{SCHULER}. At present there are new versions, but already at that
time processes such as mini--jets from resolved photons were included
approximately.  In that study we considered
different background levels, from 0.7  to 7 events/bunch
collision. The present TESLA parameters are within this range. The
change in the shape of the luminosity spectra is not essential.

Fig.~\ref{fig16} shows the flow of
particles and their energies versus pseudo--rapidity 
($\eta=-{\ln{\tan(\vartheta/2)}}$) in one $\GG \to \mathrm{hadrons}$ event at 
$ W_{\GG}=10,100$ and $500 \GEV$. Each $500 \GEV$ hadronic event produces on
the average 25 particles (neutral $+$ charged) in the range of
$-2\le \eta \le2$ ($\vartheta\ge0.27\RAD$) with a total energy
of about 15\GEV.  The average momentum of the particles is about 0.4\GEV.
Note that the flux of the particles at large angles ($\eta \approx 0$) from a
10\GEV\ \GG\ collision is only twice smaller than that from a 500\GEV\ \GG\ 
collision.

In this respect it is of interest to check the background from
different parts of the \GG\ luminosity spectra.  Fig.~\ref{fig17}
shows the distribution of particles in pseudo-rapidity for the TESLA
\GG\ luminosity spectrum at $2E_0=500\GEV$.  While the events with
$W_{\GG}<100 \GEV$ contain more than 60\% of the total
luminosity, their contribution to the number of background particles
is only about 30\%, due to the smaller energy and large longitudinal
boost of the produced system.

From figs.~\ref{fig16}, \ref{fig17} we see that the characteristics
of events at large angles (small rapidities) do not depend strongly
on the energy of the colliding photons.  Rather than using the
$W_{\GG}$ dependence for hadronic events/bunch collision (see
above), it is thus more convenient to use some ``average'' number of
central collisions  with  energy $W_{\GG} = 500\GEV$ with
equivalent background.  Fig.~\ref{fig17} allows to make a
reasonable approximation: events with $W_{\GG}>300\GEV$ are similar
to events at $W_{\GG} = 500\GEV$ and their contribution to the
luminosity and background is known. The effective average rate is
about 1.5 events per bunch collision.

The probability of an energy deposition in the detector above some
value E  is shown in Fig.~\ref{fig18}. In the left figure
the minimum angle of the detector is $\theta_{min} = 0.1\RAD$, on
the right one $\theta_{min} = 0.3\RAD$. The curves a), b), c) correspond
to 7, 2 and 0.7 hadronic events on  average per collision; the solid
curves are for $W_{\GG}=500\GEV$, the dashed  for 100\GEV. For example,
for 2 events per collision and $\theta_{min}=0.1$ the probability
of an energy deposition above 100\GEV\ is about 40\%.  This
energy is produced by many soft particles and a smooth background can
be subtracted during the jet reconstruction. More important are 
fluctuations in the background, which are discussed below. 
  
In many experimental studies the important characteristics is missing
transverse momentum. The probability to find an unbalanced
transverse momentum above some $p_t$ is shown in Fig.~\ref{fig19}
for $\vartheta_{min}=0.1$ and 0.3, for $W_{\GG}=500$ and $100\GEV$ \GG\ 
collisions. Again the 3 curves in each figure correspond to 7, 2 and 0.7
hadronic events on the average per collision. It is of interest that
the curves for $\vartheta_{min}=0.1$ and 0.3 are quite similar.
For 2 events (500\GEV) per collision the probability to get an
unbalanced $p_{\bot}\ge 5\GEV$ is about 15\%. This is comparable
with  the detector resolution.
  
While calculating $p_{\bot}$, we summed all energy depositions in
the detector, but ``interesting'' events usually have highly
energetic particles or jets. The probability
for the hadronic background adding energy to a jet is presented in
Fig.~\ref{fig20}.  We have selected 
a cell $\Delta\varphi\le 0.3$, $\Delta\eta\le0.3$, which corresponds
to a characteristic jet transverse size at $\theta = \pi/2$, and
calculated the probability of energy deposition in this region above
some energy $E$. The curves correspond to one hadronic event on the
average per bunch collision.  For other levels of background, the
probability should be multiplied by the average number of hadronic
events per collision.  
  
Note, that at the photon collider we are going to study events at
rest in the lab. system, and the jet size is just $\Delta \Omega$.
From the definition of the pseudorapidity follows 
$d\Omega = d\varphi d\eta \sin^2\vartheta$.  Therefore for obtaining the
probability of background the value given in Fig.~\ref{fig20} should
be divided by a factor of $\sin^2 \vartheta$.
  
A typical energy resolution for a 100\GEV\ jet is about 3\GEV. The
probability to have such an energy deposition at $\eta=0$ and 2
hadronic events per collision is 0.04\%.  For the 
$\H(115)\to\bbbar$ decay the optimum angular cut is $\cos \vartheta = 0.7$,
or pseudorapidity $\eta = 0.87 \approx 1$.
For such an angle the probability of 2\GEV\ energy deposition inside
a jet from the Higgs decay is 1.5\% and thus does not present a problem even
for a 10 times 
larger luminosity.

However, the probability depends very strongly on the angle. For
example, for $\eta = 2$ the probability of 2\GEV\ is
already 60\%.  So, at low angles the hadronic background can worsen the
resolution for low energy jets.

Of course, these estimates are very approximate and accurate simulation of 
certain processes is required.

\subsubsection{Large angle compton scattering and beamstrahlung}

X--ray radiation from beams can cause damage to 
multilayer dielectric mirrors. There are two main sources of such radiation~\cite{Tsit1}:

{\it Large angle Compton scattering}. The energies of these photons
are $\omega = 4\omega_0/\theta^2$ at $\theta \gg 1/\gamma$, where
$\omega_0$ is the energy of the laser photon ($\approx 1\eV$).  At a
distance $l$ the flux of photons 
$dn/ds \propto N/\gamma^2 l^2\theta^4$.  The main contribution comes from
Compton scattering on 
the low energy electrons.  The simulation for $2E_0 = 500\GEV$ gives
a power density $P \approx 10^{-7}\Watt /\CM^2$, $\omega \approx 40\KEV$ at
$\theta = 10\MRAD$ (the edge of the mirrors).

{\it Large angle beamstrahlung}. The simulation shows that X-ray
photons have a wide spectrum, $P \approx 10^{-6}\Watt /\CM^2$, 
$\omega\approx 1.5\KEV$ at $\theta = 10\MRAD$.
  
Note, that the X-ray power density on the mirrors is proportional to
$1/\theta^6$ and, if necessary, the minimum angle can be
increased, which is possible in the present scheme (Section~\ref{s5}) in which
the mirrors are placed outside the electron beams.

\subsection{The detector, experimentation issues} \label{s4.8}

The detector for experimentation at the Photon Collider could be
basically the same as for \EPEM\ collisions. Some differences are connected
only the optical system which should be placed inside the detector. 

Optimum focusing of the laser beam determines the divergence of the
laser beam at the conversion point (Section~\ref{s3.2}), it is
$\sigma_{x^{\prime}}= 0.0155$ and the angular radius
2.5$\sigma_{x^{\prime}}$ for the focusing mirror will be sufficient. As
we consider the optics situated outside the electron beams, the
required clear angle is $\pm 2\times 2.5 \times 0.0155 = \pm 78\MRAD$.

From the background consideration (previous section) follows that
the vertex detector with a length of about $\pm$ 15\CM\ length should have a
radius not smaller than 2\CM. This leaves the angular range $\pm130\MRAD$
inside the vertex for the laser beam, which is sufficient.

Beside the final focusing mirror the laser system has additional mirrors
inside the detector (Section~\ref{s5}), at angles of about $120$--$140\MRAD$.  
This does not have a major impact for the experiment as the mirrors are
situated close to the  calorimeter, their diameter is $15$--$20\CM$ and 
the thickness will be less than one radiation length.

\section{The Lasers and Optics}
\label{s5}
A key element of photon colliders is a powerful laser system which is
used for the $\EL \to\gamma$ conversion.  Lasers with the required flash
energies (several J) and pulse durations ($\approx$ 1\ps) already exist and
are used in several laboratories. The main problem is the high
repetition rate, about $10$--$15\kHz$  with 
the time structure of the electron bunches.  

The requirements of the laser system for the Photon Collider at TESLA
were discussed in Section~\ref{s3.2}. In summary, the required laser
wavelength is about 1\MKM, the flash energy 5\J, and the repetition
rate about 14\kHz. If two electron beams should be converted to
photons the average power of the laser system should be about 140\kW.
At TESLA the laser has to work only 0.5\% of the time since the
repetition rate is 5\Hz\ and duration of one train containing 3000
bunches is 1\,msec.  Thus the train structure of the LC is a very
serious complication.

In this section we will consider possible optical schemes and lasers
for the TESLA Photon Collider.

\subsection{The laser optics at the interaction region} \label{s5.1}

To overcome the ``repetition rate'' problem it is quite natural to
consider a laser system where one laser bunch is used for the 
$\EL\to \gamma$ conversion many times. Indeed, a 5\J\ laser flash contains about
$5\times 10^{19}$ laser photons and only $10^{10}$--$10^{11}$ photons are
knocked out per  collision with the electron bunch.  Below two ways
of multiple use of one laser pulse are considered for the Photon
Collider at TESLA: an optical storage ring and an external optical
cavity.

\subsubsection{The optical ``trap''}\label{s5.1.1}

\begin{figure}[p]
\centering
\hspace*{-2.1cm} \epsfig{file=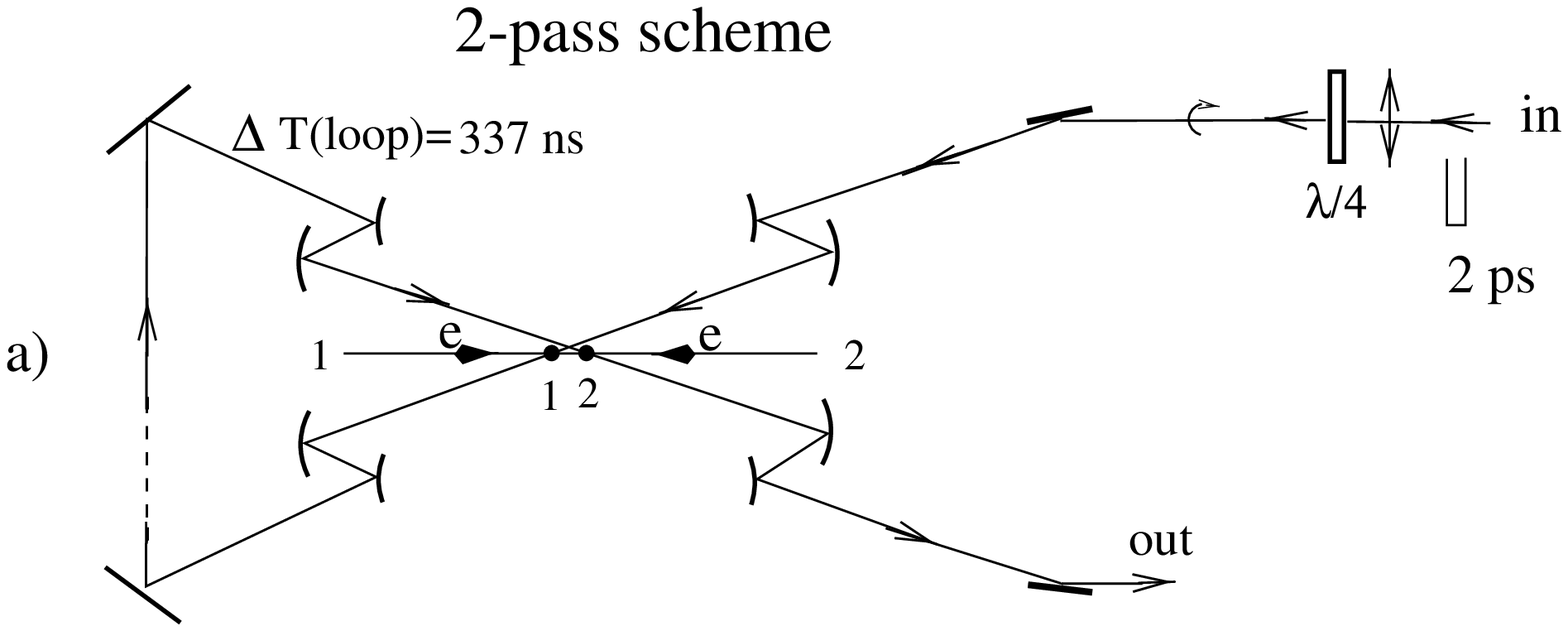,width=10.8cm,angle=0} 
\vspace*{0.3cm} 
\hspace*{-0.cm} \epsfig{file=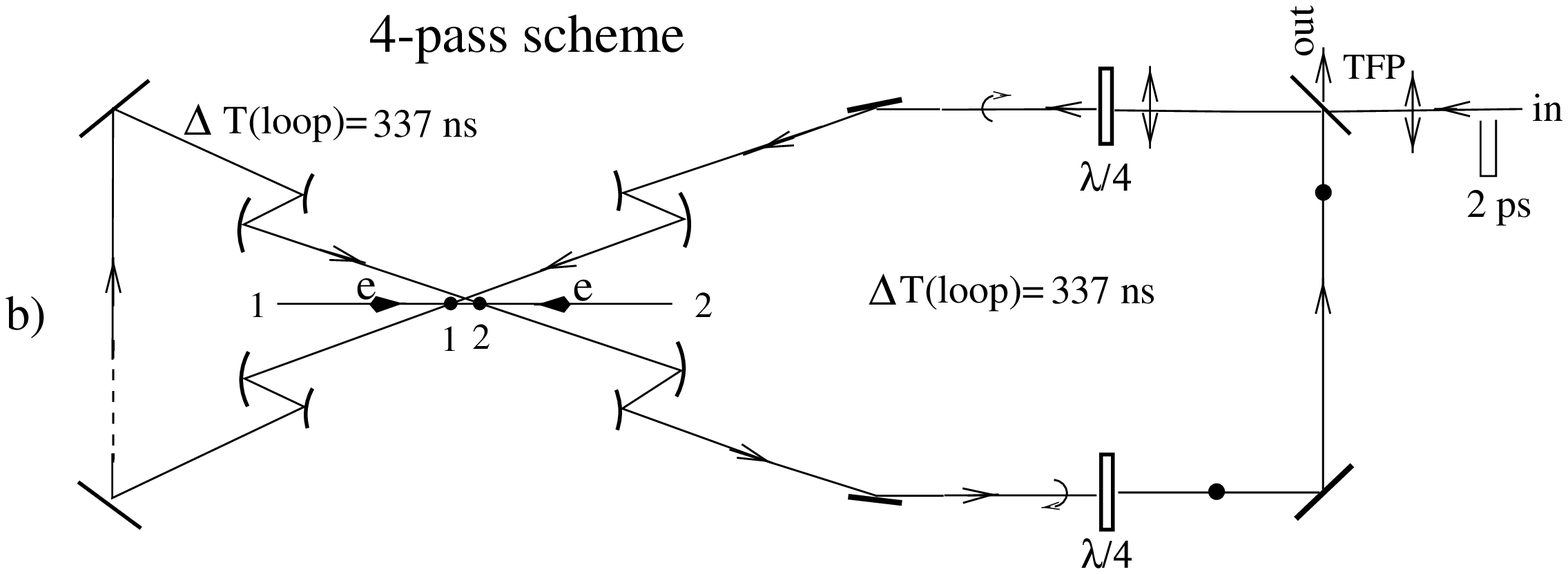,width=13.cm,angle=0} 
\vspace*{0.3cm}
\hspace*{-0.cm} \epsfig{file=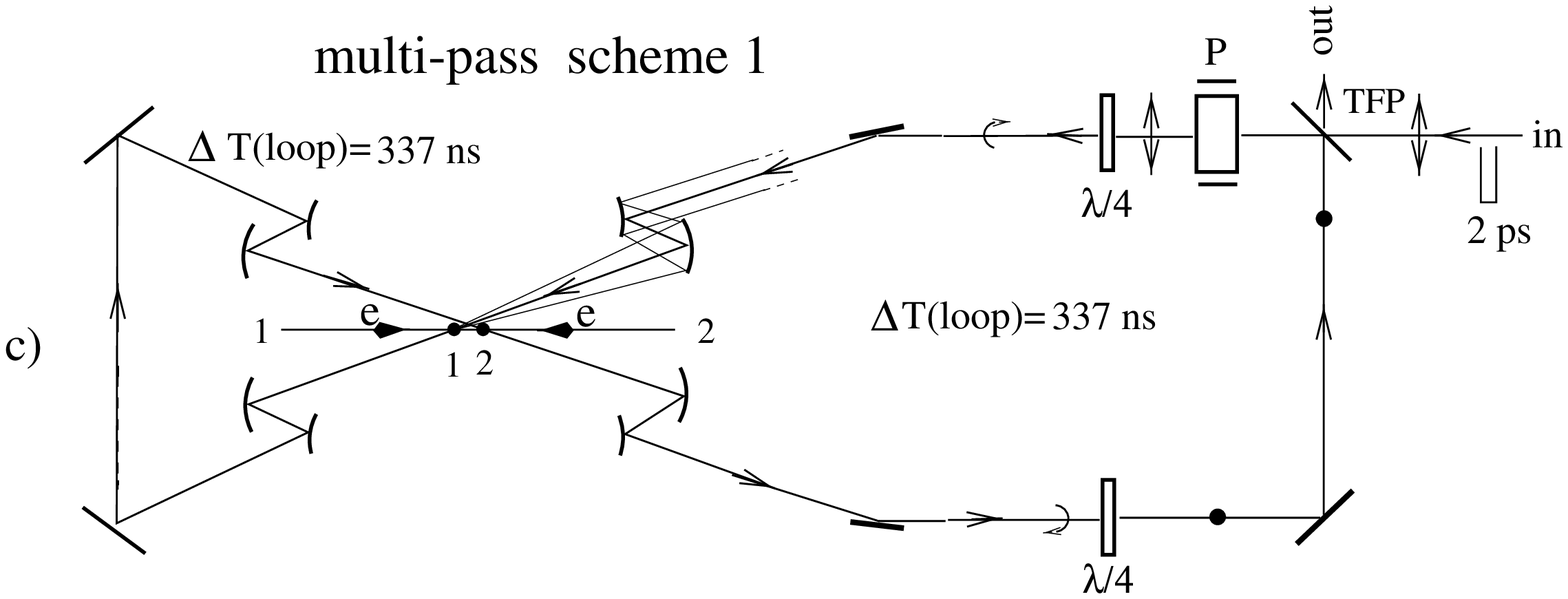,width=13.cm,angle=0} 
\vspace*{0.3cm}
\hspace*{-0.0cm} \epsfig{file=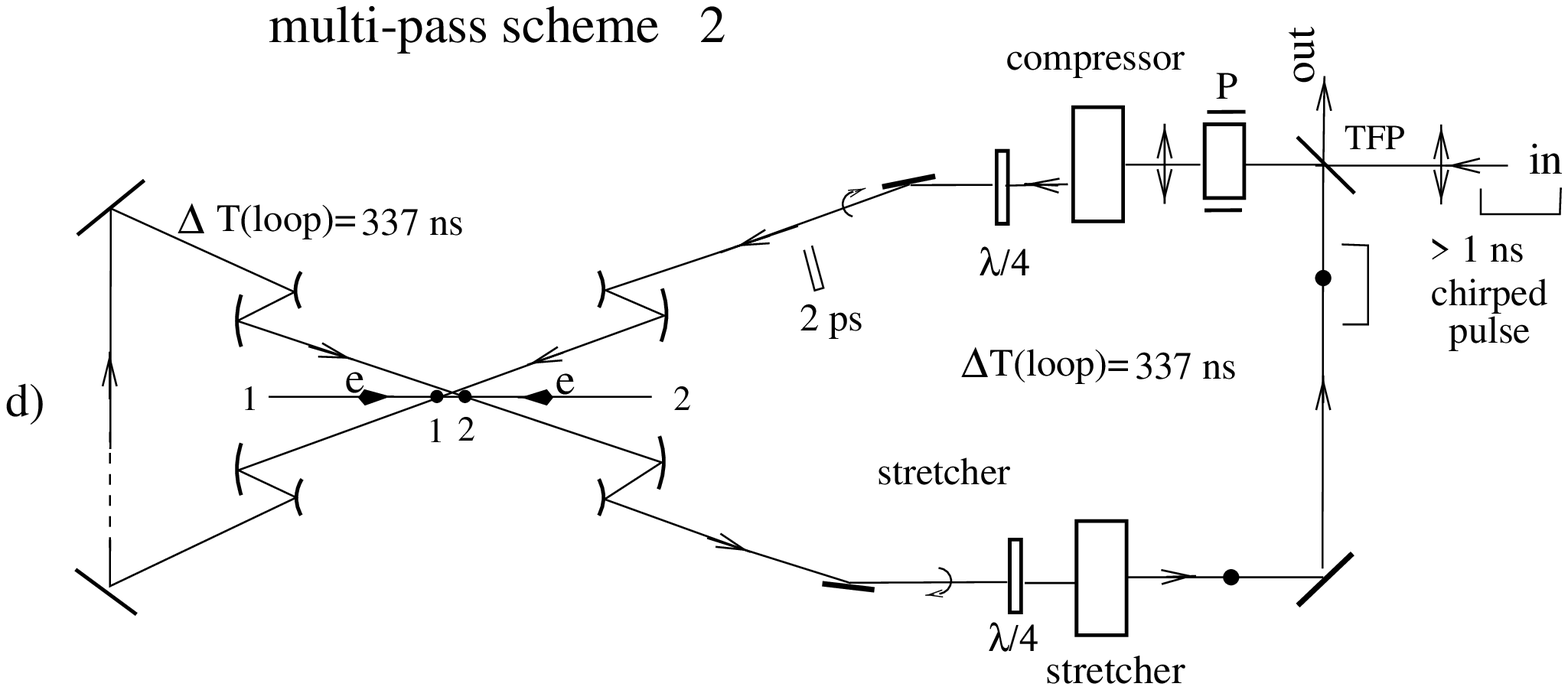,width=13.cm,angle=0} 
\caption{Optical trap: a) 2--pass optics  for $\EL \to \gamma$ conversions;  
b) 4--pass optics;
  c) optical storage ring without stretching--compression; d) optical
  storage ring with stretching--compression; P is a Pockels cell, TFP
  is a thin film polariser, thick dots and double arrows show the
  direction of polarisation.}  
\label{loop}
\end{figure}

The first approach is shown in Fig.~\ref{loop}~\cite{telnov}. 
In Fig.~\ref{loop}a the laser pulse is used twice for the $\EL \to \gamma$
conversion.  After the collision with the electron beam
(number 1) the laser beam exits from the detector and after a 337\ns\ 
loop (the interval between beam collisions at TESLA) returns back and
collides with the opposite electron beam (number 2). The second pass
does not need any special optical elements, only mirrors. This is a very
natural and simple solution. In this scheme  the laser system should generate
bunches with an interval of 337\ns.

In Fig.~\ref{loop}b the laser pulse is used for conversion four times.
In this scheme one additional optical element is used, a thin film
polariser (TFP), which is transparent for the light polarised in the
plane of the plane of the drawing and reflects light with the
orthogonal polarisation.  Directions of the polarisation during the
first cycle are shown in Fig.~\ref{loop}b. After the first cycle the
polarisation is perpendicular to the plane of the drawing and the
light is reflected from the TFP, while after the second cycle the
polarisation will be again in the plane of the drawing and the laser
pulse will escape the system via the TFP.  The laser bunches are
emitted by the laser at an average interval of 2$\times$337\ns\ but not
uniformly (337, 3$\times$337), (337, 3$\times$337), etc (see the next
paragraph). 

In Fig.~\ref{loop}c the laser pulse is sent to the interaction region
where it is trapped in an optical storage ring, which can be built
using Pockels cells (P), thin film polarisers (TFP) and 1/4--wavelength
plates ($\lambda/4$). Each bunch makes several ($n$) round trips (period
of the round trip is $2T_0$, where $T_0=337\ns$ is the interval between
bunch collisions) and then is removed from the ring.  All this can be
done by switching one Pockels cell which can change the direction of
linear polarisation by 90 degrees. The $\lambda/4$ plates are used for
obtaining the circular polarisation at the collision point.  For
obtaining linear polarisation at the IP these plates should be
replaced by 1/2 wavelength plates. A similar kind of optical trap was
considered as one of the options in the NLC Zero Design
Report~\cite{NLC}.  The number of cycles is determined by the
attenuation of the pulse and by nonlinear effects in the optical
elements.  The latter problem is very serious for Terawatt (TW) laser
pulses.  During one total loop each bunch is used for conversion twice
(see Fig.~\ref{loop}c). The laser bunch collides first with electron
beam 1 travelling to the right and after a time equal to the interval
between collisions (337\ns) it collides with beam 2 travelling to
the left.  For arbitrary number of the round trips, $n$, the laser
pulse sequence is a sum of two uniform trains with the interval
between neighbouring pulses in each train
\begin{equation}
\Delta T_t =2nT_0  
\label{DTt}
\end{equation}
and the trains are shifted by the time
\begin{equation}
\Delta T = k T_0, \;\;\;\; k=1,3, \ldots 2n-1.  
\label{Dt}
\end{equation}

In Fig.~\ref{loop}d the laser pulse is trapped in the same way as in
Fig.~\ref{loop}d, but to avoid the problems of nonlinear effects
(self--focusing) 
in the optical elements, the laser pulse is compressed using a grating
pair before collision with the electron bunch down to about $2$--$3\ps$
using grating pairs. It is then stretched again (decompressed) using
another grating pair up to the previous length of about 11\,ns just
before passing through the optical elements.

Which system is the better one, \ref{loop}b, \ref{loop}c or \ref{loop}d, is
not clear a priori.  The scheme (b) allows only 2 round trips, in
the scheme (c) the number of cycles is limited by nonlinear effects,
in the scheme (d) there is additional attenuation by the gratings used
for compression and stretching. Optical companies suggest gratings for
high powerful lasers with $R\approx 95\%$. One round trip requires four
gratings, or a 20\% loss/trip. So, the maximum number of trips for the
scheme (d) is only about two. This presents no advantage compared to
the scheme~\ref{loop}b which is much simpler and also allows two cycles,
though it is not excluded that gratings with higher reflectivity will
be available in future.

We next address the question how large the decrease of the laser energy per
round trip can be in
the scheme (c) without bunch compressor--stretchers. The minimum number
of mirrors in the scheme is about $15$--$20$. The reflectivity of multilayer
dielectric mirrors for large powers suggested by optical companies is
about 99.8\% (or better). The total loss/cycle is thus about
$3$--$4\%$. Let us add 1\% attenuation in the Pockels cell. Due to the
decrease of the laser flash energy the luminosity will vary from
collision to collision.  Calculations show that for attenuation factors of
 1.3, 1.4, 1.5 for the laser pulse , the \GG\ luminosity will only vary by 14,
 17, 21\%  (here we 
assumed that on average the thickness of the laser target is one
collision length).  For 5\% loss/turn and {\it 6 round trips} the
attenuation is 1.35, which is still acceptable.

Let us consider the problem of nonlinear effects for the scheme~\ref{loop}c.
The refractive index of the material depends on the beam intensity
\begin{equation}
 n = n_0 +n_2 I.
\end{equation}
This leads to two types of a self focusing of the laser
beam~\cite{koechner}.  The first type is a self--focusing of the beam
as a whole. The second one is self--focusing and amplification of
non--uniformities which leads to break up of the beam into a large number
of filaments with  intensities exceeding the damage level.  Both
these effects are characterised by the parameter
``B--integral''~\cite{koechner,NLC}
\begin{equation}
B= \frac{2\pi}{\lambda} \int  \Delta n dl = 
\frac{2\pi}{\lambda} n_2 I_{peak} \Delta l, 
\label{Bint}
\end{equation}
where $\Delta l$ is the thickness of the material.

If the beam has a uniform cross section then nonlinear effects do not
lead to a change of the beam profile, while for the Gaussian like
beam, $B \approx 1$ corresponds to the self--focusing angle approximately
equal to the diffraction divergence of the beam. This is not a problem
since such distortions can be easily corrected using adaptive optics
(deformable mirrors).

The second effect is more severe. Even for a uniform (in average)
distribution of the intensity over the aperture a small initial
perturbation $\delta{I_0}$ grows exponentially with a rate depending on
the spatial wave number.
The maximum rate is given in terms of the same
parameter $B$~\cite{koechner}
\begin{equation}
\delta{I} = \delta{I}_0 e^B.
\label{deltaI}
\end{equation}
This has been confirmed experimentally. To avoid amplification of
small--scale non--uniformities, the parameter $B$ should be smaller than
$3-4$~\cite{koechner,NLC}, in other words
\begin{equation}
I_{peak} < \frac{\lambda}{2n_2 \Delta l}.
\end{equation}

Now we can evaluate the relationship between the diameter and the maximum
thickness of the material.  For $A=5\J$, $\lambda = 1\MKM$,
$\sigma_{L,z}$ = 1.5\ps, a typical value of $n_2 \approx 3\times 10^{-16}\CM
^2/\Watt$ 
\footnote{it would be better to take $n_2$ for KD$^*$P  
used for Pockels cells, but we have not found it in the literature} 
and a uniform beam we get
\begin{equation}
\Delta l [\mathrm{cm}] < 0.1 S [\mathrm{cm}^2].
\end{equation}
For a beam diameter of 15\CM\ we obtain $l < 17\CM$.
For Gaussian beams the maximum thickness is about two times smaller.

Next we address the question what value to insert for 
$\Delta l$. In the scheme~\ref{loop}c the
dominant contribution to the total thickness is given by the Pockels
cell.  After the Pockels cell one can put a spatial filter (small hole
in a screen) and thus suppress the growth of spikes. $\Delta l$ in this
case is the thickness of the Pockels cell and does not depend on the
number of round trips. Moreover, the laser pulse is very short, has a
broad spectrum and the corresponding coherence length is small, about
$l_c \approx 4\pi\sigma_{L,z} \approx 0.5\CM$. The instabilities over a
uniform high intensity background develop due to the interference of
the fluctuation with the main power. However, this coherence is lost
after one coherence length. Thus, the B--integral does not characterise
the exponential growth of small scale non-uniformities once the
coherence length is much lower than $\Delta l$ (it will be suppressed
even for small values of $\Delta l$, if the material is distributed over
a long distance).  

It turns out that the problem of nonlinear effects in the scheme~\ref{loop}c is not dramatic.  The construction of a Pockels cells with an
aperture of about $10$--$15\CM$ and a switching time of 300\ns\ is not very
difficult.  Quarter-- and half--wave plates can be made thin or even
combined with mirrors (retarding mirror).

In conclusion, a very preliminary analysis shows that the optical scheme~\ref{loop}c with about 6 round trips (12 collisions with electron
beams) is a very attractive and realistic solution for the TESLA
photon collider.

\begin{figure}[!htb]
\centering
\hspace*{-0.4cm} \epsfig{file=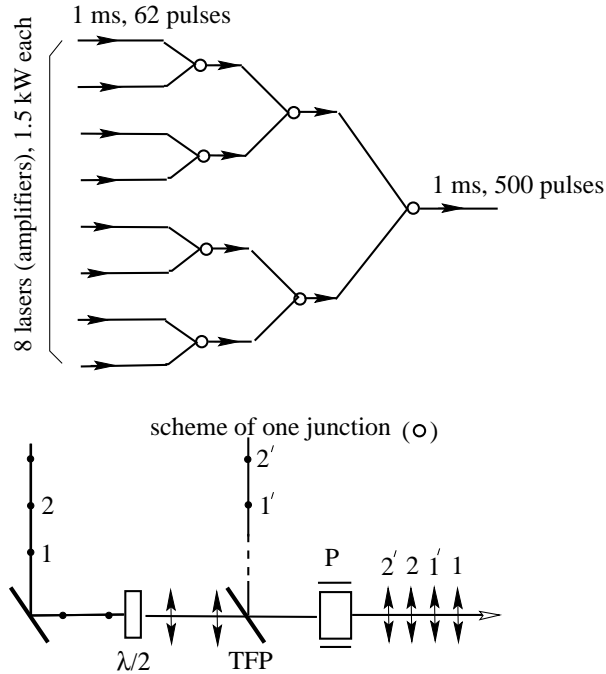,width=8cm,angle=0} 
\caption{Merging of pulses from several lasers (amplifiers)}
\label{junction}
\end{figure}

Now a few words on the laser system required for such an optical storage
ring with 6 round trips.  Schematically it is shown in
Fig.~\ref{junction}.  At the start (not shown) a low--power laser
produces a train of 1\msec\ duration consisting of 500 chirped pulses
with durations of several ns each.  Then these pulses are distributed
between 8 final amplifiers. Each of the 8 sub--trains has a duration of
1\,msec and consists of 62 pulses.  After amplification up to the
energy of 5\J\ in one pulse these sub--trains are recombined to
reproduce the initial time structure. The time spacing between bunches
in the resulting train may be equal in average (see~(\ref{Dt})) to the
6 intervals between beam collisions in TESLA in average (see~\ref{DTt}).

Due to the high average power the lasers should be based on diode
pumping. Diodes have a much higher efficiency than flash lamps. It is about
$\epsilon \approx 25\%$ for single pulses.  For pulse trains, as in our
case, the efficiency should be at least by a factor of two higher.
Moreover, diodes are much more reliable. This technology has been developed
very actively for other applications, such as  inertial fusion.

The main problem with diodes are their cost. The present cost of diode lasers
is about 5 \Euro\ per Watt~\cite{Gronberg}.  Let us estimate the required
laser power. In the case of TESLA, the duration of the pulse train
T$_0$ = 1\msec\ is approximately equal to the storage time
($\tau\approx 1$\,msec) of the most promising powerful laser crystals, such as
Yb:S-FAP. Therefore,  the storage time does not help at
TESLA. The required power of the diode pumping is
\begin{equation}
P_{diode} = \frac{A(\mbox{flash})N(\mbox{bunches})}{\epsilon T_0} = 
\frac{5\mbox{J} \times 500}{0.5 \times 10^{-3}\;} = 5\mbox{MW}. 
\label{huge}
\end{equation}
Correspondingly, the cost of such diode system will be 25\,M\Euro.  Here
we assumed a 6--fold use of one laser bunch as described above.

Moreover, the Livermore laboratory is now working on a project of inertial
confinement fusion with a high repetition rate and efficiency with the
goal of building a power plant based on fusion. This project is based
on diode pumped lasers.  According to~\cite{Payne} they are currently
working on the ``integrated research experiment'' for which ``the cost
of diodes should be reduced  to 0.5 \Euro /Watt and the cost of diodes
for fusion should be 0.07 \Euro /Watt or less.'' Thus, the perspectives of diode
pumped lasers for photon colliders are very promising. With 1 \Euro /Watt
the cost of diodes is 5\,M\Euro\ for the scheme with 6 round trips (with
Pockels cell) and 15\,M\Euro\ for 2 round trips without Pockels cell.

The average output power of all lasers in the scheme~\ref{loop}c is
about 12\kW, or 1.5\kW\ for each laser.

\subsubsection{The optical cavity} \label{s5.1.2}

One  problem with the optical storage ring at photon colliders is the
self--focusing in optical elements due to the very high laser pulse power. 
There is another  way to ``create'' a powerful laser pulse in
the optical ``trap'' without any material inside: laser pulse 
stacking in an ``external'' optical cavity~\cite{Tfrei}.

In short, the method is the following. Using a train of low energy
laser pulses one can create in the external passive cavity (with one
mirror having some small transmission) an optical pulse of the same
duration but with an energy higher by a factor of $Q$ (cavity quality factor).
This pulse circulates many times in the cavity each time
colliding with electron bunches passing the centre of the cavity. For
more details see~\cite{Tfrei}. 
   
 Such kind of cavity  would allow to drastically reduce the overall costs
of the laser system. Instead of several parallel working lasers it could be
one table--top laser feeding the external optical cavity.

\begin{figure}[!htb]
\centering
\hspace*{-0.4cm} \epsfig{file=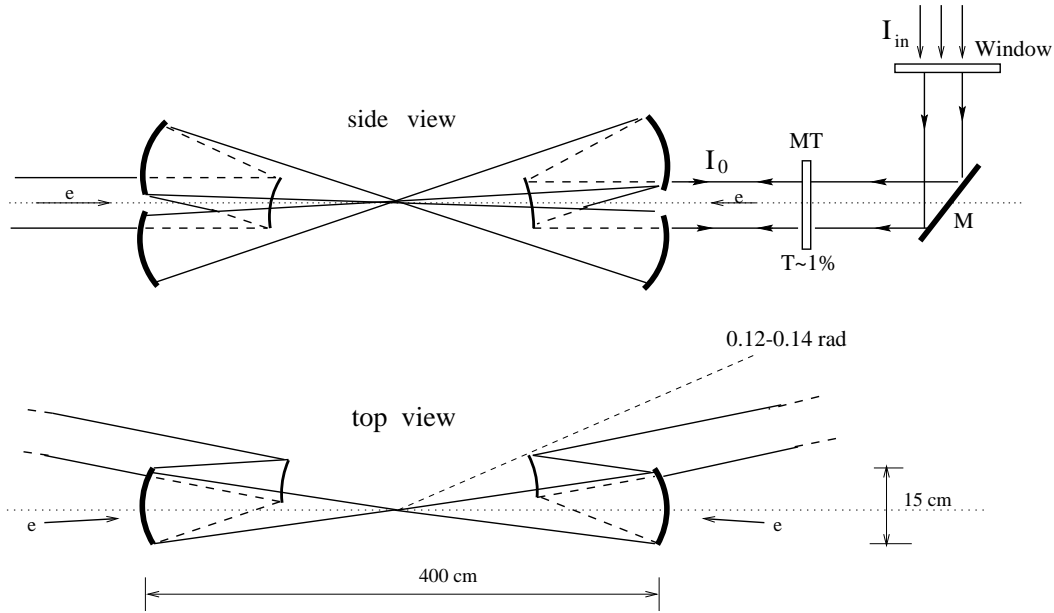,width=14cm,angle=0} 
\caption{Principle scheme of ``external'' cavity  for $\EL \to \gamma$ 
  conversion. Laser beam coming periodically from the right
  semi--transperant mirror MT excites one cavity (includes left--down
  focusing mirror, right--up focusing mirror and the MT mirror. The
  second cavity (for conversion of the opposite electron beam) is
  pumped by laser light coming from the left (not shown) and includes
  the focusing mirrors left--up and right--down. } 
\label{optics}
\end{figure}

\begin{figure}[!htb]
\centering
\hspace*{-0.4cm} \epsfig{file=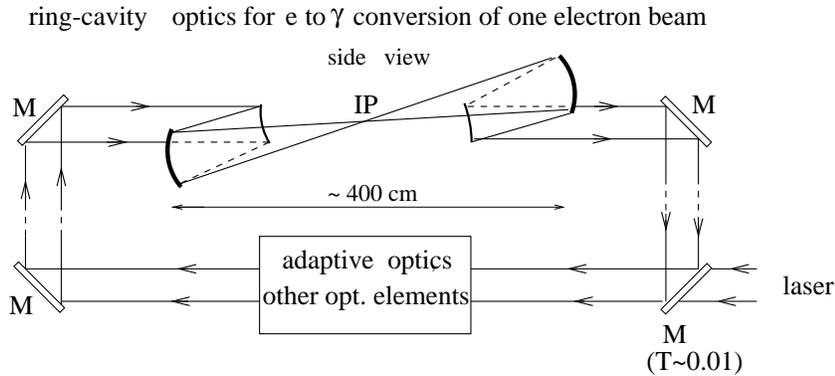,width=11cm,angle=0} 
\caption{Ring type cavity. Only the cavity for one electron beam is shown.
The top view is quite similar to that in Fig.~\ref{optics}}.
\label{ring}
\end{figure}

A possible layout of the optics scheme at the interaction region  is
shown in Fig.~\ref{optics} \cite{Tfrei,telnov}. In this variant, there
are two optical cavities (one for each colliding electron beam) placed
outside the electron beams.
Such a system has the minimum number of mirrors inside the detector.  One
of several possible problems in such a linear cavity is the back--reflection.
In a ring type cavity this problem would be much easier to solve~\cite{Will}.
A possible scheme of such a ring cavity for photon colliders is shown
in Fig.~\ref{ring}~\cite{telnov} (only some elements are shown).
 
Some technical aspects of the external cavity approach are
discussed in~\cite{Will}.  Such a cavity is operated already in
MBI(Berlin) and  $Q \approx 100$ has been demonstrated. A first
view on technical problems of the optical cavities are given below. 

The external resonant cavities have been used for comparable purposes
for many years.  A common application of those cavities is frequency
conversion of the fundamental laser wavelength into its harmonics.
Several optical laboratories have broad experience in application and
design of those optical resonant enhancement cavities.

In order to provide an effective storage of the laser radiation, the
length of the cavity has to be adjusted to an integer multiple of the
laser wavelength with sub--micrometer accuracy. This ensures that the
recirculating wave constructively interferes with the wave which is
constantly fed into the cavity. An electronic feedback system is
required for this task. Many different ways for obtaining the error
signal are described in the scientific literature. The actual control
of the resonator length is performed by means of piezoceramics which
directly drive one of the resonator mirrors.

The quality factor $Q$ of the cavity is typically limited by reflection
losses at the optical elements.  A cavity which has been operated at
the Max--Born--Institute for several years for frequency doubling
reaches a quality factor of 40 without difficulties, being determined
by a nonlinear crystal. After removing the nonlinear crystal, an
increase of the $Q$--factor to about 100 was observed.

The majority of the cavities are used with uninterrupted cw laser
radiation. Several laboratories have introduced appropriate extensions
in order to use the cavities with pulses from mode locked lasers~\cite{PTF90}. There are three major additional requirements to be
fulfilled if the cavity has to store intensive laser pulses instead of
cw radiation~\cite{TSL93}.

 One of the problems in the optical cavity is temporal broadening of the pulse
travelling in the cavity. This unfavourable effect may be caused by the
wavelength dependency of the refractive index (i.e. dispersion) which
is experienced by the pulse passing through the optical elements.
Appropriate compensation can be done using specially designed
multilayer coatings (so called ''chirped mirrors'')~\cite{MGF84},
which are now commonly used in femtosecond laser oscillators. The
chirped mirrors introduce particularly small optical losses and are
therefore preferable for high--$Q$ cavities. The maximum total thickness
of the optical elements, whose dispersion can be compensated in one
single reflection at a chirped mirror is limited to a few millimetres.

The design criteria for the resonant enhancement cavity  follows:

\begin{itemize}
\item  
  The cavity should have a ring--like geometry.
\item  
  The length of the cavity should be adjusted to the repetition rate
  of the electron bunches.
\item  
  The cavity length has to be stabilised to a very small fraction of
  the wave\-length.
\item  
  Chirped mirrors can be used to compensate for dispersion in optical
  transmissive elements of up to several millimetres thickness. However,
  nonlinear perturbation of the wavefront by self--focusing limits this
  thickness to the millimetre or sub--millimetre range.
\item 
  Deformable mirrors should be used for maintaining the phase of
  the circulating light.
\item  
  Thin glass plates should be used for protection of individual mirrors
  from electrons and gamma radiation.
\item  
  The cavity cannot contain thick vacuum windows, i.e. the whole cavity
  has to be placed in a vacuum system.
\end{itemize}

\begin{figure}[tb]
\begin{center}
\includegraphics[width=5.2243in]{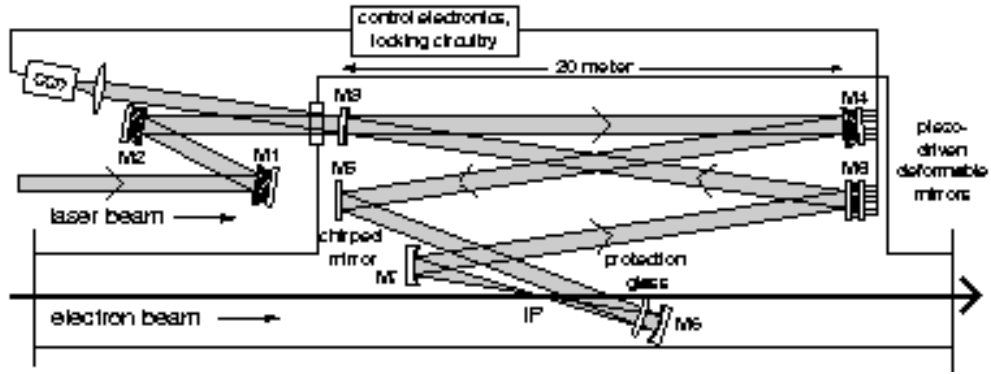}%
\caption{External ring cavity for a TESLA photon
collider. See comments in the text.}%
\label{cavity2}%
\end{center}
\end{figure}

Fig.~\ref{cavity2} shows the basic elements of a possible resonant optical
cavity for the TESLA Photon Collider (here two mirrors are missing which
would allow to remove the laser beam from the IP region without passing
the detector, as shown in Figs.~\ref{optics}, \ref{ring}).  The
laser radiation is transferred to the cavity by means of two
deformable mirrors M1 and M2. Those mirrors consist of a coated
elastic glass plate which is bent by a number of piezo actuators. The
purpose of the mirrors M1 and M2 is to adapt the incoming wavefront
to the eigenmode to be excited in the cavity within a small fraction
of the wavelength. This is essential in order to achieve constructive
interference between the pulses from the laser and the pulses
travelling inside the cavity. The actual coupling of the laser
radiation into the cavity is performed by mirror M3 which should have
a transmission of 1\% (i.e.~99\% reflectivity). All other mirrors M4
to M8 of the cavity are optimised for maximum reflectivity.

In order to maintain the phase of the circulating light wave across the
complete beam profile, the optical path length should be adjusted locally at
different positions in the beam. The required accuracy is  the order of
0.1\% of the wavelength. We propose to use the deformable mirrors M4 and M8
for this aim. The error signal for driving the individual piezo actuators of
these mirrors may be obtained by processing the image from a CCD camera
located behind the resonator mirror M3. A feedback procedure optimises the
coupling of the laser radiation into the cavity and minimises the losses of
the stored laser field by adjusting the actuators of M4 and M8 for minimal
leakage through M3. In addition, it allows for compensation of wavefront
distortions by the optical elements of the cavity and ensures that
the travelling optical wave can be focused in an optimum way

The $Q$ factor of the cavity strongly depends on the reflectivity of the
mirrors. Mirrors with multilayer coatings of reflectivity
greater than 99.9\% are already commercially available. The remaining loss in
reflection of 
high--power mirrors is mainly caused by scattering at small impurities in the
coatings. Therefore increasing the reflectivity requires to reduce the number
of scattering impurities which can only be achieved by very special and
expensive coating techniques.

A problem in the realisation of the cavity may be connected with a
gradual damage of the coatings by synchrotron radiation and scattered
electrons. This damage will lead to a slow reduction of the overall
reflectivity of the mirrors thereby reducing the overall $Q$--factor of
the cavity. The effect will be particularly important for the mirror
located downstream the electron beam (M6 in Fig.~\ref{cavity2}). In
order to avoid the damage we propose to protect this mirror with a
thin glass plate. This plate should have antireflection coatings and
easily exchangeable without misalignment of the cavity.

Taking into account these limitations we have estimated that a quality
factor of $Q=100$ should be within reach. This also complies with the
value obtained in already operating external cavities for cw lasers.
A $Q \approx 50$ would be sufficient for the photon collider
at TESLA.

Because of the high average power and the high stability, the laser has to be
laid out in MOPA (Master Oscillator -- Power Amplifier) geometry. Probably only
diode--pumped solid--state laser systems can reach the required reproducibility
of the laser parameters.
 The most promising candidate
for a laser suitable for the TESLA Photon Collider seems to be Ytterbium--doped
YAG (Yb:YAG) which has already been used to generate pulses of 0.7\ps\ duration~\cite{ASS00}. It has also been demonstrated that this material can deliver a
very high average laser power of up to 1\kW~\cite{HBM00}.

\subsubsection{Laser damage of optics}
\vspace{0.3cm}

The peak and average power in the laser system at the Photon Collider is
very large. The damage threshold for multilayer dielectric mirrors
depends on the pulse duration. The empirical scaling law is~\cite{koechner}
\begin{equation}
E_{th}[\mbox{J/cm}^2] \approx 10 \sqrt{t[\mbox{ns}]}
\end{equation} 
for pulse durations ranging from picoseconds to milliseconds.  At the
LLNL the damage threshold for 1.8\,ps single pulses of 0.7 to 2\J/cm$^2$ have
been observed on commercial multilayer 
surfaces~\cite{NLC} with an average flux on the level of $3$--$5\kW$/cm$^2$.

Comparing these numbers with the conditions at the TESLA Photon
Collider (5\J\ for 1.5\ps, 6000$\times$5\J\ for 1\msec\ and 140\kW\ average power)
one finds that the average power requirements are most demanding. With
a uniform distribution, the surface of the mirrors should be larger than
140/5 = 28$\CM^2$ and a factor of $2$--$3$ larger for Gaussian laser beams
with cut tails. So, the diameter of the laser beam on mirrors and
other surfaces should be larger than 10\CM.

\vspace{0.3cm}
{\bf Short summary on the optical schemes}
\vspace{0.3cm}

 We have considered two possible options of  laser optics for the
TESLA photon collider:

\begin{enumerate} 
  
\item 
  \underline{Optical trap (storage ring)} with about 8 diode
  pumped driving lasers (final amplifiers) with a total average power
  of about 12\kW.  Beams are merged to one train using Pockels cells and
  thin--film polarisers. Each laser pulse makes 6 round trips in the
  optical trap  colliding 12 times with the electron beams. This
  can be done now: all technologies exist.
  
\item 
  \underline{External optical cavity} is a very attractive
  approach which can additionally reduce the cost and complexity of the
  laser system.  This scheme requires very small tolerances (of the order of
  $\lambda/(2\pi Q)$, where $Q \approx 50$) and very high mirror quality.
  R\&D is required.

\end{enumerate}

\subsection{The lasers} \label{s5.2}

In this proposal we do not present a detailed scheme of a laser for the
TESLA PhotonCollider. It should be an additional R\&D.  However, we
would like to consider briefly  existing laser technologies which
allow, in principle, the laser system required for the Photon
Colliders to be built. 

Development of laser technologies is being driven by several large
programs, such as inertial fusion.  This is a fortunate situation for
photon colliders as we may benefit from the laser technology
developments of the last $10$--$15$ years which cost hundreds M\$ per 
year.  Now practically all components exist and we can just design and
build the required system. Fortunately this possibility has
appeared almost exactly in the time when the physics community is
ready for construction of the TESLA Linear Collider.  Of course,
construction of the laser system for the Photon Collider is not a
simple task and needs many efforts.  

Two kind of lasers for photon colliders are feasible now: a solid state
laser and a free electron laser (FEL). 

The technology for production of picosecond pulses with terawatt power
has been developed for solid state lasers. The wave length of
the most powerful lasers  about 1\MKM\ which is just optimum for the
TESLA Photon Collider.

A free electron laser (FEL) is also attractive because it has a
variable wave length and is based fully on the accelerator technology.
The X-ray FEL with a wave length down to 1nm is a part of the TESLA
project. The same technology can be used for the construction of an FEL with
1\MKM\ wave length for the Photon Collider. This task is much easier than
the X--ray laser.

\subsubsection{Solid state lasers}
 
In the last decade the technique of short powerful lasers
made an impressive step and has reached petawatt ($10^{15}$)
power levels and few femtosecond durations~\cite{PERRY}.
Obtaining  few joule pulses of picosecond duration is not
a problem using modern laser techniques. For photon collider
applications the main problem is the high repetition rate. 

  The success in obtaining picosecond pulses is connected with a chirped 
pulse amplification (CPA) technique~\cite{STRIC}. ``Chirped'' means
that the pulse has a time--frequency correlation. The main problem in 
obtaining short pulses is the limitation on peak power imposed
by the nonlinear refractive index. This limit on intensity is about
1\GW/$\CM^2$. The CPA technique successfully overcomes this limit.

\begin{figure}[!hbt]
\centering
\includegraphics[width=4.3in,angle=0,trim=0 0 0 0]{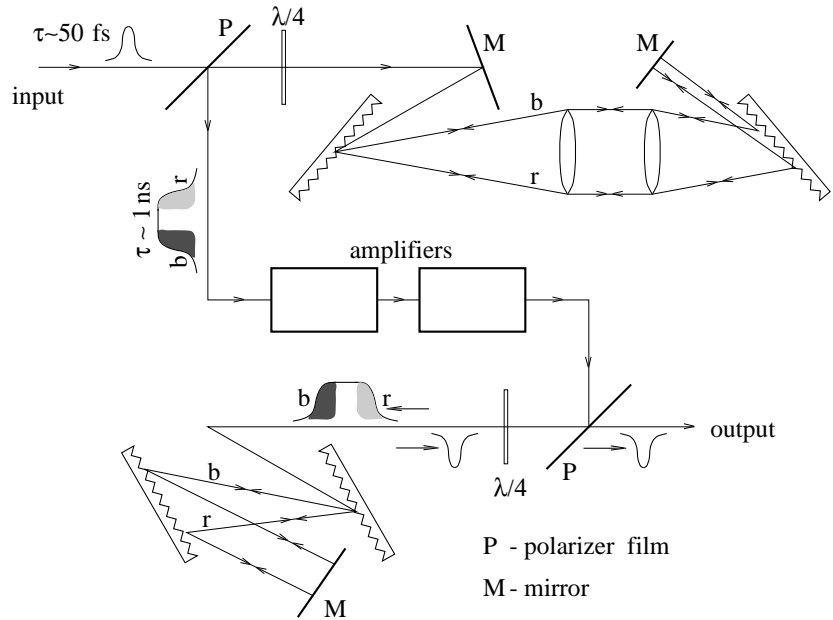}
\caption{Chirped pulse amplification. }
\label{fig23}
\end{figure}

 The principle of CPA is demonstrated in Fig.~\ref{fig23}.
A short, low energy pulse is generated in an oscillator.
Then  this pulse is stretched by a factor about $10^4$ in the grating
pair which introduces a delay proportional to the frequency. This long
 nanosecond pulse is amplified and then compressed by another grating pair 
to a pulse with the  initial or somewhat longer duration. As nonlinear
effects are practically absent, the obtained pulses have a very good  
quality close to the  diffraction limit.
  
One such laser worked since 1994 in the E--144 experiment at SLAC
which studied nonlinear QED effects in the collision of laser
photons with 50\GEV\ electrons~\cite{BAMBER}. It has a repetition rate
of 0.5\Hz, $\lambda=1.06\MKM$ (Nd:Glass), 2\J\ flash energy, 2\TW\ 
power and 1\ps\ duration. This is a table--top laser. Its parameters are
very close to our needs, only the repetition rate was low due to
overheating.  In this laser a flashlamp pumping was used.

The latter problem can be solved using another very nice technique:
diode pumping (the diode is a semiconductor laser with high
efficiency).  Since the frequency of photons from diode lasers
coincides almost with the pump frequency of the 1\MKM\ lasers they
are very efficient in converting wall plug power to laser light:
efficiencies of 10\% have been achieved.  But even more important the
heating of the laser medium with diode pumping is much lower than with
flashlamps. This gives one to two orders increase in repetition rate.
Moreover, the flashlamps have a limited lifetime of $< 10^6$ shots,
while the lifetime of diodes is many orders of magnitude higher.
 
The main problem of diodes is their cost. But it decreases very fast.
As it was mentioned, their cost is 5 \Euro /Watt, the next step in the inertial
fusion program assumes the reduction of the cost down to 0.5 \Euro /Watt
and the final goal is 0.07 \Euro /Watt. The cost of diodes for
TESLA photon colliders would be about 25 M\Euro\ already with the present 
cost and a further significant decrease is very likely.

Below is a list of laser technologies important for photon colliders:
\begin{itemize}
\item chirped--pulse technique;
\item diode pumping;
\item laser materials with high thermo--conductivity;
\item adaptive optics (deformable mirrors);
\item disk amplifiers with gas (helium) cooling;
\item large Pockels cells, polarisers;
\item high power and high reflectivity multilayer dielectric mirrors;
\item anti--reflection coatings.
\end{itemize}

Non--uniform, train structure of electron bunches at TESLA makes the
task somewhat more difficult than it would be for a uniform pulse
structure.  This leads to rather high power of pumping diodes (high
power inside one train), but as we mentioned this is not a serious problem.

However, generating a 1ms long train with 3000/6 = 500 pulses, 5\J\ 
energy each, is not the same as generation of one 2.5\kJ\ pulse (4\kJ\ 
diode pumped units are developed for laser fusion) for the same time,
because the volume of the laser crystal in the first case may be 500
times smaller. Beside, we consider 8 lasers working in parallel.

It is very convenient that the distance between electron bunches at
TESLA is large, 337\ns\ (1.4\ns\ at NLC and JLC). This time allows to
use large Pockels cells for manipulations of high power laser
pulses.

At TESLA the train is very long and storage time of laser materials
can not be used for pumping the laser medium in advance, but
on the other hand, in this case, one can use a large variety of laser
materials optimising other parameters (thermal conductivity etc.).

The development of the optimum design of the laser system for the
Photon Collider requires special R\&D. Solutions should be different for
TESLA and NLC/JLC colliders.

\subsubsection{Free electron lasers} 

   Potential features of a free electron laser (FEL) allow one to 
consider it as an ideal source of primary photons for a \GG\ 
collider. Indeed, FEL radiation is tunable and has always minimal 
(i.e.  diffraction) dispersion.  The FEL radiation is completely 
polarised either  circularly or linearly for the case of the helical or planar 
undulator, respectively.  A driving accelerator for the FEL may be a 
modification of the main linear accelerator, thus providing the 
required time structure of laser pulses.  The problem of 
synchronisation of the laser and electron bunches at the conversion 
region is solved by means of traditional methods used in accelerator 
techniques. A FEL amplifier has the potential to provide a high conversion 
efficiency of the kinetic energy of the electron beam into coherent 
radiation. At sufficient peak power of the driving electron beam the 
peak power of the FEL radiation could reach the required TW level.

\begin{figure}[!htb]
\centering
\includegraphics[width=4in,angle=0]{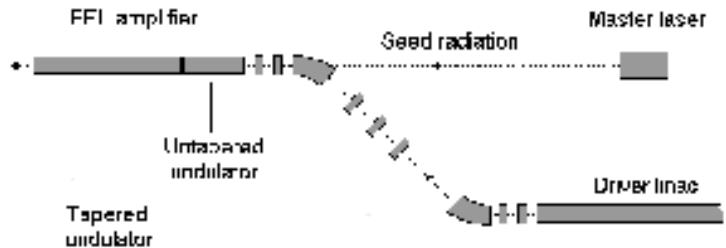}
\caption{Basic scheme of the MOPA laser system for a photon collider.} 
\label{yurkov1}
\end{figure}

The idea to use a FEL as a laser for the \GG\ collider has been
proposed in~\cite{KONDR82}. The present view on FEL systems for the
photon collisions at TESLA is discussed in~\cite{Yurkov}.
 The  FEL system is built as a master oscillator--power amplifier (MOPA) 
scheme where the low--power radiation from a Nd glass laser ($\lambda = 1 
\MKM$) is amplified in a long tapered undulator by an electron beam 
(see Fig.~\ref{yurkov1}).  The driving accelerator has the same pulse
structure as the main TESLA linac. 

The driving electron beam for the FEL is produced by the accelerator based
on TESLA technology and similar to the TTF (TESLA Test Facility)
accelerator~\cite{ttf-cdr}. Parameters of the accelerator are
presented in Table~1. The beam with a charge of 12\nC\ and normalised
emittance of 30$\pi\, \MM\, \MRAD$ is generated in the photoinjector,
accelerated in superconducting modules with the gradient $20-25\mbox{MV}/\M$
and compressed down to a 2\ps\ duration in the bunch compressors. Note
that the emittance is not a critical parameter for the considered FEL.
The number of bunches per macropulse is about 3 times lower
than that in the TESLA train, but as discussed in the previous section one
laser bunch can be used several times for $\EL\to\G$ conversion.

\begin{table}[htb]
\center{
\begin{tabular} { l l }
\hline 
Energy                           & 1.5\GEV \\  
Charge per bunch                 & 12\nC \\  
Peak current                     & 2.4\kA \\  
Bunch length (RMS)               & 0.6\MM \\  
Normalised emittance             & 30$\pi\MM\MRAD$ \\  
Energy spread (RMS)              & 1\MEV \\  
Repetition rate                  & 5\HZ \\  
Macropulse duration              & 800\MKS \\ 
\# of bunches per macropulse     & 1130 \\  
Bunch spacing                    & 708\ns \\  
Average beam power               & 102\kW \\  
\hline 
\end{tabular}
}
\caption{Parameters of the driving accelerator
\label{table1}}
\end{table}

\begin{table}[htbp]
\center{
\begin{tabular} { l l }
\hline 
\underline{Undulator} \\   
\hspace{5pt} Type                             & Helical \\
\hspace{5pt} Period                           & 10\CM \\  
\hspace{5pt} Magnetic field (entr./exit)      & 1.4\T / 1.08\T \\ 
\hspace{5pt} Total length                     & 60\M \\
\hspace{5pt} Length of untapered section      & 10.7\M \\
\hspace{5pt} Beam size in the und. (RMS)      & 230\MKM \\
\underline{Radiation} \\   
\hspace{5pt} Wavelength                       & 1\MKM \\
\hspace{5pt} Dispersion                       & Dif. limit \\
\hspace{5pt} Pulse energy                     & 2.2\J \\
\hspace{5pt} Pulse duration (HWHM)            & 1.6\ps \\
\hspace{5pt} Repetition rate                  & 5\Hz \\  
\hspace{5pt} Macropulse duration              & 800\MKS \\ 
\hspace{5pt} \# of pulses per macropulse      & 1130 \\  
\hspace{5pt} Peak output power                & 0.7\TW \\
\hspace{5pt} Average power                    & 12.5\kW \\
\hspace{5pt} Efficiency                       & 12.2\% \\ 
\hline 
\end{tabular}}
\caption{Parameters of the FEL amplifier
\label{table2}}
\end{table}

  The peak power of the master laser with the wavelength of 1\MKM\ is 
assumed to be $1\MW$ with a pulse duration of several picoseconds, so that 
the average power will be below 0.1\Watt. This means that only a small 
fraction of the power can be taken from the 2\Watt\ of infrared radiation 
generated in the laser system of the photoinjector. Then this radiation can 
be transported to the undulator entrance.  The problem of 
synchronisation of electron and optical bunches is therefore solved 
naturally.
\begin{figure}[htb]
\centering
\epsfig{file=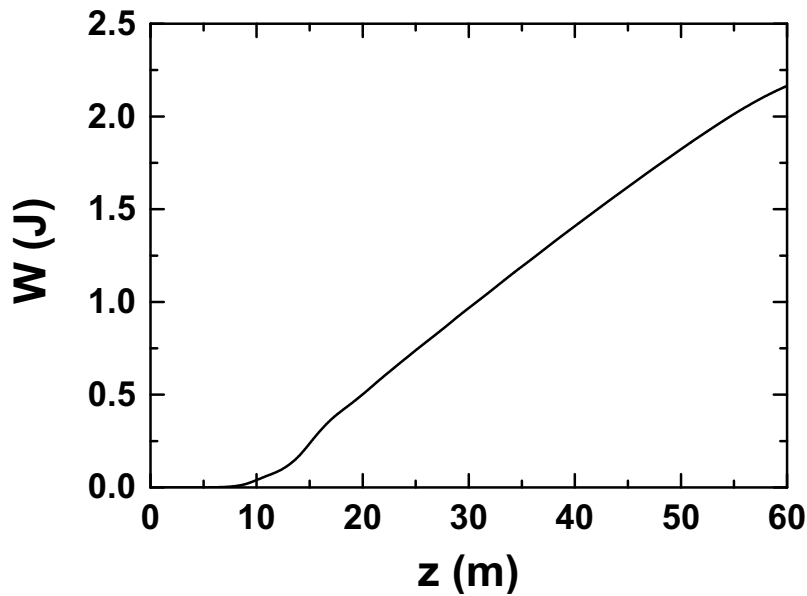,width=0.7\textwidth}
\caption{Energy in the radiation pulse versus the undulator length.}
\label{fig:pz-1}
\end{figure}

To obtain reasonable luminosity of the \GG\ collider at TESLA,
the energy in the radiation pulse at the FEL amplifier exit should be
above 2\J\ and the peak power should reach sub--terawatt level. For the
chosen parameters of the electron beam this means that the FEL efficiency
must exceed 10\%. In an FEL amplifier with a uniform undulator the
efficiency is limited by saturation effects and is below 1\% in the
considered case.  Saturation of the radiation power in the FEL
amplifier occurs due to the energy loss by the particles which fall
out of the resonance with the electromagnetic wave. Nevertheless,
effective amplification of the radiation is possible in the nonlinear
regime by means of using a tapered undulator. In this case a large
fraction of particles is trapped in the effective potential of the
interaction with the electromagnetic wave and is decelerated.

Parameters of the FEL amplifier with the tapered undulator are
presented in Table~\ref{table2}. The tapering can be done by decreasing  the
magnetic field at fixed undulator period. The undulator is helical to
provide polarised radiation and is superconducting. The resonance is maintained
by decreasing  the magnetic field at fixed period of the undulator.

The dependence of the radiated energy versus the undulator length is
shown in Fig.~\ref{fig:pz-1}. The efficiency 12.2\%, reached in the
end of the undulator,  corresponds to 2.2\J\ in the optical pulse.

Use of a free electron laser as a source of primary
photons for the \GG\ collider at TESLA seems to be natural
solution.  TESLA already includes an integrated X--ray FEL facility.
Powerful VUV radiation has been produced at DESY in a SASE FEL with
15\M long undulator~\cite{prl-sase}. The FEL for the photon colliders 
is simpler than the X--ray FEL.

Scale and cost of the FEL facility for the Photon Collider can be
estimated in a simple way. It requires a 1.5\GEV\ linear accelerator
similar to the main TESLA accelerator and a  60\M\ long undulator.

\vspace{3mm}
{\bf Summary on lasers}
\vspace{3mm}

We have considered briefly two kinds of lasers for the photon
collider at TESLA: a solid state laser and a FEL.  Both approaches are
technically feasible.  However, the first one looks somewhat more
attractive, because it might be a large room--size system, while a FEL
includes a 160\M\ long accelerator (with wiggler) which would be a large
facility.  For energies $2E_0 \geq 800\GEV$ where longer laser wave
length will be required, a FEL may be the best choice.

\section{Summary}\label{s6}

The Photon Collider presents a unique opportunity to study \GG\ and
\GE\ interactions at high energies and luminosities, which can
considerably enrich the physics program of the \EPEM\ linear collider
TESLA. The parameters of the super--conducting collider TESLA: the
energy, the interval between electron bunches are particularly suited
for design and performance of the Photon Collider. 

This novel option requires only one new additional element: the
powerful laser, which can be built using modern laser technologies.
The optimum laser wave length for TESLA is about 1\MKM, which is
exactly the region of the most powerful developed solid state lasers.

The second interaction region and the detector may be very similar to those
for \EPEM\ collisions  and  can be also be used for study
of \EMEM\ or \EPEM\ interactions.


\clearpage




\begin{flushleft}

\end{flushleft}
\end{document}